\documentclass[aps,
	prd, 
	onecolumn, 
	a4paper, 
	10pt, 
	preprintnumbers,  
	tightenlines, 
	notitlepage,
	floatfix,
	nofootinbib,
	longbibliographyprd,
	superscriptaddress]{revtex4-2}

\input{CQGformat.input}

\usepackage[utf8]{inputenc}
\usepackage[dvipsnames]{xcolor}
\usepackage{booktabs}

\usepackage{amssymb}
\usepackage{bm}
\usepackage{bbm}
\usepackage{IEEEtrantools}
\usepackage{microtype}
\usepackage[toc,title,page,titletoc]{appendix}
\usepackage{amsmath}
\usepackage{amsthm}
\usepackage{mathrsfs}
\usepackage{enumitem}
\usepackage[T1]{fontenc}
\usepackage{xfrac}
\usepackage{lmodern}
\usepackage{hyperref}
\usepackage{mathbbol}

\usepackage[normalem]{ulem}
\usepackage{comment}

\usepackage{tikz}
\usetikzlibrary{shapes.geometric, arrows.meta, backgrounds,
  fit, graphs, quotes}

\tikzset{
  node distance=2cm,
  io/.style={trapezium, align=center, trapezium left angle=70,trapezium right angle=-70,minimum height=1cm, text centered, draw=black},
  process/.style={rectangle, align=center, minimum width=.75cm, minimum height=.75cm, text centered, draw=black},
  point/.style={circle,inner sep=0pt,minimum size=1pt, fill=black},
  op/.style={circle, minimum size=2pt, inner sep=0pt, text centered, draw=black},
  >={stealth},
  every new ->/.style={thick},
  flow/.style={thick, YellowOrange},
  line/.style={draw,thick,-stealth}
}

\renewcommand{\S}{{\mathcal S}}
\newcommand{\E}{{\mathcal E}}
\newcommand{\T}{{\mathcal T}}
\newcommand{\Lie}{{\mathcal L}}
\renewcommand{\O}{{\mathcal O}}
\renewcommand{\P}{\textrm{P}}
\renewcommand{\L}{\textrm{L}}
\def\RR{\textrm{R}}
\newcommand{\SDW}{\textrm{s}}
\newcommand{\RDW}{\textrm{r}}

\newcommand{\sI}{{\mathscr I}}
\newcommand{\sH}{{\mathscr H}}

\newcommand{\emm}{m}
\newcommand{\ess}{s}

\def\mb{\bar{m}}

\def\GHPwt{\ensuremath{\circeq}}

\def\rhb{\bar\rho}

\def\tab{\bar\tau}

\def\tap{\tau'}

\def\tabp{\bar{\tau}'}

\font\ec=ecrm0800 at 12pt

\def\th{\hbox{\ec\char'336}}
\def\edth{\hbox{\ec\char'360}}

\def\edthp{\hbox{\ec\char'360}'}

\def\Dbar{\tilde \edth}
\def\Nbar{\tilde \th'}

\newcommand{\chand}[2]{\,{}_{#1} \mathcal{L}_{#2}{}}

\def\tah{{\tau}^\circ}
\def\tabh{{\bar\tau}^\circ}
\def\Psih{ {\Psi}^\circ}

\def\Oh{{\Omega}^\circ}
\def\rhoph{{\rho}^{\prime \circ} }

\newcommand{\nphantom}[1]{\settowidth{\dimen0}{$#1$}\hspace*{-\dimen0}}

\renewcommand{\Re}{\operatorname{Re}}

\def\l{\ensuremath{\ell}}
\def\lm{{\ensuremath{\ell \emm}}}
\newcommand{\Y}[1]{\, {}_{#1}Y^{\phantom{\textrm{p}}}_{\ell \emm}}
\newcommand{\bY}[1]{\, {}_{#1}\bar{Y}^{\phantom{\textrm{p}}}_{\ell \emm}}
\newcommand{\Yp}[1]{\, {}_{#1}Y^\textrm{p}_{\ell \emm}}
\newcommand{\bYp}[1]{\, {}_{#1}\bar{Y}^\textrm{p}_{\ell \emm}}

\newcommand{\C}[1]{\lambda_{\ell,#1}}
\def\thA{\ensuremath{\theta,\varphi}}

\newcommand{\swsy}[1]{\,{}_{#1}Y_{\ell\, \emm}}
\newcommand{\swsY}[3]{\,{}_{#1}Y_{#2\, #3}}

\newcommand\widebar[1]{\mathop{\overline{#1}}}
\newcommand{\preindex}[2]{\,{}_{#2}#1}
\newcommand{\half}{\dfrac{1}{2}}

\newcommand\pd{\partial}
\newcommand{\dbd}[1]{\dfrac{\partial}{\partial #1}}

\newcommand{\lie}[1]{\mathcal{L}_#1}

\newcommand{\sforall}{\quad \forall}
\def\W{\zeta}
\def\Dr{\Delta r}

\def\Drm{\ensuremath{\Delta r_-}}
\def\ori{\textrm{S}}
\def\rp{\ensuremath{r_\textrm{p}}}

\def\rhom{\rho_\textrm{max}}

\def\dOm{\ensuremath{\delta^2(\Omega-\Omega_\textrm{p})}}
\def\Teff{T^\textrm{R}}

\newcommand{\nls}{\phantom{\frac{1}{\rho}}}

\def\ben{\begin{equation}}
\def\een{\end{equation}}
\def\bena{\begin{eqnarray}}
\def\eena{\end{eqnarray}}
\def\non{\nonumber \\}

\def\beq{\begin{equation}}
\def\eeq{\end{equation}}
\def\beqs{\begin{subequations}}
\def\eeqs{\end{subequations}}
\def\bal{\begin{align}}
\def\eal{\end{align}}
\def\no{\nonumber \\}

\definecolor{caribbeangreen}{rgb}{0.0, 0.8, 0.6}

\begin{document}

\title{New metric reconstruction scheme for gravitational self-force calculations}

\author{Vahid Toomani$^{1, 2}$, Peter Zimmerman$^3$, Andrew Spiers$^4$, Stefan Hollands$^{1, 2}$, Adam Pound$^4$, Stephen~R.~Green$^3$}

\address{$^1$ Institut f\"{u}r Theoretische Physik, Universit\"{a}t Leipzig
Br\"{u}derstrasse 16, D-04103 Leipzig, Germany}

\address{$^2$ Max Planck Institute for Mathematics in Sciences (MiS), Inselstra{\ss}e 22, 04103 Leipzig, Germany}

\address{$^3$ Max Planck Institute for Gravitational Physics (Albert Einstein Institute)
Am M\"{u}hlenberg 1, 14476 Potsdam, Germany}

\address{$^4$ School of Mathematical Sciences and STAG Research Centre, University of Southampton, Southampton, SO17 1BJ, UK}

\date{\today}

\begin{abstract}
Inspirals of stellar-mass objects into massive black holes will be important sources for the space-based gravitational-wave detector LISA. Modelling these systems requires calculating the metric perturbation due to a point particle orbiting a Kerr black hole. Currently, the linear perturbation  is obtained with a metric reconstruction procedure that puts it in a ``no-string'' radiation gauge which is singular on a surface surrounding the central black hole. Calculating dynamical quantities in this gauge involves a subtle procedure of ``gauge completion'' as well as cancellations of very large numbers. The singularities in the gauge also lead to pathological field equations at second perturbative order. In this paper we re-analyze the point-particle problem in Kerr using the corrector-field reconstruction formalism of Green, Hollands, and Zimmerman (GHZ). We clarify the relationship between the GHZ formalism and previous reconstruction methods, showing that it provides a  simple formula for the ``gauge completion''. We then use it to develop a new method of computing the metric in a more regular gauge: a Teukolsky puncture scheme. This scheme should ameliorate the problem of large cancellations, and by constructing the linear metric perturbation in a sufficiently regular gauge, it should provide a first step toward second-order self-force calculations in Kerr. Our methods are developed in generality in Kerr, but we illustrate some key ideas and demonstrate our puncture scheme in the simple setting of a static particle in Minkowski spacetime.
\end{abstract}

\maketitle

\section{Introduction}

When the gravitational-wave detector LISA launches in the early 2030s, one of its key sources will be extreme-mass-ratio inspirals (EMRIs)~\cite{Babak:2017tow,Barausse:2020rsu}, in which stellar-mass compact objects spiral into massive black holes in galactic nuclei~\cite{Amaro-Seoane:2020zbo}. Currently, the most viable method of modelling these systems is with gravitational self-force theory~\cite{Barack:2018yvs,Pound:2021qin}, an asymptotic approximation in the limit $m/M\ll 1$, where $m$ and $M$ are the companion's and black hole's respective masses. At leading order in this approximation, the companion can be approximated as a point mass moving on a geodesic, and the self-force method reduces to a venerable problem~\cite{Zerilli:1971wd} in general relativity: finding the linear metric perturbation generated by a point mass orbiting a black hole. The metric perturbation then exerts a self-force on the particle, accelerating it away from geodesic motion.

To be sufficiently accurate for LISA science, self-force calculations must be carried beyond linear perturbation theory, to second order in $m/M$~\cite{Barack:2018yvs,Pound:2021qin}. They also need to be carried out in the spacetime of an astrophysically realistic, spinning, Kerr black hole. This is now one of the central challenges in EMRI modelling. While the foundations of second-order self-force theory are well understood~\cite{Pound:2012nt,Upton:2021oxf}, and concrete calculations are underway in Schwarzschild spacetime~\cite{Pound:2019lzj,Miller:2020bft,Warburton:2021kwk}, there have not yet been any second-order self-force calculations in Kerr.

Computations of the metric perturbation in Kerr face the significant obstacle that the perturbative Einstein equation in Kerr is not separable in any known basis of functions, unlike in Schwarzschild. Traditionally, in linear perturbation theory this obstacle is avoided by instead solving the (fully separable) Teukolsky equation for the Weyl scalar $\psi_0$ (or $\psi_4$). In vacuum, $\psi_0$ contains nearly all the invariant information in a linear metric perturbation~\cite{Wald:1973jmp}, and from $\psi_0$ the metric perturbation can be reconstructed in a radiation gauge using a procedure due to Chrzanowski~\cite{Chrzanowski:1975wv} and Cohen and Kegeles~\cite{Kegeles:1979an} (CCK). The CCK reconstruction procedure can only be applied in vacuum regions~\cite{Ori:2002uv,Price:2006ke}, but it can nevertheless be used in first-order self-force calculations by carrying it out separately in the two regions $r<r_\textrm{p}(t)$ and $r>r_\textrm{p}(t)$, where $r$ is the Boyer-Lindquist radial coordinate and $r_\textrm{p}(t)$ is the particle's time-dependent orbital radius. The resulting metric perturbation is in a ``no-string radiation gauge''~\cite{Pound:2013faa,Merlin:2016boc,vandeMeent:2017fqk}. This approach, initiated in Refs.~\cite{Keidl:2010pm,Shah:2010bi} and formulated in detail in Refs.~\cite{Pound:2013faa,Merlin:2016boc,vandeMeent:2015lxa}, underlies almost all numerical first-order self-force calculations in Kerr (e.g.,~\cite{Shah:2012gu,vandeMeent:2015lxa,Colleoni:2015ena,vandeMeent:2016pee,vandeMeent:2016hel,vandeMeent:2017bcc}) as well as analytical, weak-field calculations (e.g.,~\cite{Kavanagh:2016idg,Bini:2018ylh,Bini:2019lkm,Antonelli:2020aeb}). Most prominently, van de Meent has used it to compute the first-order self-force on fully generic, inclined and eccentric bound orbits in Kerr~\cite{vandeMeent:2017bcc}. These implementations have been in the frequency domain, but work is also ongoing in the time domain~\cite{Barack:2017oir,Long:2021ufh}.

While it has enabled this progress at first order, the no-string gauge comes with a major drawback: it is singular over the entire surface $r=r_\textrm{p}(t)$~\cite{Pound:2013faa}, which we denote $\mathscr{S}_\textrm{p}$. Unlike in a regular gauge such as the Lorenz gauge, where the point-particle singularity is confined to the particle's worldline~\cite{Pound:2021qin}, in the no-string gauge the metric perturbation suffers from both jump discontinuities and Dirac-delta singularities over $\mathscr{S}_\textrm{p}$. These singularities represent a significant barrier to second-order calculations. The second-order metric perturbation is sourced by quadratic combinations of the first-order perturbation, and if the first-order perturbation contains distributional singularities, these quadratic combinations are ill defined. Without a well-defined field equation to start from, it is unclear how the second-order metric perturbation can be found. Even when restricted to first-order applications, the no-string construction has shortcomings that we will outline below.


In this paper we present a method for obtaining the metric perturbation in a more regular gauge. The basic approach (following~\cite{Keidl:2006wk}) is to split the perturbation into two parts: a puncture $h^\mathrm{P}_{ab}$ (the most singular part) and a residual field $h^\mathrm{R}_{ab}=h_{ab}-h^\mathrm{P}_{ab}$. The puncture is based on a local expansion (to some finite order) of the singular part of $h_{ab}$, with compact spatial support imposed by hand. Such expansions are known analytically at the level of the metric, and can be applied in a regular gauge (such as the Lorenz gauge) that confines the singularity to the worldline. The puncture field approximately solves the linearized Einstein equation with a point-particle stress-energy tensor, however errors are introduced by the finite order of the expansion and the imposition of compact spatial support. The idea of the ``puncture scheme''~\cite{Keidl:2006wk} is to correct this solution using the residual field, which satisfies the linearized Einstein equation with a less singular ``effective source''.

In calculating the residual field, however, one is faced with the same challenges encountered earlier in solving the linearized Einstein equation about Kerr---but now with an extended source. Our key insight is that this can be accomplished for $h^\mathrm{R}_{ab}$ using the ``corrector tensor'' reconstruction formalism recently developed by Green, Hollands, and Zimmerman~\cite{Green:2019nam} (GHZ). Like the CCK procedure, the GHZ procedure begins at the level of Weyl scalars, and reconstructs from these the metric perturbation. However, unlike the CCK procedure, the GHZ procedure accounts for nonvanishing stress-energy by adding a ``corrector tensor'', determined from the stress-energy via ordinary differential equations. (As part of this work, we calculate analytic expressions for the solution of these equations.) The puncture scheme gives a residual field with a ``softened'' string singularity compared to the CCK procedure, which we gauge-transform to lie in a neighborhood of $\mathscr{S}_\textrm{p}$ (the ``shadowless'' gauge). Our final metric perturbation therefore contains a singular puncture piece on the particle worldline, and a regular piece around the sphere. This provides a metric perturbation suitable as input for second-order calculations, and as we explain below, our approach offers other potential advantages over the current no-string construction.

We divide our presentation into three parts. In the first part, Sec.~\ref{sec:flat space illustration}, we review  CCK reconstruction, the no-string solution, and the GHZ corrector-field formalism. Using the simple model problem of a static point mass in flat spacetime, we carry out the GHZ procedure and illuminate its relationship to previous methods, highlighting (i) the half-string singularity structure in the GHZ solution, (ii) how it can be transformed to the no-string gauge and how it provides additional information beyond current no-string calculations, and (iii) why the gauge singularities in both the half-string solution and no-string solution render the second-order field equations ill defined. In the second part, comprising Secs.~\ref{sec:Kerr} and~\ref{sec:puncture}, we develop our Teukolsky puncture scheme. Section~\ref{sec:Kerr} first extends the flat-spacetime calculations to the case of a generic, spatially compact source in Kerr spacetime. Section~\ref{sec:puncture} then draws on that general treatment to formulate the puncture scheme. In the final part, Sec.~\ref{sec:return to flat space}, we return to the flat-spacetime model problem to demonstrate the scheme, showing that it yields the correct value for the type of quasi-invariant quantity typically calculated in self-force applications. We conclude with a summary discussion in Sec.~\ref{sec:discussion}.

To maintain focus on the core issues, we relegate some technical aspects of our treatment to appendices. In particular, \ref{app:GHP and Held} summarizes the Geroch-Held-Penrose (GHP) and Held refinements of the Newman-Penrose (NP) formalism; we rely on these throughout our analysis, but we strive as far as possible to present our final results in a form that does not rely on the reader being intimately familiar with them. 

{\bf Notations and conventions:} We adopt a mostly negative metric signature, geometric units, and Wald's notation~\cite{Wald:1984rg} for abstract indices ($a,b,c,\ldots$). A glossary of commonly used symbols is provided at the end of the paper.

\section{Metric reconstruction with a particle source: guide to the literature with an illustration in Minkowski space}
\label{sec:flat space illustration}

Our overarching context is an asymptotic expansion of the metric, of the form $g_{ab}+\varepsilon h_{ab}+\varepsilon^2 j_{ab}+O(\varepsilon^3)$, where $g_{ab}$ is a Kerr metric of mass $M$ and angular momentum $J=Ma$, and $\varepsilon$ is a formal parameter that counts powers of the mass ratio $m/M$. $h_{ab}$ satisfies the linearized Einstein equation,
\begin{equation}\label{eq:Einstein}
\mathcal E_{ab}(h) = T_{ab},
\end{equation}
with a point-mass source 
\begin{equation}\label{eq:Tab}
T^{ab} = \mu\int v^a v^b \delta^4(x,x_p(\tau)) d \tau.
\end{equation}
Here we have absorbed the usual $8\pi$ into the mass $\mu=8\pi m$. $\tau$ is the particle's proper time, $x_\textrm{p}(\tau)$ is its worldline, $v^a$ is its four-velocity, and $\delta^4(x,x_\textrm{p}(\tau))$ is the covariant delta function defined so that $\int \delta^4(x,y) f(y) dV = f(x)$, where $dV$ is the covariant integration element. For applications to EMRIs, at leading order $x_\textrm{p}$ is approximated as a bound geodesic in $g_{ab}$. 
Given the first-order perturbation, at points away from the particle, the second-order perturbation should then satisfy the second-order vacuum Einstein equation,
\begin{equation}\label{eq:2nd Einstein}
\mathcal E_{ab}(j) = - G^{(2)}_{ab}(h,h),
\end{equation}
where $G^{(2)}_{ab}(h,h)\sim \partial h\partial h+h\partial\partial h$; see Eq.~(4) of Ref.~\cite{Pound:2021qin}. Our primary goal is to solve Eq.~\eqref{eq:Einstein} for $h_{ab}$ in a gauge that is sufficiently well behaved for us to ultimately solve Eq.~\eqref{eq:2nd Einstein} for $j_{ab}$. 

Moreover, we wish to avoid solving Eq.~\eqref{eq:Einstein} directly and instead reconstruct $h_{ab}$ from a solution to the spin-weight-($+2$) Teukolsky equation,
\begin{equation} 
\label{eq:teuk}
{\mathcal O} \psi_0 = \S^{ab}T_{ab},
\end{equation}
where  $\psi_0={\cal T}^{ab}h_{ab}$. ${\mathcal O}$, $\S^{ab}$, and ${\cal T}^{ab}$ are all linear, second-order differential operators given (along with ${\cal E}_{ab}$) in~\ref{sec:operators}. We do not need their explicit form for the present discussion, but we note that since Eq.~\eqref{eq:teuk} holds for any $h_{ab}$, it implies the operator identity~\cite{Wald:1978vm}
\begin{equation}\label{eq:SE=OT}
{\cal O}{\cal T}^{cd} = \mathcal{S}^{ab}{\cal E}_{ab}{}^{cd}
\end{equation}
and its adjoint ${\cal T}^\dagger_{ab}{\cal O}^\dagger = {\cal E}_{ab}{}^{cd}\mathcal{S}^\dagger_{cd}$, where we have used the fact that ${\cal E}_{ab}$ is self-adjoint and defined ${\cal E}_{ab}{}^{cd}h_{cd}={\cal E}_{ab}(h)$ to make the index structure clear.

In this section, we review existing methods of solving Eq.~\eqref{eq:Einstein} using metric reconstruction and demonstrate the GHZ method. This will set the stage for the remainder of the paper.

\subsection{Metric reconstruction and completion procedures}
\label{sec:CCK and Ori}

\subsubsection{CCK-Ori reconstruction}

The traditional CCK reconstruction procedure specializes to vacuum, wherein we seek to solve $\mathcal E_{ab}(h) = 0$ with a given initial/boundary condition on $h_{ab}$. From the solution to the vacuum Teukolsky equation, ${\mathcal O}\psi_0 = 0$, with the corresponding initial/boundary conditions, one first obtains a spin-weight-$(-2)$ Hertz potential $\Phi$ by solving the inversion relation
\begin{equation} 
\label{eq:th4}
\th^4 \widebar{\Phi} = -2 \psi_0,
\end{equation}
where an overline indicates complex conjugation. For an appropriate choice of tetrad, the operator $\th$, defined in Eq.~\eqref{eq: GHP ops}, reduces to a derivative along an outgoing principle null vector $l^a$, meaning Eq.~\eqref{eq:th4} is a fourth-order ordinary differential equation (ODE) along outgoing null curves. Then, from $\Phi$ one obtains a metric perturbation
\begin{equation}\label{eq:CCK}
\hat{h}_{ab} = 2\Re(\S^\dagger_{ab} \Phi),
\end{equation}
which satisfies the ingoing radiation gauge (IRG) conditions $\hat{h}_{ab}l^b=0=g^{ab}\hat{h}_{ab}$.

For $\hat{h}_{ab}$ to be a solution to the Einstein equation with the given initial/boundary conditions, $\Phi$ must satisfy the same boundary conditions as $\psi_0$: e.g., outgoing wave boundary conditions at infinity or ingoing wave boundary conditions at the horizon. Given this condition, Eq.~\eqref{eq:th4} enforces that $\Phi$ is also a solution to the adjoint Teukolsky equation, ${\mathcal O}^\dagger\Phi = 0$~\cite{Ori:2002uv}. As first pointed out by Wald~\cite{Wald:1978vm}, it then follows from the adjoint of the identity~\eqref{eq:SE=OT} that
\begin{equation}
{\cal E}_{ab}(\hat h) = 2\Re (\mathcal{T}^\dagger_{ab} \mathcal{O}^\dagger\Phi) = 0.
\end{equation}
Moreover, it can be shown that ${\cal T}^{ab}\Re(\mathcal{S}^\dagger_{ab}\Phi)=-\frac{1}{4}\th^4\overline\Phi$, meaning Eq.~\eqref{eq:th4} enforces that $\hat h_{ab}$ has the same Weyl scalar as $h_{ab}$: ${\cal T}^{ab}\hat h_{ab} = {\cal T}^{ab}h_{ab}=\psi_0$. A theorem due to Wald~\cite{Wald:1973jmp} then implies that (up to gauge) $\hat{h}_{ab}$ can only differ from $h_{ab}$ by a perturbation $\dot{g}_{ab}$ toward another Kerr metric, which can be absorbed into redefinitions of the background parameters $M$ and $a$.

In Ref.~\cite{Ori:2002uv}, Ori provided the first analysis of the CCK procedure in the nonvacuum case. He obtained two explicit, closed-form solutions to Eq.~\eqref{eq:th4} in terms of modes of $\psi_0$ and modes of the Teukolsky source $\mathcal{S}^{ab}T_{ab}$, for any spatially compact source. However, these solutions to Eq.~\eqref{eq:th4} do not provide a solution to the Einstein equation. Assuming $\psi_0$ satisfies retarded boundary conditions, and restricting to the case of a point mass, Ori showed that if one solves Eq.~\eqref{eq:th4} subject to outgoing wave conditions at infinity, obtaining a solution $\Phi^+$, then $\Phi^+$ becomes singular along a string emanating from the particle to the horizon; likewise if one solves Eq.~\eqref{eq:th4} subject to ingoing wave conditions at the horizon, obtaining a solution $\Phi^-$, then $\Phi^-$ becomes singular along a string emanating from the particle to infinity. Furthermore, these solutions to Eq.~\eqref{eq:th4} fail to satisfy ${\mathcal O}^\dagger\Phi^\pm = 0$ along the string, meaning the perturbations 
\begin{equation}\label{eq:Ori hab}
\hat{h}^\pm_{ab} = 2\Re(\S^\dagger_{ab} \Phi^\pm) 
\end{equation}
also fail to satisfy the vacuum Einstein equation there. In short, CCK reconstruction fails in nonvacuum. This failure is associated with the fact that no solution to the nonvacuum Einstein equation~\eqref{eq:Einstein} can satisfy the IRG conditions unless $T_{ab} l^a l^b=0$~\cite{Price:2006ke}, and no solution can be put in the particular CCK form~\eqref{eq:CCK} unless $T_{ab}l^b=0$~\cite{Green:2019nam}.

\subsubsection{The no-string solution}

The no-string construction to some extent overcame the problem in Ori's method by trading the string singularities for a singularity on a sphere at the particle's orbital radius. Let $r_\textrm{p}(u)$ be the orbital radius as a function of retarded time, and let $\mathscr{S}_\textrm{p}$ be the surface $r=r_\textrm{p}(u)$. Construct $\hat h^+_{ab}$ outside $\mathscr{S}_\textrm{p}$ and $\hat h^-_{ab}$ inside $\mathscr{S}_\textrm{p}$. In their respective regions, each of them is a vacuum solution, and from Wald's theorem, each can only differ from the complete solution by a perturbation of the form $\dot{g}_{ab}$. The no-string solution is then obtained by adding these ``completion'' terms and gluing the two vacuum solutions together at $\mathscr{S}_\textrm{p}$:
\begin{equation}\label{eq:no-string hab}
h^\textrm{N}_{ab} = (\hat h^+_{ab}+\dot g^+_{ab})\,\Theta^+ + (\hat h^-_{ab}+\dot g^-_{ab})\,\Theta^- + D_{ab}\,\delta(\mathscr{S}_\textrm{p}), 
\end{equation}
where $\Theta^\pm \equiv \Theta\left(\pm[r-r_\textrm{p}(u)]\right)$ are Heaviside functions, and $\delta(\mathscr{S}_\textrm{p})\equiv\delta(r-r_\textrm{p}(u))$ in Kerr-Newman coordinates. To satisfy the nonvacuum Einstein equation, the mass and angular momentum in $h^\textrm{N}_{ab}$ must jump by the particle's energy $E$ and azimuthal orbital angular momentum $L$ when crossing $\mathscr{S}_{\mathrm{p}}$, implying the completion terms must satisfy~\cite{Merlin:2016boc,vandeMeent:2017fqk}
\begin{equation}
\dot g^+_{ab} - \dot g^-_{ab} = \frac{\partial g_{ab}}{\partial M}E + \frac{\partial g_{ab}}{\partial J}L.
\end{equation}
Typically, $\dot g^-_{ab}$ is absorbed into $g_{ab}$, leaving only $\dot g^+_{ab}$. 

The field ~\eqref{eq:no-string hab} is highly singular on $\mathscr{S}_{\mathrm{p}}$, containing both a jump discontinuity and a Dirac delta there. Additionally, while the delta function is generically required in order to satisfy the Einstein equation~\cite{Pound:2013faa,Shah:2015nva}, the current no-string reconstruction and completion procedure does not provide a ready means of finding the coefficient $D_{ab}$, which has only ever been calculated in the case of a static particle in Minkowski spacetime~\cite{Pound:2013faa}. This coefficient is not needed for first-order self-force calculations because the self-force and other physical quantities can be found by taking limits to the particle from off of $\mathscr{S}_\textrm{p}$~\cite{Pound:2013faa}. But as we will establish below, the time dependence of $D_{ab}$ provides an important diagnostic on the behaviour of the solution. 

In general, one further adjustment is made to Eq.~\eqref{eq:no-string hab}: a gauge perturbation generated by a discontinuous vector $\xi^a =\xi^a_- \Theta^-$ is added to ensure that coordinate frequencies have the same meaning on either side of $\mathscr{S}_\textrm{p}$~\cite{Shah:2015nva,vandeMeent:2017fqk,Bini:2019xwn}. This condition is enforced by imposing continuity of the stationary, axisymmetric pieces of the Boyer-Lindquist components $h_{ab}t^a t^b$ and $h_{ab}t^a \varphi^b$ across $\mathscr{S}_\textrm{p}$, for example. Such a gauge transformation, sometimes called the ``gauge completion''~\cite{vandeMeent:2016hel}, is required to compute dynamical effects on the particle's orbit. Since the gauge perturbation $\Lie_\xi g_{ab}$ includes a derivative of $\Theta^-$, it introduces another delta function, altering the value of the unknown coefficient $D_{ab}$.

Finally, we note one more crucial aspect of the no-string solution. In the frequency domain, a point-particle source is not one-dimensional. Instead, it fills the orbital libration region between the particle's minimum and maximum orbital radius. Applying CCK reconstruction is impossible inside the libration region, and no-string calculations therefore rely on the method of extended homogeneous solutions~\cite{vandeMeent:2015lxa,vandeMeent:2017bcc}, in which the vacuum solutions outside the libration region are analytically extended into it~\cite{Barack:2008ms}. In practice, this method can involve cancellations of very large numbers~\cite{vandeMeent:2017bcc}, introducing the computational expense of arbitrary-precision arithmetic. We comment further on this in Sec.~\ref{sec:discussion}.

\subsubsection{The GHZ formalism}

A critical weakness of the current no-string construction is that it cannot be applied to problems with spatially extended sources, which limits its utility even in the context of first-order self-force applications. For example, it is inapplicable in a puncture scheme, one of the standard methods of self-force theory~\cite{Wardell:2015kea}. With the GHZ formalism, we now have far more flexibility in tackling nonvacuum problems. The GHZ procedure supplements Ori's reconstructed metric perturbations~\eqref{eq:Ori hab} with a {\em corrector tensor} $x^\pm_{ab}$:\footnote{GHZ also derived the adjoint Teukolsky equation satisfied by Ori's Hertz potential, $\O^\dagger\Phi = \eta$, where the source is the solution to $\Re({\cal T}^\dagger_{ab}\eta) = T_{ab} - {\cal E}_{ab}(x)$. An alternative approach, presented by GHZ, is to solve these equations rather than Eqs.~\eqref{eq:teuk} and \eqref{eq:th4}. In this paper we  opt to use Eqs.~\eqref{eq:teuk} and \eqref{eq:th4} because (i) doing so allows us to exploit the existing solutions to those equations, and (ii) the source $\eta$ in the adjoint Teukolsky equation for $\Phi$ has the undesirable property of being spatially noncompact even if $T_{ab}$ is not.} 
\begin{equation}\label{eq:GHZ}
h^\pm_{ab} = 2\Re(\S^\dagger_{ab} \Phi^\pm) + x^\pm_{ab}.
\end{equation}
Unlike the CCK-Ori perturbation $\hat h^\pm_{ab}$, these perturbations satisfy ${\cal E}_{ab}(h^\pm)=T_{ab}$. The corrector tensor is chosen to satisfy $x^\pm_{ab}l^b=0$ but not the tracefree condition $g^{ab}x^\pm_{ab}=0$; because of its nonzero trace, it can be made to satisfy the specific pieces of the Einstein equation that $\hat{h}^\pm_{ab}$ cannot, $[T_{ab}-{\cal E}_{ab}{}^{cd}(x^\pm_{cd})]l^b = 0$. Remarkably, these pieces of the Einstein equation can be put in the form of a sequence of three ODEs along the integral curves of $l^a$, given in~\eqref{eq:xmmb}--\eqref{eq:xnn} below. $x^+_{ab}$ is the solution that vanishes at infinity $\mathscr{I}^+$ (and everywhere outside $\mathscr{S}_\textrm{p}$ for a point particle); $x^-_{ab}$ is the solution that vanishes at the horizon $\mathscr{H}^-$ (and everywhere inside $\mathscr{S}_\textrm{p}$ for a point particle). Therefore, in the GHZ procedure, the only partial differential equation one must solve is the Teukolsky equation for $\psi_0$. Ori's solution to the ODE~\eqref{eq:th4} then provides the modes of $\Phi$ in closed form in terms of modes of $\psi_0$, and straightforward integration of three more ODEs yields $x_{ab}$.

As we illustrate below, the GHZ solutions $h^\pm_{ab}$ for a point mass are precisely the half-string solutions characterized in Ref.~\cite{Pound:2013faa}. The GHZ procedure provides the first systematic method of obtaining these solutions. Starting from them, we will show how to (i) find a gauge transformation to the no-string solution~\eqref{eq:no-string hab}, including the Dirac-delta coefficient $D_{ab}$, (ii) use $D_{ab}$ to determine the ``gauge completion'' that settles the coordinate frequencies, and (iii) extend the no-string construction to generic spatially compact sources. Although our description works with $\psi_0$ and an IRG metric perturbation, we note that an essentially identical construction applies if we work with $\psi_4$ and an ORG (outgoing radiation gauge) perturbation, with $l^a$ replaced by an ingoing principal null vector $n^a$. We also note that as originally formulated, the GHZ formalism is restricted to sources that are compact in time, which excludes our case of interest. \ref{sec:noncompact sources} describes how we remove this limitation.

\subsection{Model problem: static particle in flat spacetime}\label{sec:static particle}

To give the reader a more concrete grasp of the form of the GHZ solution, we follow Refs.~\cite{Keidl:2006wk,Pound:2013faa} by considering the simplest possible scenario: a static point particle in flat spacetime. This simple example captures the essential features of the solution, and it will serve as a model for our treatment of the general problem in Kerr. We specifically calculate the half-string solution $h^-_{ab}$, which is regular at the origin; the regular-at-infinity solution $h^+_{ab}$ can be easily found in the same way.

We adopt retarded coordinates $(u,r,\theta,\varphi)$ and  the tetrad
\begin{subequations}
\label{eq:NP tet flat}
\begin{align}
l &= \dbd{r}, \label{eq:lNP flat} \\
n &= \dbd{u}-\frac12 \dbd{r}, \label{eq:nNP flat}  \\
m &=\frac{1}{\sqrt{2}r}\left( \dbd{\theta} + i \csc{\theta} \dbd{\varphi} \right) \label{eq:mNP flat}
\end{align}
\end{subequations}
for the calculations in this section. We place the particle at a static position $x^a_\textrm{p}(u) = (u, \rp, \theta_\textrm{p}, \varphi_\textrm{p})$, with four-velocity $v = \dbd{u}$, such that the only nonzero frame components of the stress-energy~\eqref{eq:Tab} are
\begin{align}\label{eq:frame Tla}
T_{ll} = 2T_{ln} = 4T_{nn} = \frac{\mu}{\rp^2} \delta(r-r_\textrm{p}) \dOm,
\end{align}
where $\dOm \equiv\delta(\cos\theta-\cos\theta_\textrm{p})\delta(\varphi-\varphi_\textrm{p})$. 

We calculate the metric perturbation as a sum of spin-weighted spherical harmonics $\swsy{s}(\thA)$, making frequent use of the identities in~\ref{sec:spin-weighted harmonics} and the spin-raising and lowering operators $\Dbar$ and $\Dbar'$ given in Eq.~\eqref{eq:Hop}.  For convenience we define the shorthand $\Yp{s} \equiv \Y{s}(\theta_\textrm{p},\varphi_\textrm{p})$, and we note in advance that all sums over the mode number $m$ run from $-\l$ to $+\l$. Since the calculations in this section are straightforward, we keep them terse and focus on their conclusions. For readers wishing to reproduce them, we note that with our choice of tetrad, the only nonvanishing GHP and NP spin coefficients are
$\rho = -1/r$, $ \rho' = 1/(2r)$, and $\beta=\beta'= \cot\theta/(2\sqrt{2} r)$.

\subsubsection{CCK-Ori metric reconstruction}\label{sec:CCK-Ori flat}

We begin by solving the Teukolsky equation~\eqref{eq:teuk} for $\psi_0$, assuming a stationary solution.
With the standard mode decomposition
$\psi_0 =  \sum_{\l\geq 2, m}\,  \psi_0^{\lm}(r) \, \Y{2}(\thA) $,
the radial equation reads
\begin{equation}\label{eq:psi0 eqn flat}
r^{-6} \pd_r (r^6 \pd_r \psi_0^{\lm} ) -(\l-2)(\l+3)  \psi_0^{\lm} = {}_2S_{\lm} \delta(r-\rp),
\end{equation}
where ${}_s S_{\lm}=\mu \bYp{} \C{-s}/(4r_\textrm{p}^4)$, and $\lambda_{\l,s}$ is defined in Eq.~\eqref{eq:Ys}. The solution that is regular at the origin and infinity is 
\begin{equation}\label{eq:psi0lm sol}
    \psi_0^{\lm} = -\frac{{}_2C_{\lm}(r)}{4r^2} \qquad \text{with}\quad {}_s C_{\lm}(r) \equiv \frac{4r_\textrm{p}^4\, {}_s S_{\lm}}{2\l+1}\frac{r_<^\l}{r_>^{\l+1}}.
\end{equation}
Here we have defined $r_{<} \equiv \min{(r,\rp)}$ and  $r_{>} \equiv \max{(r,\rp)}$.

Next, we find the regular-at-the-origin Hertz potential $\Phi = \sum_{\l\geq2,m}\Phi_{\lm}(r) \, \Y{-2}(\thA)$. Appealing to Eq.~\eqref{eq:bar flm} and $(-1)^m\overline{{}_s S_{\l,-m}}={}_s S_{\lm}$, we can write $\bar\psi^\lm_0=\psi^\lm_0$ and integrate Eq.~\eqref{eq:th4} as $\Phi_{\lm} = -2\int^r_0 d r_4\int^{r_4}_{0}d r_3\int^{r_3}_{0} d r_2 \int^{r_2}_{0}d r_1 \psi^{\lm}_{0}(r_1)$. We then obtain
\begin{equation}\label{eq:Phi modes}
\Phi_{\lm} = r^2\, {}_{-2}C_{\lm}(r) + \Phi^{\ori}_{\lm}(\Delta r)\Theta^+,
\end{equation}
where $\Delta r\equiv r-r_\textrm{p}$. The first term corresponds to the no-string Hertz potential  $\Phi^\textrm{N} = \Phi^+\Theta^+ + \Phi^-\Theta^-$ used in current self-force calculations.
In each of the two regions $r\neq r_\textrm{p}$, it satisfies the vacuum Teukolsky equation and the inversion relation $\partial_r^4 \widebar{\Phi} ^\pm = -2\psi_0$, and it is regular at $r=0$ and $\infty$. The second term, $\Phi^\ori_{\lm}$, contains the half-string extending from $r=r_\textrm{p}$ to infinity. It is a homogeneous solution to $\partial_r^4 \widebar{\Phi}^\ori = 0$ given by the third-order polynomial
\begin{equation}\label{eq:PhiS modes}
\Phi^{\ori}_{\lm} = 4r_\textrm{p}^2\,{}_{-2}S_{\lm}\sum_{j=1}^3 b^\l_j (\Delta r)^j,
\end{equation}
with constant coefficients $b_1^\l=r_\textrm{p}^2$, $b_2^\l=r_\textrm{p}$, and $b_3^\l=\l (\l+1)/6$.

Equation~\eqref{eq:PhiS modes} is precisely the flat-spacetime reduction of the singular Hertz potential derived by Ori~\cite{Ori:2002uv}. The fact that it is singular can be deduced from its large-$\l$ behavior: as pointed out in Ref.~\cite{Pound:2013faa}, the sum $\sum_{\lm}\Phi^{\ori}_{\lm}\,\Y{-2}$ diverges not only on the string, but at every point in the region $r>r_\textrm{p}$.\footnote{To see this, note that $\C{2}\sim \l^{-2}$ and $b^\lm_3\sim \l^2$ imply $\Phi^\ori_{\lm}\sim \Yp{}$. Letting $\gamma$ be the angle between $(\theta_\textrm{p},\varphi_\textrm{p})$ and $(\thA)$, and appealing to Eq.~\eqref{eq:addition theorem}, we then have $ \sum_m \Phi^\ori_{\lm}{}_{-2}Y_{\lm}\sim \l P_\l(\cos\gamma) \sim \l^{1/2}$ for $\gamma\neq0$ or $\sim \l$ for $\gamma=0$.} Based on this, Ref.~\cite{Pound:2013faa} concluded that this term was ill defined. However, that conclusion was incorrect. While the sum is ill defined as an ordinary function, it is well defined as a distribution, and in the region $r>r_\textrm{p}$ it satisfies the equation identified by GHZ, $\O^\dagger\Phi^\ori = \eta$, where $\eta$ is supported only on the string at $(\thA)=(\theta_\textrm{p},\varphi_\textrm{p})$. We explicitly evaluate the sum $\sum_{\lm}\Phi^{\ori}_{\lm}\,\Y{-2}$ in~\ref{sec:mode sums} and find that the result is smooth everywhere except on the string. 

The last step in the CCK-Ori procedure is to reconstruct the IRG metric $\hat h_{ab}=2\Re(\S^\dagger_{ab}\Phi)$. The result again splits into no-string and string pieces,
\begin{align}\label{eq:hhat}
   \hat h_{ab} = \hat{h}^\textrm{N}_{ab} + \hat{h}^\ori_{ab}\,\Theta^+.
\end{align}
Away from $\mathscr{S}_\textrm{p}$, these satisfy $\hat{h}^\textrm{N}_{ab}=2\Re(\S^\dagger_{ab}\Phi^\textrm{N})$ and $\hat{h}^\ori_{ab}=2\Re(\S^\dagger_{ab}\Phi^\ori)$. The nonzero frame components of the no-string piece are given by
\begin{subequations} \label{eqs:hhat}
   \begin{align}
         \hat{h}^\textrm{N}_{nn} &= - \sum_{\ell\geq2,m} {}_0C_{\lm}(r)\Y{}, \label{eq:hhatnn in}\\ 
         \hat{h}^\textrm{N}_{n\mb} &= \frac{1}{\sqrt{2}}\sum_{\ell\geq2,m}{}_{-1}C_{\lm}(r)\left[\l\Theta^- -(\l+1)\Theta^+\right]\Y{-1}, \label{eq:hhatnmb in}\\
          \hat{h}^\textrm{N}_{\mb\mb} &= - \sum_{\l\geq 2, m}(\l+2)(\l-1){}_{-2}C_{\lm}(r)\Y{-2}, \label{eq:hhatmbmb in}
    \end{align}
\end{subequations}
and their complex conjugates. $\hat{h}^\textrm{N}_{ab}$ is a solution to the linearized vacuum Einstein equation in the regions $r\neq r_\textrm{p}$ and is regular at $r=0$ and $\infty$. The nonzero frame components of the string piece are
   \begin{subequations} \label{eqs:hhat S}
        \begin{align} 
          \hat{h}_{nn}^\ori &=  \sum_{\l\geq 2, m} {\cal F}_{nn}\left(\bYp{} \Y{}\right), \label{eq:hhatnn out} \\
          \hat{h}_{n\mb}^\ori &=  \frac{4r_\textrm{p}^4}{\sqrt{2}r}  \sum_{\l\geq 2, m}{}_{-1}S_{\lm}\,\Y{-1} + \sum_{l\geq 2, m}{\cal F}_{n\mb}\left(\bYp{} \Y{}\right)
          \label{eq:hhatnmb out},\\
          \hat{h}_{\mb\mb}^\ori &= -\frac{4r_\textrm{p}^3\Dr}{r} \sum_{\l\geq 2, m}(\l+2)(\l-1){}_{-2}S_{\lm}\,\Y{-2}, \label{eq:hhatmbmb out}
        \end{align}    
    \end{subequations}
and their complex conjugates. 

We have suggestively written two of the terms as linear differential operators ${\cal F}_{ab}$ acting on $\sum \bYp{}\Y{}$, with ${\cal F}_{ab}$ given in Eq.~\eqref{eq:Fab}. By virtue of the completeness relation~\eqref{eq:completeness}, these quantities can be written in terms of $\dOm$, and they will cancel terms in the corrector tensor $x_{ab}$. The other terms in $\hat{h}^\ori_{ab}$, which we have written in terms of the source modes ${}_s S_{\lm}$, will remain in the final half-string metric $h_{ab} = \hat h_{ab}+x_{ab}$. We discuss their singularity structure below.

\subsubsection{Corrector tensor}

We now complete the GHZ procedure by calculating the corrector tensor $x_{ab}$. Its only nonzero components are $x_{m\mb}$, $x_{nm}=\bar{x}_{n\mb}$, and $x_{nn}$, and they satisfy the hierarchical sequence of ODEs~\eqref{eq:xmmb}--\eqref{eq:xnn}, which here reduce to
\begin{align}
\pd_r (r^2 \pd_r x_{m \mb} ) &=   r^2 T_{ll},\label{eq:xR_mmb eq flat}\\
\frac{1}{2}\pd_r [r^{-2} \pd_r(r^2 x_{nm}) ] &= \frac{1}{2r} \pd_r \Dbar  x_{m\mb},\label{eq:xnm flat}\\
- r^{-2} \pd_r(r x_{nn}) &= -\frac{1}{2}\left( \frac{2}{r^2}\Dbar'\Dbar + \pd_r^2 + \frac{4}{r}\pd_r + \frac{2}{r^2} \right) x_{m\mb} \nonumber \\
&\quad + \frac{1}{2r} \left(\pd_r+\frac{3}{r} \right) (\Dbar' x_{nm} + \Dbar x_{n\mb} ) + T_{ln}.\label{eq:xnn flat}
\end{align}
These can be integrated in sequence for the three components. 

Integrating from the origin (i.e., imposing $x_{ab}|_{r=0}=0$), we find 
\begin{equation}\label{eq:x inner outer}
x_{ab} = x_{ab}^\textrm{S} \,\Theta^+ \qquad\text{with}\quad x^\textrm{S}_{ab} = - {\cal F}_{ab}[\dOm].
\end{equation}
This is made up of angular operators acting on $\dOm$, with a rational dependence on $r$; again refer to the definition~\eqref{eq:Fab} of ${\cal F}_{ab}$. We highlight, in particular, 
\begin{equation} \label{eq:x frame}
x^\textrm{S}_{m\mb} = \frac{\mu\Dr} {rr_\textrm{p}}  \dOm,
\end{equation}
which represents the only contribution to the trace $g^{ab}h_{ab}$.
Unlike $\hat h^\ori_{ab}$, which has support over the entire region $r>r_\textrm{p}$, $x_{ab}$ is supported only on the string. This is because $\Phi$ is constructed through radial integrations of $\psi_0$, which is nonzero everywhere, while $x_{ab}$ is constructed through radial integrations of $T_{ab}$, which is supported only at the particle.

\subsubsection{Total GHZ solution and transformation to the no-string gauge}\label{sec:S*Phi+x metric}

The total  metric  $h_{ab}=\hat{h}_{ab} + x_{ab}$ once again divides into no-string and half-string pieces,
\begin{equation}\label{eq:GHZ flat}
h_{ab} = \hat{h}^\textrm{N}_{ab} + h^\textrm{S}_{ab}\,\Theta^+,
\end{equation}
where $h^\textrm{S}_{ab} = \hat{h}^\textrm{S}_{ab} + x^\textrm{S}_{ab}$. After decomposing the angular delta functions in $x^\textrm{S}_{ab}$ using Eq.~\eqref{eq:completeness}, we find a near cancellation between the terms involving ${\cal F}_{nn}$ and ${\cal F}_{n\mb}$ in $\hat h^\textrm{S}_{ab}$ and those in $x^\textrm{S}_{ab}$. However, the cancellation is inexact because $\hat h^\textrm{S}_{ab}$ does not contain $\l=0,1$ modes. We are therefore left with the nonzero components
\begin{subequations} \label{eqs:h out}
        \begin{align}
            h^\textrm{S}_{nn} & = x^{\ell=0,1}_{nn},\\
            h^\textrm{S}_{n\mb} & = \frac{4r_\textrm{p}^4}{\sqrt{2}r}  \sum_{\l\geq 2, m}{}_{-1}S_{\lm}\,\Y{-1} +x^{\ell=0,1}_{n\mb},\label{eq:hnmb out}
        \end{align}    
\end{subequations}
$h^\textrm{S}_{\mb\mb} = \hat h^\textrm{S}_{\mb\mb}$, $h^\textrm{S}_{m\mb} = x^\textrm{S}_{m \mb}$, and their complex conjugates. 

The surviving pieces of $\hat{h}^\textrm{S}_{ab}$ exhibit a power-law divergence on the string, which may be found by evaluating the sums over $\l$ using the method in~\ref{sec:mode sums}. This half-string singularity has the form discussed in previous literature and displayed in Table I (locally near the particle in a generic  spacetime) and Eqs.~(181)--(183) (in flat spacetime) of Ref.~\cite{Pound:2013faa}. Explicitly, it diverges as $1/\gamma^2$, where $\gamma$ is the angle between $(\theta_{\mathrm{p}},\varphi_{\mathrm{p}})$ and $(\theta,\varphi)$. $x_{ab}$ has made two net additional contributions to the total metric: the $\ell=0,1$ modes; and the delta function~\eqref{eq:x frame} in the trace $g^{ab}h_{ab}$ (mentioned in text in Ref.~\cite{Pound:2013faa} but not explained). Neither of these can ever arise from the Hertz potential because $\Phi$ can only contain modes with $\l \geq |s|=2$ and because $g^{ab}\S^\dagger_{ab}\Phi=0$.

How can we bring the metric~\eqref{eq:GHZ flat} to the no-string form~\eqref{eq:no-string hab} currently used in self-force calculations? From the general argument reviewed in Sec.~\ref{sec:CCK and Ori}, we know that in the vacuum region $r>r_\textrm{p}$, up to gauge, $h_{ab}$ can only differ from $\hat h^\textrm{N}_{ab}$ by $\l=0,1$ perturbations. It follows that the string field is pure gauge except its $\l=0,1$ piece, and we can write it as
\begin{equation}
h^\textrm{S}_{ab} = x^{\l=0,1}_{ab} + \lie{{\xi^\textrm{S}}}g_{ab}
\end{equation}
for some  vector $\xi^\textrm{S}$. A no-string solution $h^\textrm{N}_{ab} = h_{ab}-\Lie_{\xi}g_{ab}$, with $\xi^a = {\xi^\textrm{S}}^a \Theta^+$, is then
\begin{equation}\label{eq:h no string}
h^\textrm{N}_{ab} = \hat h^\textrm{N}_{ab}  + x^{\l=0,1}_{ab}\Theta^+ - 2{\xi^\textrm{S}}^c \,  g_{c(a}r_{b)} \,\delta(\Dr),
\end{equation}
where $r_adx^a=dr$. It is straightforward to check that
\begin{subequations}\label{eq:xi modes}
\begin{align}
\xi^\textrm{S}_l = 2\xi^\textrm{S}_n &= \mu\sum_{\l\geq2,m}\C{1}^2 \bYp{} \Y{},\\
\xi^\textrm{S}_{\mb} &= - \frac{\mu\Delta r }{\sqrt{2}\rp}\sum_{\l\geq2,m}\C{1} \bYp{} \Y{-1}.
\end{align}
\end{subequations}
This vector is unique up to the addition of Killing vectors (which trivially contribute nothing to $\lie{{\xi^\textrm{S}}} g_{ab}$).

Equation~\eqref{eq:h no string} is in a no-string gauge but not yet in the form used in practice. We can see from Eq.~\eqref{eq:Fab} that $x_{ab}\sim r$ at large $r$, implying that~\eqref{eq:h no string} is not in an asymptotically flat gauge. More precisely, $x^{\l=0}_{ab}\sim r^0$ and $x^{\l=1}_{ab}\sim r$. Removing these gauge artifacts will put our no-string solution in the form used in current calculations. It will also provide a new means of determining the gauge completion alluded to in Sec.~\ref{sec:CCK and Ori}. 

Consider $\l=0$. From Eqs.~\eqref{eq:x inner outer}, \eqref{eq:Fab}, \eqref{eq:completeness}, and \eqref{eq:addition theorem}, the nonzero components are
\begin{equation}\label{eq:x ell=0}
x^{\l=0}_{nn} \Theta^+ = x^{\l=0}_{m\bar m} \Theta^+ = \frac{\mu\Delta r}{4\pi r r_\textrm{p}} \Theta^+ = \left(-\frac{2m}{r}+2\alpha_0\right) \Theta^+,
\end{equation}
where we have used $\mu=8\pi m$ and defined $\alpha_0 \equiv m/r_\textrm{p}$. The first term in parentheses is a mass perturbation due to the particle's mass. The second term is a gauge perturbation corresponding to a rescaling of time and radius, $u\to (1-\alpha_0)u$ and $r\to (1+\alpha_0)r$, equivalent to a gauge transformation generated by
\begin{equation}\label{eq:Xi ell=0}
\Xi_{\l=0} =\alpha_0 u\, \dbd{u} - \alpha_0 r\, \dbd{r}. 
\end{equation}
However, if we restrict this transformation to the region $r>r_\textrm{p}$, then it introduces yet another kind of poor behaviour in the metric perturbation: the coefficient of $\delta(\Delta r)$ in Eq.~\eqref{eq:h no string} picks up a term proportional to $\Xi^a$, which grows linearly with time. This indicates that the time coordinate is discontinuous at $r=r_\textrm{p}$, and it destroys the manifest stationarity of the spacetime. We can only avoid this pathology by applying at least the first term in the transformation~\eqref{eq:Xi ell=0} for all $r$; if we introduce any $r$ dependence into the linear-in-$u$ term in $\Xi^a$, we necessarily introduce linear-in-$u$ terms into the metric perturbation.\footnote{To see that this is true, note that $\xi^a = \beta_0 u\, t^a + f^a(r)$ generates the most general transformation that preserves manifest stationarity and spherical symmetry. Here $\beta_0$ is a constant,  $t = \dbd{u}$ is the timelike Killing vector, $f = f^u \dbd{u} + f^r \dbd{r}$, and by ``manifest stationarity'' of a metric perturbation we mean $\lie{t} h_{ab}=0$.} Simply applying Eq.~\eqref{eq:Xi ell=0} for all $r$, we keep our solution independent of $u$, and we obtain a total monopole perturbation $\tilde h^{\l=0}_{ab} = x^{\l=0}_{ab} \Theta^+ - \Lie_{\Xi_{\l=0}} g_{ab}$ with nonzero components
\begin{equation}
\tilde h^{\l=0}_{nn} = \tilde h^{\l=0}_{m\bar m}  = -\frac{2m}{r}\Theta^+ - 2\alpha_0\Theta^-.
\end{equation}

The treatment of the $\l=1$ modes is very similar. The large-$r$ behaviour $x^{\l=1}_{ab}\sim r$ is associated with a uniform acceleration of the coordinate system, and removing it requires a translation $\Xi^a_{\l=1}\sim u^2$. Preserving manifest stationarity of the solution again requires applying this transformation for all $r$; we provide the details in \ref{sec:dipole}.  

With this, our final no-string solution is brought to the form
\begin{equation}\label{eq:h no string + l=1,2}
\tilde h^\textrm{N}_{ab} = \hat h^\textrm{N}_{ab}  + \tilde h^{\l=0,1}_{ab} - 2\xi^{\textrm{S}}_{(a} r_{b)}\,\delta(\Dr),
\end{equation}
where $\tilde h^{\l=0,1}_{ab} = x^{\l=0,1}_{ab} \Theta^+ - \lie{{\Xi_{\l=0,1}}} g_{ab}$. As expected,  our result recovers the no-string solution found in Ref.~\cite{Pound:2013faa} (where the coefficient of $\delta(\Delta r)$ was found by explicitly solving the Einstein equation). In our new construction, the gauge perturbation $\lie{{\Xi_{\l=0,1}}} g_{ab}$ for $r<r_\textrm{p}$ has played the role of the gauge completion used in no-string self-force calculations in Kerr. In those past calculations, the gauge completion was determined by imposing continuity of certain metric components (or other fields) at $r=r_\textrm{p}$. The GHZ procedure has given us an alternative, more manifestly desirable criterion: the coefficient of the delta function must respect the stationarity of the spacetime. We extend that criterion to the EMRI scenario in the next section.

\subsubsection{Lessons for the Kerr problem}

Our analysis of the simple model problem has illuminated the  relationship between the GHZ procedure and previous reconstruction methods, and it implies several lessons that carry over to the realistic EMRI problem: 

\begin{itemize}
\item The complete half-string solution~\eqref{eq:GHZ} contains both a power-law divergence on the string, arising from Ori's Hertz potential, and an angular delta function supported on the string, arising from the GHZ corrector tensor. Both pieces grow polynomially with $r$ at large $r$. There are nontrivial cancellations of delta-function content between the Hertz term and the corrector tensor, but there will always remain a delta function in the trace $g^{ab}h_{ab}$. 
\item The string delta function can be obtained in closed form from the stress-energy tensor. But the power-law divergence arises from the Hertz potential, which will only be obtainable as a sum of modes in Kerr. The sum converges as a distribution but diverges (at all points with $r>r_\textrm{p}$) as an ordinary function. While previous literature was incorrect in deeming this sum inadmissible, we can still draw the same conclusion: it is not obvious that this piece can be calculated in Kerr, where only a finite number of modes can be computed. 
\item The GHZ procedure provides a new route to the no-string solution~\eqref{eq:no-string hab}, via an explicit transformation from the half-string solution. This new route determines the coefficient $D_{ab}$ of the radial delta function. If the transformation puts the no-string solution in an asymptotically flat gauge, then part of the transformation must be extended to all $r<r_\textrm{p}$ to prevent $D_{ab}$ from growing with time. This provides a greatly simplified means of obtaining the ``gauge completion'' for the no-string solution. In the context of a bound geodesic in Kerr, ensuring that $D_{ab}$ does not grow with time will ensure that the entire metric perturbation respects the tri-periodicity of the orbit. \item The metric perturbation, whether in the half-string or no-string gauge, is too singular to obtain a distributionally well-defined second-order field equation. At vacuum points, the second-order metric perturbation $j_{ab}$ should satisfy Eq.~\eqref{eq:2nd Einstein}. The source term $G^{(2)}_{ab}(h,h)\sim h\partial^2 h + \partial h\partial h$ is ill defined on the string in the half-string gauge and on $\mathscr{S}_\textrm{p}$ in the no-string gauge. In either case, products of delta functions arise on the singular surface. In the half-string case, the power-law divergence at the string is likewise too singular for $G^{(2)}_{ab}$ to be well defined.
\end{itemize}

\section{GHZ and the ``shadowless'' solution for spatially extended sources in Kerr}
\label{sec:Kerr}

In this section, we extend our results from flat spacetime to Kerr. We also generalize the analysis to allow for a stress-energy tensor $T_{ab}$ that is supported in a spatially extended (but bounded) region $r_\textrm{min} \leq r \leq r_\textrm{max}$, where $r_\textrm{min/max}$ are radii in the Kerr exterior; this generalization is essential for our puncture scheme in subsequent sections. However, because the calculations follow the flat-spacetime template, readers uninterested in the technical details may skip to the summary in Sec.~\ref{sec:Kerr summary}.

We allow $T_{ab}$ to be singular at the boundaries of its support, $r=r_\textrm{min}$ and $r=r_\textrm{max}$. Specifically, we allow it (and $\psi_0$) to contain $\delta(r-r_\textrm{min/max})$ and $\delta'(r-r_\textrm{min/max})$. This will allow $h_{ab}$ to be discontinuous at these surfaces. That level of singularity is too strong to obtain a well-behaved source in the second-order Einstein equation, but it is useful in the simple demonstration of our puncture scheme in Sec.~\ref{sec:return to flat space}, and it is likely to be useful for the puncture scheme in Kerr if restricted to first order.

Many of our results will be presented in a coordinate-independent form, but wherever we refer to coordinate quantities, we adopt  outgoing Kerr-Newman coordinates $(u,r,\theta,\varphi_*)$. These are related to Boyer-Lindquist coordinates $(t,r,\theta,\varphi)$ by
\begin{subequations}
\label{eq:EF}
\begin{align}
\label{eq:rstar}
u=& \ t-r_* \equiv t-r-\frac{r_+^2+a^2}{r_+-r_-}\ln \left(\frac{r-r_+}{r_+}\right) + \frac{r_-^2+a^2}{r_+-r_-}\ln \left(\frac{r-r_-}{r_-}\right),\\
\varphi_*=& \ \varphi-\varphi_0(r) \equiv \varphi -  \frac{a}{r_+-r_-}\ln
\frac{r-r_+}{r-r_-},
\end{align}
\end{subequations}
where $r_\pm = M\pm \sqrt{M^2-a^2}$ are the inner and outer horizon radii. Similarly, when we have call to refer to a specific tetrad, we adopt the Kinnersley tetrad, which has Kerr-Newman components
\begin{subequations}
\label{eq:Kintet up}
\begin{align}
l =&\ \frac{\partial}{\partial r},\\
n =&\ \frac{r^2 + a^2}{\Sigma} \frac{\partial}{\partial u} - \frac{\Delta}{2\Sigma} \frac{\partial}{\partial r} + \frac{a}{\Sigma} \frac{\partial}{\partial \varphi_*},\\
m =&\ \frac{1}{\sqrt{2} \Gamma} \left( ia\sin\theta \frac{\partial}{\partial u} + \frac{\partial}{\partial \theta} + i\csc\theta \frac{\partial}{\partial \varphi_*} \right),
\end{align}
\end{subequations}
where
$\Gamma=r+ia\cos\theta$, $\Delta=r^2-2Mr+a^2$, and $\Sigma=r^2+a^2\cos^2\theta$.
With this choice, the nonzero spin coefficients are given by Eqs.~\eqref{eq:GHP coeffs in Kinn} and~\eqref{eq:NP coeffs in Kinn}\par
Our calculations will rely heavily on Held's variant of the GHP formalism, reviewed in~\ref{sec:Held}. We write all differential operators in terms of the Held operators $\th$, $\Nbar$, $\Dbar$, and $\Dbar'$ defined in Eq.~\eqref{eq:Hops}, which reduce to Eq.~\eqref{eq:Hop} in the Kinnersley tetrad and Kerr-Newman coordinates. We then write all background quantities in terms of  the spin coefficient $\rho$ and a set of ``Held scalars'' $\Omega^\circ$, $\Psi^\circ$, $\rho'^\circ$, and $\tau^\circ$, given in terms of the GHP coefficients in Eqs.~\eqref{eq:OmegaH} and \eqref{eq:Held coeffs} [or explicitly in the Kinnersley tetrad in Eq.~\eqref{eq:Held coeffs in Kinn}]. All quantities adorned with a ${}^\circ$ are annihilated by the GHP derivative $\th$. This allows us to solve differential equations of the form $\th F = G$ in a coordinate-independent way. With our choice of tetrad and coordinates, $\th=l^a \partial_a =\dbd{r}$ when acting on a scalar, reducing the differential equation to a radial ODE along outgoing null rays. But we can instead write it as an ODE in $\rho$ using the fact that, when acting on an expression in Held form,
\begin{equation}\label{eq:thorn as d/drho}
\th=l^a \nabla_a = (l^a \nabla_a\rho) \dbd{\rho} = \rho^2 \dbd{\rho},
\end{equation}
where the last equality follows from Eq.~\eqref{eq:GHP eqs}. Here we have adopted the Kinnersley tetrad in the intermediate equations, but the result $\th=\rho^2\frac{\partial}{\partial\rho}$ is valid in any tetrad aligned with the principal null directions. Using Eq.~\eqref{eq:thorn as d/drho}, we can straightforwardly write the solution to $\th F = G$ as $F = \int d \rho\, G/\rho^2$, without specifying a tetrad or coordinate system.

As we did in flat space, we find the GHZ solution $h^-_{ab}$ that is regular in the region $r<r_\textrm{min}$. The solution $h^+_{ab}$ can be found straightforwardly in the same way. We then find the transformation to a ``shadowless'' gauge that generalizes the no-string gauge.

\subsection{CCK-Ori reconstruction}

We first describe the construction of $\Phi$, and reconstruction of $\hat h_{ab}$, for a spatially compact source. Our description draws heavily from Ref.~\cite{Ori:2002uv}.

Assume we have solved the Teukolsky equation~\eqref{eq:teuk} for $\psi_0$ with retarded boundary conditions. Outside the source, $\psi_0$ is a vacuum solution, which we label $\psi^+_0$ for $r>r_\textrm{max}$ and $\psi^-_0$ for $r<r_\textrm{min}$. We can obtain the Hertz potential from $\psi_0$ through four integrations of $\th^4 \widebar{\Phi} = -2\psi_0$ [Eq.~\eqref{eq:th4}]. If we integrate from the past horizon, then we obtain the solution $\Phi^-$. As shown by Ori (see also Theorem 5 of GHZ), $\Phi^-$ satisfies the vacuum equation ${\cal O}^\dagger\Phi^-=0$ along each integral curve of $l^a$ except those curves that pass through the source. On those curves, $\Phi^-$ is a nonvacuum solution both inside the source and in its ``shadow'': like the string in the particle case, this is the region filled by integral curves of $l^a$ that start from the source and extend to future null infinity. So, in particular, $\Phi^-$ is a vacuum solution in the entire region $r<r_\textrm{min}$. Analogously, if we instead integrate Eq.~\eqref{eq:th4} from future null infinity, we obtain a field $\Phi^+$ that is vacuum in the region $r>r_\textrm{max}$.

In the region $r>r_\textrm{max}$, the two particular solutions $\Phi^+$ and $\Phi^-$ can only differ by a solution to the homogeneous equation $\th^4\Phi = 0$. Using Eq.~\eqref{eq:thorn as d/drho}, we can write that solution as
\begin{equation}\label{ddef}
\Phi^\ori = d_0^\circ + \frac{d_1^\circ}{\rho} +  \frac{d_2^\circ}{\rho^2} + \frac{d_3^\circ }{\rho^3}
\end{equation}
for some Held scalars $d^\circ_i \circeq \{i-4,i\}$, which are annihilated by $\th$; this generalizes Eq.~\eqref{eq:PhiS modes}. Similarly generalizing Eq.~\eqref{eq:Phi modes}, we therefore have
\begin{equation}\label{eq:Hertz Kerr}
\Phi = \Phi^\textrm{N} + \Phi^\textrm{M} \Theta_\textrm{M}  +  \Phi^\ori \Theta^+_\textrm{M},
\end{equation}
where it is understood that $\Phi\equiv\Phi^-$. $\Theta^+_\textrm{M} \equiv \Theta(r-r_\textrm{max})$ and $\Theta^-_\textrm{M} \equiv \Theta(r_\textrm{min}-r)$ have support in the vacuum regions $r>r_\textrm{max}$ and $r<r_\textrm{min}$, and $\Theta_\textrm{M} = (1-\Theta^-_\textrm{M} - \Theta^+_\textrm{M})$ has support inside the ``matter region'' $r_\textrm{min} < r < r_\textrm{max}$. The first term in Eq.~\eqref{eq:Hertz Kerr} is
\begin{equation}\label{eq:PhiN Kerr}
\Phi^\textrm{N} = \Phi^+\Theta^+_\textrm{M} + \Phi^-\Theta^-_\textrm{M},
\end{equation}
the generalization of the no-string Hertz potential. The second term in Eq.~\eqref{eq:Hertz Kerr}, $\Phi^\textrm{M}$, is the potential inside the matter region, which vanished in the point-particle limit. We will refer to $\Phi^\textrm{N}$ as the shadowless Hertz potential, which is vacuum everywhere in its domain $\{r<r_\textrm{min}\}\cup\{r>r_\textrm{max}\}$; to $\Phi^\textrm{S}$ as the \emph{shadow potential}, which is nonvacuum in the source's shadow; and to $\Phi^\textrm{M}$ as the matter potential.

From $\Phi$, we can obtain the reconstructed field $\hat h_{ab}=2\Re(\mathcal{S}^\dagger_{ab}\Phi)$ in the form
\begin{equation}
\hat h_{ab} = \hat h^\textrm{N}_{ab} + \hat h^\textrm{M}_{ab} \Theta_\textrm{M} + \hat h^\textrm{S}_{ab}\Theta^+_\textrm{M},\label{eq:hhat Kerr}
\end{equation}
where 
\beq\label{eq:hhatN Kerr}
\hat h^\textrm{N}_{ab} = 2\Re(\mathcal{S}^\dagger_{ab}\Phi^+)\Theta^+_\textrm{M} + 2\Re(\mathcal{S}^\dagger_{ab}\Phi^-)\Theta^-_\textrm{M},
\eeq
$\hat h^\textrm{M}_{ab} = 2\Re(\mathcal{S}^\dagger_{ab}\Phi^\textrm{M})$, and $\hat h^\textrm{S}_{ab} = 2\Re(\mathcal{S}^\dagger_{ab}\Phi^\textrm{S})$. The shadow field $\hat h^\textrm{S}_{ab}$ can be explicitly evaluated to obtain
\begin{subequations}\label{eq:hhatS Kerr}
\begin{align}
\hat h^\ori_{nn} &= 2 \Re \Big\{ -2 \rho \bar{\rho} ^2 \tau^\circ \left( \tilde{\edth} d^\circ_0 - \tau^\circ d^\circ_1 \right) - \bar{\rho}^2 \left( \tilde{\edth}^2 d^\circ_0 - 2 \tau^{\circ 2} d^\circ_2 \right)\nonumber \nls\\
&\quad- \frac{\bar{\rho}^2}{\rho} \left( \tilde{\edth}^2 d^\circ_1 - 2 \tau^\circ \tilde{\edth} d^\circ_2 \right) - \frac{\bar{\rho}^2}{\rho^2} \left( \tilde{\edth}^2 d^\circ_2 - 4 \tau^\circ \tilde{\edth} d^\circ_3 \right) - \frac{\bar{\rho}^2}{\rho^3} \left( \tilde{\edth}^2 d^\circ_3 \right) \Big\},\label{Sdnn} \nls\\
\hat h^\ori_{nm} &= - \rho \bar{\rho} \left( 2 \tilde{\edth}' \bar{d}^\circ_0 - 2 \bar{\tau}^\circ \bar{d}^\circ_1 + \Omega^{\circ 2} \tilde{\edth}' \bar{d}^\circ_2 + \Omega^{\circ 3} \tilde{\edth}' \bar{d}^\circ_3 \right)\nonumber \nls\\
&\quad- \rho^2 \bar{\rho} \left( \Omega^\circ \tilde{\edth}' \bar{d}^\circ_0 + \Omega^{\circ 2} \tilde{\edth}' \bar{d}^\circ_1 + \Omega^{\circ 3} \tilde{\edth}' \bar{d}^\circ_2 + \Omega^{\circ 4} \tilde{\edth}' \bar{d}^\circ_3 \right)\nonumber \nls\\
&\quad- \bar{\rho} \left( \tilde{\edth}' \bar{d}^\circ_1 + \Omega^\circ \tilde{\edth}' \bar{d}^\circ_2 - 2 \bar{\tau}^\circ \bar{d}^\circ_2 + \Omega^{\circ 2} \tilde{\edth}' \bar{d}^\circ_3 \right) + \frac{1}{\bar{\rho}} \left( \tilde{\edth}' \bar{d}^\circ_3 \right),\label{Sdnm} \nls\\
\hat h^\ori_{mm} &= 2 \bar{d}^\circ_2 + 2 \bar{\rho} \bar{d}^\circ_1.\label{Sdmm} \nls
\end{align}
\end{subequations}
This expression determines the shadow field up to the four Held scalars $d^\circ_i$. It appears here for the first time.

The scalars $d^\circ_i$ can be determined from junction conditions at $r=r_\textrm{max}$. These conditions read 
\begin{align}\label{eq:junction}
\th^n \Phi^\textrm{S} +\th^n \Phi^+ &= \th^n \Phi^\textrm{M} + [\th^n\Phi]\quad \text{at }r=r_{\textrm{max}}    
\end{align}
for $n=0,\ldots,3$. Here  
$[K]\equiv (\lim_{r\to r_\textrm{max}^+}K) - (\lim_{r\to r_\textrm{max}^-}K)$ denotes the jump in a quantity $K$ across the surface. 
The jumps can readily be expressed in terms of $\psi_0$. Given our assumptions on $\psi_0$, in a neighbourhood of $r=r_\textrm{max}$ it has the form $\psi_0=A^\circ\delta'(r-r_\textrm{max})+B^\circ\delta(r-r_\textrm{max})$ plus a piecewise smooth function. $\th^4\overline{\Phi}=-2\psi_0$ then implies $[\Phi]=[\th\Phi]=0$, $[\th^2\Phi]=-2\bar A^\circ$, and $[\th^3\Phi]=-2\bar B^\circ$. The four conditions~\eqref{eq:junction} then reduce to equations for the four scalars $d^\circ_i$ upon substituting Eq.~\eqref{ddef} for $\Phi^\textrm{S}$. Note that because we have assumed $\psi_0$ contains at most a $\delta'$ at the boundaries, $\th^2\Phi$ contains at most a jump discontinuity there, implying that no delta functions can arise in Eq.~\eqref{eq:hhat Kerr}.

\subsubsection{Mode construction}

The pieces of $\Phi$ can be found more directly if we adopt a standard decomposition into spheroidal harmonics ${}_s S_{\lm\omega}$, with\footnote{We note that Ori instead decomposed into a more typical Boyer-Lindquist Fourier basis $e^{im\varphi-i\omega t}$, meaning our radial functions are related to his by ${}_s R_{\lm\omega}={}_s R^\textrm{Ori}_{\lm\omega}e^{im\varphi_0-i\omega r_*}$, with $r_*$ and $\varphi_0(r)$ defined in Eq.~\eqref{eq:EF}.}
\beq\label{eq:psi0 mode decomp}
\psi_0 = \int d\omega \sum_{\lm} {}_2 R_{\lm\omega}(r){}_2 S_{\lm\omega}(\theta)e^{im\varphi_*-i\omega u},
\eeq
and
\beq\label{eq:Phi mode decomp}
\Phi = \int d\omega \sum_{\lm} {}_{-2} R_{\lm\omega}(r){}_{-2} S_{\lm\omega}(\theta)e^{im\varphi_*-i\omega u}.
\eeq
In terms of the radial modes, the inversion relation $\th^4 \widebar\Phi = -2\psi_0$  reduces to
\beq\label{eq:inversion modes}
(-1)^m\frac{\partial^4}{\partial r^4}\, {}_{-2}\bar R_{\l,-m,-\omega} = -2\ {}_2 R_{\lm\omega}.
\eeq
Here, to express the modes of $\bar\Phi$ in terms of the modes of $\Phi$, we have taken the complex conjugate of Eq.~\eqref{eq:Phi mode decomp} and used the identity ${}_s \bar S_{\lm\omega} = (-1)^{m+s} {}_{-s} S_{\l,-m,-\omega}$. The inversion relation~\eqref{eq:inversion modes} is to be solved subject to the junction conditions at $r=r_\textrm{min}$ and $r=r_\textrm{max}$, 
\begin{subequations}\label{eq:junction modes}
\begin{align}
\partial_r^n \, {}_{-2}R^\textrm{M}_{\lm\omega} &= \partial_r^n \, {}_{-2}R^-_{\lm\omega} + [\partial_r^n \, {}_{-2}R_{\lm\omega}] &\ \text{at } r=r_\textrm{min},\label{eq:rmin junction}\\
\partial_r^n \, {}_{-2}R^\textrm{S}_{\lm\omega} &= \partial_r^n \, {}_{-2}R^\textrm{M}_{\lm\omega}-\partial_r^n \, {}_{-2}R^+_{\lm\omega}+[\partial_r^n \, {}_{-2}R_{\lm\omega}] &\text{at } r=r_\textrm{max}\label{eq:rmax junction}
\end{align}
\end{subequations}
for $n=0,\ldots,3$. As in the discussion below~\eqref{eq:junction}, the jumps $[\partial_r^n \, {}_{-2}R_{\lm\omega}]$ can be expressed in terms of the coefficients of any radial delta functions $\delta^{(k)}(r-r_\textrm{min/max})$ in ${}_2 R_{\lm\omega}$.

First consider $\Phi^\textrm{N}$. If we impose retarded boundary conditions on $\psi_0$, then in the vacuum regions $r>r_\textrm{max}$ and $r<r_\textrm{min}$, the radial function can be written as multiples of the standard radial basis functions\footnote{For consistency we adopt Ori's unusual nomenclature for the radial functions. What is here called ``down'' is normally called ``in'' and ``out'' is normally called ``up''} ${}_2 R^\textrm{out/down}_{\lm\omega}(r)$,
\beqs\label{eq:2Rpm}
\begin{align}
{}_2 R^+_{\lm\omega}(r) &= A^+_{\lm\omega}\ {}_2 R^\textrm{out}_{\lm\omega}(r) \qquad r>r_\textrm{max},\\
{}_2 R^-_{\lm\omega}(r) &= A^-_{\lm\omega}\ {}_2 R^\textrm{down}_{\lm\omega}(r) \qquad r<r_\textrm{min},
\end{align}
\eeqs
where $A^\pm_{\lm\omega}$ are constants, and where the basis solutions behave as
\beqs\label{eq:Routdown}
\begin{align}
{}_s R^\textrm{out}_{\lm\omega} &\sim r^{-1-2s} \hphantom{\Delta^{-s} e^{-ik r_*-i\omega r_*+im\varphi_0(r)} \qquad}\nphantom{r^{-1-2s}}\text{for }r\to\infty,\\
{}_s R^\textrm{down}_{\lm\omega} &\sim \Delta^{-s} e^{-ik r_*-i\omega r_*+im\varphi_0(r)} \qquad\text{for }r\to r_+ 
\end{align}
\eeqs
with $k\equiv \omega-ma/(2Mr_+)$. In the vacuum regions, the radial mode of $\Phi^+$ must likewise be a constant multiple of ${}_{-2}R^\textrm{out}_{\lm\omega}$, and that of $\Phi^-$ must be a constant multiple of ${}_{-2}R^\textrm{down}_{\lm\omega}$; as Ori showed, this ensures that $\Phi^\pm$ satisfy $\th^4\overline{\Phi}^\pm=-2\psi^\pm_0$, and they are necessarily the unique solutions because any homogeneous solution takes the form~\eqref{ddef}, which violates $\O^\dagger\Phi=0$ and therefore cannot be in $\Phi^\pm$ in the vacuum regions. The constants of proportionality between ${}_{-2}R^\pm_{\lm\omega}$ and ${}_{-2}R^\textrm{out/down}_{\lm\omega}$ can  be easily found by evaluating $\th^4\overline{\Phi}^-=-2\psi^-_0$ at the horizon and $\th^4\overline{\Phi}^+=-2\psi^+_0$ at large $r$ [in the mode form~\eqref{eq:inversion modes}], taking advantage of the simple forms of the basis solutions there. The result, expressed with our conventions, is~\cite{Ori:2002uv} 
\beqs\label{eq:-2Rpm}
\begin{align}
{}_{-2} R^+_{\lm\omega}(r) &= -2(-1)^m C^+_{\l,-m,-\omega}\,A^+_{\l,-m,-\omega}\ {}_{-2} R^\textrm{out}_{\l,-m,-\omega}(r)\qquad r>r_\textrm{max},\\
{}_{-2} R^-_{\lm \omega}(r) &= -2(-1)^m C^-_{\l,-m,-\omega}\, A^-_{\l,-m,-\omega}\ {}_{-2} R^\textrm{down}_{\l,-m,-\omega}(r)\qquad r<r_\textrm{min},
\end{align}
\eeqs
where
\begin{align}
C^+_{\lm\omega} = \frac{16\omega^2}{p} \qquad\text{and}\qquad
C^-_{\lm\omega} = \frac{1}{(\omega+2iq)(\omega+iq)\omega (\omega-iq)}.
\end{align}
Here $q=r_+-r_-$, $p$ is given by Eq. (21) in Ref.~\cite{Ori:2002uv}, and we have assumed that ${}_{2} R^\textrm{out/down}_{\lm\omega}$ and ${}_{-2} R^\textrm{out/down}_{\lm\omega}$ have the same normalization [i.e., the same constant coefficient in front of the leading-order asymptotic forms in Eq.~\eqref{eq:Routdown}]. After obtaining the modes of $\psi_0$, one can therefore immediately obtain the modes of the shadowless potential $\Phi^\textrm{N}$. This is the procedure used to obtain the no-string potential in Refs.~\cite{vandeMeent:2015lxa, vandeMeent:2016pee, vandeMeent:2017bcc} (though beginning from modes of $\psi_4$ rather than $\psi_0$). 

Next consider $\Phi^\textrm{M}$ in the matter region $r_\textrm{min}<r< r_\textrm{max}$. Its radial modes ${}_{-2}R^\textrm{M}_{\lm\omega}$ are given by the solution to the inversion relation~\eqref{eq:inversion modes} with the boundary conditions~\eqref{eq:rmin junction} at $r=r_\textrm{min}$. Since ${}_{-2}R^-_{\lm\omega}$ satisfies the vacuum Teukolsky equation, we can use the vacuum equation to express $\partial^2_r\,{}_{-2}R^-_{\lm\omega}$ and $\partial^3_r\,{}_{-2}R^-_{\lm\omega}$ in terms of ${}_{-2}R^-_{\lm\omega}$ and $\partial_r\,{}_{-2}R^-_{\lm\omega}$. The boundary conditions therefore only involve ${}_{-2}R^-_{\lm\omega}$ and its first derivative.

Finally consider $\Phi^\textrm{S}$ in the region $r>r_\textrm{max}$. It is convenient to write Eq.~\eqref{ddef} as $\Phi = \sum_{i=0}^3 B^\circ_i(u,\theta,\varphi_*)r^i$, such that
\begin{equation}
{}_{-2}R^\textrm{S}_{\lm\omega} = \sum_{i=0}^3 B^{\lm\omega}_i r^i,
\end{equation}
where $B^{\lm\omega}_i$ are the (spin-weight $-2$) coefficients in the spheroidal-harmonic expansion of $B^\circ_i$. Equation~\eqref{eq:rmax junction} immediately becomes four linear, algebraic equations for the four coefficients $B^{\lm\omega}_i$. The coefficients $d^\circ_i$ in Eq.~\eqref{ddef} can then be calculated from $B^\circ_i$ using $r= -(\rho+\rhb)/(2\rho \rhb) = -\frac{1}{2}(\Omega^\circ+2/\rho)$, which determines
\beqs\label{eq:d from B}
\begin{align}
d^\circ_0 &= B^\circ_0 - \frac{1}{8} \Omega^\circ (4 B^\circ_1 -2 B^\circ_2\Omega^\circ + B^\circ_3 \Omega^{\circ\, 2}),\\
d^\circ_1 &= -B^\circ_1 + B^\circ_2 \Omega^\circ - \frac{3}{4} B^\circ_3 \Omega^{\circ\,2},\\
d^\circ_2 &= B^\circ_2 - \frac{3}{2}B^\circ_3 \Omega^\circ,\\
d^\circ_3 &= -B^\circ_3.    
\end{align}
\eeqs
This completes the construction of $\Phi$. In what follows, we will use $\Phi^\textrm{S}$ in the form~\eqref{ddef} with the understanding that $d^\circ_i$ can be obtained from Eq.~\eqref{eq:d from B}.

In Ref.~\cite{Ori:2002uv}, Ori provided an alternative method of constructing $\Phi$ directly from the source for $\psi_0$, without requiring $\psi_0$ itself. Here, to emphasize the connection with the no-string reconstruction method that has been employed in practice, we have instead focused on starting from $\psi_0$ to construct the ``shadowless'' field in the same manner as one constructs the no-string field, and on elucidating what gets added to that field. We have not attempted the alternative path of dividing Ori's solution into the three constituents in Eq.~\eqref{eq:Hertz Kerr}.

\subsection{Corrector tensor}

We now turn to the corrector tensor. The nonzero components of $x_{ab}$ satisfy a hierarchy of equations that may be compactly written in GHP form. They read~\cite{Green:2019nam}\footnote{Here we have made some simplifications to the equation of GHZ and corrected a minor typo.}
\begin{equation}
\label{eq:xmmb}
\begin{split}
 \rho^2 \th \left( \frac{\rhb}{\rho^3} \th \left[\frac{\rho}{\rhb} x_{m\mb} \right] \right)&= T_{ll}
\end{split}
\end{equation}
for $x_{m\mb}$, 
\begin{multline}\label{eq:xnm}
\frac{\rho}{2(\rho+\rhb)} 
\th\left( (\rho+\rhb)^2\th \frac{x_{nm}}{\rho(\rho+\rhb)}\right) = T_{lm} - \half\{(\th+\rho-\rhb)(\edth+\tab'-\tau) + 2\tab'(\th-2\rho) \\
-
(\edth-\tau-\tab')\rhb +2\rho\tau\}x_{m\mb}
\end{multline}
for $x_{nm}$, and
\begin{multline}
\label{eq:xnn}
 \half (\rho+\rhb)^2\th \left( \frac{1}{\rho+\rhb}x_{nn} \right) = T_{ln} - \Re\{ \edthp \edth  + (\tau'-\tab)\edthp + \left[\edth(\tab+\tau')\right] - 2 \tau'\tau - 2 \Psi_2 \nls\\
\qquad\qquad\qquad\qquad\qquad\qquad\qquad\qquad  -2 \rhb\rho' + 2(\rho' \th + \rho\th') -  \th'\th \}  x_{m\mb}  \nls\\
\qquad\qquad\ -\Re  \{[ (\th-2\rho)(\edthp-\tab) + (\tap+\tab)(\th+\rhb) \nls\\
-2(\edthp-\tap)\rho-2\tab\th] x_{nm} \} \nls
\end{multline}
for $x_{nn}$. \par
We solve these equations in sequence by putting all quantities in Held form and using Eq.~\eqref{eq:thorn as d/drho}. Integrating outward from the past horizon, for $x_{m\mb}$ we find
\begin{align}\label{eq:xmmbar Kerr}
x_{m\mb} = \frac{\rhb}{\rho} \int_{\rho_\textrm{min}^-}^\rho
d \rho_2(1+\rho_2\Omega^\circ)\int_{\rho_\textrm{min}^-}^{\rho_2} d \rho_1 \rho_1^{-4} T_{ll}.
\end{align}
To include the support of delta functions at $r_{\rm min}$, the integrals run from $\rho^-_\textrm{min} = \rho(r_\textrm{min})-0^+$. Also, the complex contour corresponding to increasing $r$ along the real contour $r_\textrm{min}<r<r_\textrm{max}$ at fixed angles is understood here and below. For $x_{nm}$ we get
\begin{align}\label{eq:xnm Kerr}
x_{nm} = 2\rho(\rhb+\rho) 
\int_{\rho_\textrm{min}^-}^\rho  \frac{d \rho_2}{ \rho_2^{4} } \left(\frac{1+\rho_2\Omega^\circ}{2+\rho_2\Omega^\circ}\right)^2\int_{\rho_\textrm{min}^-}^{\rho_2} \frac{d \rho_1} {\rho_1^{2}} \frac{2+\rho_1\Omega^\circ}{1+\rho_1\Omega^\circ}(T_{lm}+ \mathcal{N} x_{m\mb}),
\end{align}
where $\mathcal{N}$ is the differential operator on the right-hand side of \eqref{eq:xnm}. Finally for $x_{nn}$ we get
\begin{equation}\label{eq:xnn Kerr}
x_{nn} = 2(\rho+\rhb)\int_{\rho_\textrm{min}^-}^\rho
\frac{d \rho_1}{\rho_1^4} \left(\frac{1+\rho_1\Omega^\circ}{2+\rho_1\Omega^\circ}\right)^2 \left[T_{ln} + \Re(\mathcal U x_{m\mb}) + \Re(\mathcal V x_{nm} )\right],
\end{equation}
where $\mathcal U$ and $\mathcal V$ are the differential operators on the right-hand side of Eq.~\eqref{eq:xnn}.

Following the pattern of Eq.~\eqref{eq:Hertz Kerr}, we write this solution as\footnote{We assume for simplicity that the integrand in Eq.~\eqref{eq:xnn Kerr} does not contain a $\delta'(r-r_\textrm{min/max})$. If it does, then $x_{nn}$ will contain a delta function at $r=r_\textrm{min/max}$.}
\begin{equation}
x_{ab} = x^\textrm{M}_{ab}\Theta_\textrm{M} + x^\textrm{S}_{ab}\Theta^+_\textrm{M},
\end{equation}
generalizing Eq.~\eqref{eq:x inner outer}.  The corrector tensor $x^\textrm{M}_{ab}$ inside the source is given by Eqs.~\eqref{eq:xmmbar Kerr}--\eqref{eq:xnn Kerr} with $r_\textrm{max}>r>r_\textrm{min}$. In the source-free region $r > r_\textrm{max}$, the integrals in Eqs.~\eqref{eq:xmmbar Kerr}--\eqref{eq:xnn Kerr} can be evaluated explicitly to find 
\begin{subequations}\label{eq:xS Kerr}
\begin{align}
x^\textrm{S}_{m\bar{m}} &= a^\circ \left( \frac{\rho}{\bar{\rho}} + \frac{\bar{\rho}}{\rho} \right) + b^\circ \left( \rho + \bar{\rho} \right),\label{eq:xmmb gen sol} \nls\\
x^\textrm{S}_{nm} &= \bar{\rho} \Omega^\circ \tilde{\edth} a^\circ + \frac{\left( \rho + \bar{\rho} \right)^2}{\rho} \tau^\circ a^\circ + \bar{\rho} \tilde{\edth} b^\circ + \left( \rho + \bar{\rho} \right)^2 \tau^\circ b^\circ + \frac{2 \rho - \bar{\rho}}{\rho^2} c^\circ + \rho \left( \rho + \bar{\rho} \right) e^\circ,\label{eq:xnm gen sol}\\
x^\textrm{S}_{nn} &=  2 \Re \bigg\{ \half \frac{\rho + \bar{\rho}}{\rho^2} \tilde{\th}' a^\circ + (1 + \rho \bar{\rho} \Omega^{\circ 2}) \tilde{\edth}'\tilde{\edth} a^\circ + 2 \left( \rho^2 + 2 \rho \bar{\rho} \right) \tau^\circ \bar{\tau}^\circ a^\circ\nonumber \nls\\
&\quad+ \left( \frac{\rho^2}{\bar{\rho}} + \frac{3}{2} \rho \bar{\rho} \Omega^\circ \right) \Psi^\circ a^\circ - 2 \rho^{\prime \circ} a^\circ + \tilde{\th}' b^\circ + 2 \rho \bar{\rho} \bar{\tau}^\circ \tilde{\edth} b^\circ + 2 \rho^2 \bar{\rho} \tau^\circ \bar{\tau}^\circ b^\circ + \rho^2 \Psi^\circ b^\circ\nonumber \nls\\
&\quad+ \rho^2 \tilde{\edth}' e^\circ + 2 \rho^2 \bar{\rho} \bar{\tau}^\circ e^\circ - \frac{\bar{\rho}}{\rho} \left( 4 \Omega^\circ + \frac{1}{\rho} \right) \tilde{\edth}' c^\circ + 4 \frac{\bar{\rho}}{\rho} \bar{\tau}^\circ c^\circ + \left( \rho + \bar{\rho} \right) f^\circ \bigg\}.\label{eq:xnn gen sol} \nls
\end{align}
\end{subequations}
The quantities $a^\circ, b^\circ, c^\circ, e^\circ, f^\circ$ are integration ``constants'' with GHP weights listed in Table~\ref{table:GHPWeights}. To show their general structure, we give the first two explicitly here:
\begin{subequations}\label{eq:corrector constants}
\begin{align}
a^\circ &= \half\int_{\rho^-_\textrm{min}}^{\rho^+_\textrm{max}}
d \rho_2 (1+\rho_2\Omega^\circ) \int_{\rho^-_\textrm{min}}^{\rho_2} \frac{
 d \rho_1 }{\rho_1^{4}} T_{ll}  - \frac{\rho_\textrm{max}}{2} (1+ \tfrac12 \rho_\textrm{max} \Omega^\circ)\int^{\rho^+_\textrm{max}}_{\rho^-_\textrm{min}} \frac{d \rho_1} {\rho_1^{4}} T_{ll}, \\
b^\circ &= \half \int^{\rho^+_\textrm{max}}_{\rho^-_\textrm{min}}\frac{d \rho_1}{ \rho_1^{4}} T_{ll} - a^\circ \Omega^\circ,
\end{align}
\end{subequations}
where $\rho^+_{\rm max}=\rho(r_{\rm max})+0^+$. The remaining three are given in Eqs.~\eqref{eq:cHeld}, \eqref{eq:eHeld}, and \eqref{eq:fHeld} of \ref{app:integrations} . These functions are annihilated by $\th$; in Kerr-Newman coordinates and the Kinnersley frame, they are functions of $(u,\theta,\varphi_*)$. We see from the integral expressions that $x^\textrm{S}_{ab}$ is only nonzero in the shadow of the source [i.e., for values of $(u,\theta,\varphi_*)$ where $T_{ab}$ is nonzero].
\begin{table}[t]
\addtolength{\tabcolsep}{-0.35pt}
\caption{\label{table:GHPWeights}GHP weights $\{p,q\}$ of the Held scalars $a^\circ,\ldots,g^\circ_{\dot a}$ appearing in the GHZ and shadowless solutions. The unlisted scalars $\xi^\circ_a$ have the weights $\xi^\circ_l \circeq \{1,1\}$, $\xi^\circ_n \circeq \{-1,-1\}$, and $\xi^\circ_m \circeq \{2, 0\}$.}
\vspace{0.35em}
\begin{indented}
\item[]
\begin{tabular}{@{} c c c c c c c c@{}}
\toprule
$a^\circ$   & $b^\circ$   & $c^\circ$   & $d^\circ_i$  & $e^\circ$   & $f^\circ$    & $g^\circ_{\dot M}$ & $g^\circ_{\dot a}$ \\
\midrule
$\{0,0\}$ & $\{-1,-1\}$ & $\{1,-1\}$ & $\{i-4,i\}$ & $\{-2,-4\}$ & $\{-3,-3\}$ &  $\{-3,-3\}$  & $\{-2,-4\}$\rule{0pt}{1.1em}\rule[-0.65em]{0pt}{0pt}\\
\bottomrule
\end{tabular}
\end{indented}
\end{table}

\subsection{Total GHZ solution and passage to the shadowless gauge}

Combining the results of the previous two sections, we obtain the total solution
\begin{equation}\label{eq:GHZ Kerr}
h_{ab} = \hat h^\textrm{N}_{ab} + h^\textrm{M}_{ab}\Theta_\textrm{M} + h^\textrm{S}_{ab}\Theta^+_\textrm{M},
\end{equation}
which we divide into the ``no-shadow'' field $\hat h^\textrm{N}_{ab}$ outside the source, the ``shadow field'' $h^\textrm{S}_{ab} = \hat h^\textrm{S}_{ab} + x^\textrm{S}_{ab}$, and the field inside the source, $h^\textrm{M}_{ab} = \hat h^\textrm{M}_{ab} + x^\textrm{M}_{ab}$. $\hat h^\textrm{S}_{ab}$ is given in Eq.~\eqref{eq:hhatS Kerr}, $x^\textrm{S}_{ab}$ in~\eqref{eq:xS Kerr}, $\hat h^\textrm{M}_{ab}$ by $2\Re({\cal S}^\dagger_{ab}\Phi^\textrm{M})$, and $x^\textrm{M}_{ab}$ in Eqs.~\eqref{eq:xmmbar Kerr}--\eqref{eq:xnn Kerr} (with $r_\textrm{max}>r>r_\textrm{min}$). Unlike in the flat-spacetime result~\eqref{eq:GHZ flat}, there is no straightforward way to find cancellations between the reconstructed field $\hat h^\textrm{S}_{ab}$ and the corrector field $x^\textrm{S}_{ab}$ (or between $\hat h^\textrm{M}_{ab}$ and $x^\textrm{M}_{ab}$).

We transform this solution to a shadowless gauge following the familiar argument. The shadow field must be pure gauge up to a perturbation toward another Kerr solution, $\dot g_{ab}$, meaning
\begin{equation}\label{eq:xidet}
h^\textrm{S}_{ab} = \dot g_{ab} + \lie{\xi} g_{ab}.
\end{equation}
We solve this equation for $\xi^a$ by putting all quantities in Held form. Doing so will enable us to write each component of the equation in the form of a polynomial in $\rho$,
\begin{equation}\label{eq:rho polynomial}
\sum_n \rho^n {X^{(n) \, \circ}}_{ab} = 0.
\end{equation} 
Such an equation implies that each coefficient must vanish: ${X^{(n) \, \circ}}_{ab} = 0$. (Proof: divide by the highest power of $\rho$, 
then successively apply $\th$ using $\th \rho^{-1} = -1$.)

To arrive at equations of the form~\eqref{eq:rho polynomial}, we must put $\dot g_{ab}$ and $\lie{\xi} g_{ab}$ in Held form. This will be facilitated by adopting the IRG condition
\begin{equation}\label{eq:gdot IRG}
\dot g_{ab}l^b = 0 = \dot g_{ab}m^a\bar m^b.
\end{equation}

\subsubsection{Held form of $\dot g_{ab}$}\label{app:zero modes}

We can use the Held formalism to write a compact expression for $\dot{g}_{ab}$ in the IRG. We start by calculating it in the Kerr-Newman gauge as
\begin{equation}\label{dgZ}
\dot g^\textrm{KN}_{bc} =  \dot M \partial_M g^{M,a}_{bc} + \dot a \partial_a g^{M,a}_{bc},
\end{equation}
where $g^{M,a}_{bc}$ is the Kerr metric written as a function of $M$ and $a$ in outgoing Kerr-Newman coordinates. Here $\dot M$ represents a perturbation to the mass at fixed spin parameter $a$, and $\dot a$ represents a perturbation to the spin parameter at fixed mass $M$. Since the black hole angular momentum is $J=Ma$, $\dot g^\textrm{KN}_{bc}$ contains an angular momentum perturbation $\dot J = a\dot M + M\dot a$. \par
We then transform to the IRG, defining
\begin{equation}\label{dg}
\dot g_{ab} = \dot g^\textrm{KN}_{ab} + \lie{\zeta} g_{ab},
\end{equation}
with a gauge vector $\zeta^a$ chosen such that $\dot g_{ab}l^b = 0 = \dot g_{ab} m^a \mb^b$. With $\lie{\zeta} g_{ab}$ written as in Eqs.~ \eqref{Liexigla}--\eqref{Liexigma}, a lengthy calculation reveals that $\zeta^a$ must be chosen as
\begin{equation}
\zeta^b = - \frac{\dot a}{\rho^{\circ \prime} a} \Re \left\{ \tau^\circ \bar{\tau}^\circ \rho l^b - \bar{\tau}^\circ m^b \right\} .
\end{equation}
The components of $\dot g_{ab}$ are then given by
\begin{subequations}\label{eq:gdot}
\begin{align}
\dot g_{nn} ={}& \left( \rho + \bar{\rho} \right) g^\circ_{\dot M} + 2 \rho \bar{\rho} \left( \rho + \bar{\rho} \right) \Re \left\{ \half \Omega^\circ \tilde{\edth}' g^\circ_{\dot a} + \bar{\tau}^\circ g^\circ_{\dot a} \right\}, \nls\\
\dot g_{nm} ={}& \rho \left( \rho + \bar{\rho} \right) g^\circ_{\dot a}, \nls\\
\dot g_{mm} ={}& \dot g_{nl} = \dot g_{ml} = \dot g_{ll} = \dot g_{\mb m} = 0, \nls
\end{align}
\end{subequations}
where
\begin{align}\label{def: gdot components}
g^\circ_{\dot M} &= \Psi^\circ \frac{\dot M}{M} \circeq \left\{ -3, -3 \right\},\\ 
g^\circ_{\dot a} &=  \frac{\tau^\circ}{2 \rho^{\circ \prime}} \Psi^\circ \frac{\dot a}{a} \circeq \left\{ -2, -4 \right\}.
\end{align}
In the Kinnersley tetrad, the Held scalars reduce to $g^\circ_{\dot M}=\dot M$ and $g^\circ_{\dot a}=\frac{i}{\sqrt{2}} M \dot{a} \sin\theta$.\par
The perturbations $\dot M$ and $\dot a$ can be evaluated using Abbott-Deser integrals~\cite{Abbott:1981ff}, which define gauge-invariant conserved charges for linear perturbations. We consider a spherical shell  $\Sigma_t$ around the black hole, defined by $\{t=\textrm{const.}, r_1<r<r_2\}$, where $r_2>r_\textrm{max}>r_\textrm{min}>r_1$; here $t$ can be any time function. Following the conventions in Ref.~\cite{Dolan:2012jg}, given a Killing field $X^a$ of the background spacetime, we define the charge
\begin{equation}\label{eq:Q def}
Q_X(\Sigma_t) = \int_{\Sigma_t} T^{ab} X_b d \Sigma_a,
\end{equation}
where $d\Sigma_a$ is the future-directed surface element on $\Sigma_t$. Stokes' theorem, together with $\nabla_a(T^{ab}X_b)=0$, shows that $Q_X$ is conserved: $Q_X(\Sigma_{t_2})=Q_X(\Sigma_{t_1})$ for any $t_1$ and $t_2$. \par
To relate $Q_X$ to our $\dot M$ and $\dot a$, we express $Q_X$ as the difference between two 2-surface integrals. We define the 2-form
\begin{equation}
F_{ab}(h) =  X^c \nabla_{[a} \gamma_{b]c} + \gamma_{c[a}\nabla_{b]}X^c
-X_{[a} \nabla^c \gamma_{b]c}, 
\end{equation}
where $\gamma_{ab} \equiv h_{ab} - \half g_{ab} g^{cd}h_{cd}$. The Einstein equation implies $T^{ab} X_b=\nabla_b F^{ab}$, and so Stokes' theorem implies
\begin{equation}
Q_X = \frac{1}{2}\int_{\partial\Sigma_2} F^{ab}(h) d \Sigma_{ab} - \frac{1}{2}\int_{\partial\Sigma_1}F^{ab}(h) d \Sigma_{ab},
\end{equation}
where $\partial\Sigma_1$ and $\partial\Sigma_2$ are the inner and outer spherical boundaries at $r=r_1$ and $r=r_2$. These integrals are to be evaluated with the perturbation $\hat h^\textrm{N}_{ab}$ at $\partial\Sigma_1$ and with $\hat h^\textrm{N}_{ab}+h^\textrm{S}_{ab}$ at $\partial\Sigma_2$. A lemma due to van de Meent~\cite{vandeMeent:2017fqk} establishes that the integrals receive no contribution from $\hat h^\textrm{N}_{ab}$, which implies
\begin{equation}
Q_X = \frac{1}{2}\int_{\partial\Sigma_2} F^{ab}(h^\textrm{S}) d \Sigma_{ab} = \frac{1}{2}\int_{\partial\Sigma_2} F^{ab}(\dot g) d \Sigma_{ab}.
\end{equation}
In the last equality we have appealed to the fact that the integral vanishes for gauge perturbations~\cite{Dolan:2012jg}. Finally, evaluation of the last integral with $X^a=t^a$ and $X^a=\varphi^a$ yields~\cite{Dolan:2012jg}
\begin{equation}\label{eq:Mdot adot}
Q_t = -\dot M\qquad \text{and} \qquad Q_\varphi = \dot J.
\end{equation}
Therefore $\dot M$ and $\dot J$ can be evaluated from Eq.~\eqref{eq:Q def}. $\dot a$ is then given in terms of these by $\dot a = (\dot J - a\dot M)/M$.

\subsubsection{Held form of $\lie{\xi} g_{ab}$}\label{sec:Held Lie}

Equations~\eqref{Liexigla}--\eqref{Liexigma} express $\lie{\xi} g_{ab}$ in Held form, but they leave $\xi^a$ itself as an arbitrary function of $\rho$. We now use the gauge condition to determine the dependence on $\rho$. 

Given $h^\textrm{S}_{ab}l^b=0=\dot g_{ab}l^b$, Eq.~\eqref{eq:xidet} implies 
\begin{equation}
l^b\lie{\xi} g_{ab} = 0. 
\end{equation}
This equation has been considered in the past by GHZ and Ref.~\cite{PriceThesis}, for example. It can be treated as yet another sequence of ODEs along the integral curves of $l^a$, with the left-hand sides given in Eq.~\eqref{Liexigla}. Using Eq.~\eqref{eq:thorn as d/drho} and integrating with respect to $\rho$, we find the general solution
\begin{subequations}\label{eq:residual IRG}
\begin{align}
\xi_l &= \xi^\circ_l, \nls\\
\xi_n &= \xi^\circ_n + \half \left( \frac{1}{\rho} + \frac{1}{\bar{\rho}} \right) \tilde{\th}' \xi^\circ_l + \rho \bar{\rho} \tau^\circ \bar{\tau}^\circ \xi^\circ_l + \half \left( \Psi^\circ \rho + \bar{\Psi}^\circ \bar{\rho} \right) \xi^\circ_l + \rho \bar{\tau}^\circ \Omega^\circ \xi^\circ_m\nonumber \nls\\
&\quad - \bar{\rho} \tau^\circ \Omega^\circ \xi^\circ_{\bar{m}} - \rho \bar{\tau}^\circ \tilde{\edth} \xi^\circ_l - \bar{\rho} \tau^\circ \tilde{\edth}' \xi^\circ_l, \nls\\
\xi_m &= \frac{1}{\bar{\rho}} \xi^\circ_m + \bar{\rho} \tau^\circ \xi^\circ_l - \tilde{\edth} \xi^\circ_l. \nls
\end{align}
\end{subequations}
This determines $\xi^a$ up to the three Held scalars $\xi^\circ_l \circeq \left\{ 1, 1 \right\}$, $\xi^\circ_n \circeq \left\{ -1, -1 \right\}$, and $\xi^\circ_m \circeq \left\{ 2, 0 \right\}$. 

We now substitute Eq.~\eqref{eq:residual IRG} into the remaining components of $\lie{\xi} g_{ab}$, given in Eqs.~\eqref{Liexigna}--\eqref{Liexigma}. The results are
\begin{align}
\left( \lie{\xi} g \right)_{m\bar{m}} ={}& \half \left( \frac{1}{\rho} + \frac{1}{\bar{\rho}} \right)^2 \rho \bar{\rho} \tilde{\th}' \xi^\circ_l + \left( {\rho'}^\circ \bar{\rho} + \bar{\rho}^{\prime \circ} \rho \right) \xi^\circ_l - \bar{\rho} \tilde{\edth} \tilde{\edth}' \xi^\circ_l - \rho \tilde{\edth}' \tilde{\edth} \xi^\circ_l\nonumber \nls\\
&+ \left( \rho + \bar{\rho} \right) \xi^\circ_n + \frac{\rho}{\bar{\rho}} \tilde{\edth}' \xi^\circ_m - \left( \rho + \bar{\rho} \right) \bar{\tau}^\circ \xi^\circ_m + \frac{\bar{\rho}}{\rho} \tilde{\edth} \xi^\circ_{\bar{m}} - \left( \rho + \bar{\rho} \right) \tau^\circ \xi^\circ_{\bar{m}}, \nls \label{HeldLiexigmmb} \nls\\
\left( \lie{\xi} g \right)_{mm} ={}& 2 \tilde{\edth} \xi^\circ_m + 4 \bar{\rho} \tau^\circ \xi^\circ_m - 2 \bar{\rho} \tilde{\edth} \tilde{\edth} \xi^\circ_l,\label{HeldLiexigmm} \nls\\
\left( \lie{\xi} g \right)_{nm} ={}& \half \bar{\rho} \left( \rho + \bar{\rho} \right) \tau^\circ \left( \rho^{\prime \circ} + \bar{\rho}^{\prime \circ} \right) \xi^\circ_l + 3 \bar{\rho} \tau^\circ \tilde{\th}' \xi^\circ_l + \rho \bar{\rho}^2 \Omega^{\circ 2} \tau^\circ \tilde{\th}' \xi^\circ_l - \bar{\rho} \rho^{\prime \circ} \tilde{\edth} \xi^\circ_l\nonumber \nls\\
&+ \rho \left( \rho + \bar{\rho} \right) \Psi^\circ \tilde{\edth} \xi^\circ_l - 2 \rho \bar{\rho} \bar{\tau}^\circ \tilde{\edth}^2 \xi^\circ_l - \half \bar{\rho} \Omega^\circ \tilde{\edth} \tilde{\th}' \xi^\circ_l - \half \bar{\rho} \left( \rho + \bar{\rho} \right) \tau^\circ \left( \tilde{\edth} \tilde{\edth}' + \tilde{\edth}' \tilde{\edth} \right) \xi^\circ_l\nonumber \nls\\
&+ \bar{\rho} \tilde{\edth} \xi^\circ_n + \bar{\rho} \left( \rho + \bar{\rho} \right) \tau^\circ \xi^\circ_n + 2 \rho \bar{\rho} \tau^\circ \bar{\tau}^\circ \xi^\circ_m + \rho \bar{\rho}^2 \tau^\circ \bar{\tau}^\circ \Omega^\circ \xi^\circ_m + \frac{1}{\bar{\rho}} \tilde{\th}' \xi^\circ_m- \bar{\rho} \Psi^\circ \xi^\circ_m\nonumber \nls\\
&+ \half \bar{\rho} \left( \Psi^\circ + \bar{\Psi}^\circ \right) \xi^\circ_m + 2 \rho \bar{\rho} \bar{\tau}^\circ \Omega^\circ \tilde{\edth} \xi^\circ_m - \rho \left( \rho + \bar{\rho} \right) \Omega^\circ \Psi^\circ \xi^\circ_m - \bar{\rho}^2 \Omega^\circ \tau^\circ \tilde{\edth} \xi^\circ_{\bar{m}}\nonumber \nls\\
&+ \half \bar{\rho} \left( \rho^{\prime \circ} + \bar{\rho}^{\prime \circ} \right) \Omega^\circ \xi^\circ_m + \bar{\rho} \bar{\tau}^\circ \tilde{\edth} \xi^\circ_m + \rho \tau^\circ \tilde{\edth}' \xi^\circ_m - \bar{\rho} \left( \rho + \bar{\rho} \right) {\tau^\circ}^2 \xi^\circ_{\bar{m}},\label{HeldLiexignm} \nls
\end{align}
and 
\begin{align}\label{HeldLiexignn}
\left( \lie{\xi} g \right)_{nn} ={}& 2 \Re \bigg\{ \frac{1}{\rho} \tilde{\th}^{\prime 2} \xi^\circ_l - \rho \bar{\rho} \Omega^\circ \bar{\tau}^\circ \tilde{\edth} \tilde{\th}' \xi^\circ_l - 2 \rho^2 \bar{\rho} \bar{\tau}^{\circ 2} \tilde{\edth}^2 \xi^\circ_l - \rho^2 \bar{\rho} \tau^\circ \bar{\tau}^\circ \left( \tilde{\edth}' \tilde{\edth} + \tilde{\edth} \tilde{\edth}' \right) \xi^\circ_l\nonumber \nls\\
&+ \dfrac{3}{2} \frac{\rho}{\bar{\rho}} \left( \rho + \bar{\rho} \right) \Psi^\circ \tilde{\th}' \xi^\circ_l - \rho^2 \Omega^\circ \Psi^\circ \tilde{\th}' \xi^\circ_l - \rho^{\prime \circ} \tilde{\th}' \xi^\circ_l + 2 \rho \left( \rho + \bar{\rho} \right) \tau^\circ \bar{\tau}^\circ \tilde{\th}' \xi^\circ_l\nonumber \nls\\
&+ \rho^2 \bar{\rho} \Omega^\circ \rho^{\prime \circ} \bar{\tau}^\circ \tilde{\edth} \xi^\circ_l - \rho \left( \rho + \bar{\rho} \right) \rho^{\prime \circ} \bar{\tau}^\circ \tilde{\edth} \xi^\circ_l + 2 \rho^2 \bar{\rho} \bar{\tau}^\circ \Psi^\circ \tilde{\edth} \xi^\circ_l + \rho \bar{\rho} \left( \rho + \bar{\rho} \right) \bar{\rho}^{\prime \circ} \tau^\circ \bar{\tau}^\circ \xi^\circ_l\nonumber \nls\\
&- 2 \rho^2 \bar{\rho}^2 \Omega^\circ \rho^{\prime \circ} \tau^\circ \bar{\tau}^\circ \xi^\circ_l + \rho^2 \rho^{\prime \circ} \Psi^\circ \xi^\circ_l + \tilde{\th}' \xi^\circ_n + 2 \rho \bar{\rho} \bar{\tau}^\circ \tilde{\edth} \xi^\circ_n + 2 \rho \bar{\rho}^2 \tau^\circ \bar{\tau}^\circ \xi^\circ_n\nonumber \nls\\
&+ \rho^2 \Psi^\circ \xi^\circ_n + 2 \rho \Omega^\circ \bar{\tau}^\circ \tilde{\th}' \xi^\circ_m + 2 \rho^2 \bar{\rho} \Omega^\circ \tau^\circ \bar{\tau}^\circ \tilde{\edth}' \xi^\circ_m + \rho \left( \rho + \bar{\rho} \right) \Omega^\circ \rho^{\prime \circ} \bar{\tau}^\circ \xi^\circ_m\nonumber \nls\\
&- \rho^2 \bar{\rho} \Omega^{\circ 2} \bar{\rho}^{\prime \circ} \bar{\tau}^\circ \xi^\circ_m - \rho \bar{\rho} \Omega^\circ \left( \rho \Psi^\circ - \bar{\rho} \bar{\Psi}^\circ \right) \bar{\tau}^\circ \xi^\circ_m + 2 \rho^2 \bar{\rho}^2 \Omega^\circ \tau^\circ \bar{\tau}^{\circ 2} \xi^\circ_m \bigg\}, \nls
\end{align}
where $\xi^\circ_{\bar{m}} \equiv \bar{\xi}^\circ_m$.

\subsubsection{Independent equations in Eq.~\eqref{eq:xidet}}\label{subsec:Solve:eq:xidet}

We now gather the ingredients in Eq.~\eqref{eq:xidet}: Eqs.~\eqref{eq:hhatS Kerr} and \eqref{eq:xS Kerr} for $h^\textrm{S}_{ab}$, Eq.~\eqref{eq:gdot} for $\dot g_{ab}$, and Eqs.~\eqref{HeldLiexigmmb}--\eqref{HeldLiexignn} for ${\cal L}_\xi g_{ab}$. Since $l$ components of Eq.~\eqref{eq:xidet} trivially vanish, the nontrivial components are $m\mb$, $mm$, $mn$, and $nn$. We put these components in the form~\eqref{eq:rho polynomial} by (i) dividing the entire equation by a sufficiently high power of $\bar\rho$  to eliminate all positive powers of $\bar\rho$, and then (ii) using Eq.~\eqref{eq:OmegaH} to replace $\frac{1}{\bar\rho}$ with $\frac{1}{\rho}+\Omega^\circ$.\par
By examining coefficients of powers of $\rho$, we identify seven independent equations:
\begin{itemize}
\item From the $m\mb$ component, which does not involve $\hat h^\textrm{S}_{ab}$, we find
\begin{align}
\label{HeldSDaggerXLiexigmmb:eq1}
a^\circ &= \tilde{\th}' \xi^\circ_l + \half \tilde{\edth} \xi^\circ_{\bar{m}} + \half \tilde{\edth}' \xi^\circ_m, \nls\\
\label{HeldSDaggerXLiexigmmb:eq2}
b^\circ &= -\half \left( \tilde{\edth} \tilde{\edth}' + \tilde{\edth}' \tilde{\edth} - \rho^{\prime \circ} - \bar{\rho}^{\prime \circ} \right)\xi^\circ_l + \xi^\circ_n + \half \Omega^\circ \left( \tilde{\edth}' \xi^\circ_m - \tilde{\edth} \xi^\circ_{\bar{m}} \right) - \bar{\tau}^\circ \xi^\circ_m - \tau^\circ \xi^\circ_{\bar{m}}.
\end{align}
\item From the $mm$ component, which does not involve $x_{ab}$, we find
\begin{align}
\label{HeldSDaggerXLiexigmm:eq1}
\bar{d}^\circ_2 &= \tilde{\edth} \xi^\circ_m, \nls\\
\label{HeldSDaggerXLiexigmm:eq2}
\bar d_1^\circ &= - \tilde{\edth} \tilde{\edth} \xi^\circ_l + 2 \tau^\circ \xi^\circ_m. \nls
\end{align}
\item From the $nm$ component we find
\begin{align}
\label{HeldSDaggerXLiexignm:eq1}
\tilde{\edth}' \bar{d}^\circ_3 + c^\circ &= \tilde{\th}' \xi^\circ_m, \nls\\
\label{HeldSDaggerXLiexignm:eq6}
E^\circ &= \Psi^\circ \tilde{\edth} \xi^\circ_l - \Omega^\circ \Psi^\circ \xi^\circ_m + g^\circ_{\dot a}, \nls
\end{align}
where
\begin{equation}
E^\circ \equiv - \tilde{\edth}' \bar{d}^\circ_0 - \Omega^\circ \tilde{\edth}' \bar{d}^\circ_1 - \Omega^{\circ 2} \tilde{\edth}' \bar{d}^\circ_2 - \Omega^{\circ 3} \tilde{\edth}' \bar{d}^\circ_3 + e^\circ.
\end{equation}
\item From the $nn$ component we find
\begin{equation}
\label{HeldSDaggerXLiexignn:eq3}
F^\circ = \dfrac{3}{2} \left( \Psi^\circ + \bar{\Psi}^\circ \right) \tilde{\th}' \xi^\circ_l + \Omega^\circ \bar{\tau}^\circ \tilde{\th}' \xi^\circ_m - \Omega^\circ \tau^\circ \tilde{\th}' \xi^\circ_{\bar{m}}
+ g^\circ_{\dot M},
\end{equation}
where
\begin{align}
F^\circ &\equiv 2 \Re \Big\{ - \half \tilde{\edth}^{\prime 2} \bar{d}^\circ_1 - \Omega^\circ \tilde{\edth}^{\prime 2} \bar{d}^\circ_2 + \bar{\tau}^\circ \tilde{\edth}' \bar{d}^\circ_2  - \dfrac{3}{2} \Omega^{\circ 2} \tilde{\edth}^{\prime 2} \bar{d}^\circ_3 + 4 \Omega^\circ \bar{\tau}^\circ \tilde{\edth}' \bar{d}^\circ_3 + \dfrac{1}{4} \Omega^{\circ 2} \tilde{\th}' a^\circ \nonumber \nls\\
&\qquad\qquad + \half \Psi^\circ a^\circ
+ \dfrac{3}{2} \Omega^{\circ 2} \tilde{\edth}' c^\circ - 2 \Omega^\circ \bar{\tau}^\circ c^\circ + f^\circ \Big\}. \nls
\end{align}
\end{itemize}

All other equations that follow from Eq.~\eqref{eq:xidet} turn out to be expressible as combinations of these seven. We provide the complete list of equations in~\ref{app:dependent equations}. The seven independent equations are sufficient to determine $\xi^\circ_l$, $\xi^\circ_n$, and $\xi^\circ_m$ (up to the addition of Killing vectors) in terms of $a^\circ, b^\circ, c^\circ, d^\circ_i, e^\circ, f^\circ, g^\circ_{\dot M}$, and $g^\circ_{\dot a}$. Since there are more equations than there are unknowns, the equations will also imply nontrivial relationships between $a^\circ, \ldots, g^\circ_{\dot a}$.

In the next two sections, we outline how to solve Eqs.~\eqref{HeldSDaggerXLiexigmmb:eq1}--\eqref{HeldSDaggerXLiexignm:eq6} and \eqref{HeldSDaggerXLiexignn:eq3} for $\xi^\circ_l$, $\xi^\circ_n$, and $\xi^\circ_m$. Our method must be sufficiently general to cover the case of a particle on a bound geodesic. In that case, the particle's orbit has three (generically incommensurate) frequencies of motion~\cite{Barack:2018yvs}, and $T_{ab}$ has a discrete frequency spectrum containing all harmonics of the orbital frequencies. In a well-behaved gauge, the metric perturbation has this same set of discrete frequencies. The presence of zero-frequency modes  in this spectrum will imply that, just as we found in flat spacetime, a piece of $\xi^a$ necessarily grows with $u$. To preserve the metric's discrete frequencies, that growing piece of $\xi^a$ must be extended throughout the spacetime. 

We split $\xi^a = Y^a + Z^a$, where the temporal Fourier transform of $Y^a$
has support at $\omega=0$, and where $Z^a$ has a vanishing time average, 
\begin{equation}\label{Averop}
\langle Z_a \rangle \equiv \lim_{T\to\infty}\frac{1}{2T}\int_{-T}^T Z_a du = 0.
\end{equation}
Here and below, it is understood that $du$ integrals are performed in 
the above choice of frame and coordinates -- if we wanted to write invariant expressions we would need to use the (GHP-) Lie-derivative \cite{edgar2000integration}. It follows immediately that 
$l^a \mathcal{L}_Y g_{ab} = 0 = l^a \mathcal{L}_Z g_{ab}$, i.e. that the 
corresponding gauge perturbations are in IRG. Consequently, 
the gauge vector fields $Y^a, Z^a$ may be written separately as in \eqref{eq:residual IRG} in terms of $Y_{l}^\circ, Y_{n}^\circ, Y_{m}^\circ$ and $Z_{l}^\circ, Z_{n}^\circ, Z_{m}^\circ$, respectively.
We further write $Y_a = \Xi_a + \langle \xi_a \rangle$ as a sum of a growing-in-$u$
piece $\Xi_a$ and a constant-in-$u$ piece $\langle \xi_a \rangle$. These pieces will not separately produce pure gauge perturbations in IRG and so separately cannot be written as in \eqref{eq:residual IRG}, but we can
nevertheless write
\begin{equation}\label{eq:xi circ decomp}
\xi_{l,n,m}^\circ(u,\theta,\varphi_*) = \Xi_{l,n,m}^\circ(u,\theta,\varphi_*) + \langle \xi_{l,n,m}^\circ \rangle (\theta,\varphi_*) + Z_{l,n,m}^\circ (u,\theta,\varphi_*).
\end{equation}
Here $\Xi_{l,n,m}^\circ + \langle \xi_{l,n,m}^\circ \rangle = Y^\circ_{l,n,m}$ corresponds to the decomposition into growing-in-$u$
and constant-in-$u$ pieces at the level of Held scalars. The determination of $Y_a = \Xi_a + \langle \xi_a \rangle$ (and the corresponding Held-scalars) will be referred to as the ``stationary sector'', and 
that of $Z_a$ as the ``oscillatory sector''.

\subsubsection{Stationary sector}\label{sec:stationary sector}

Our first task will be to determine the growing piece $\Xi^a$ in the gauge vector $Y^a$. Let $\Xi^a$ be a gauge vector which is polynomial-in-$u$ in retarded Kerr-Newman coordinates. The gauge perturbation $\Lie_\Xi g_{ab}$ can only preserve the frequencies of the metric perturbation if $\Lie_\Xi g_{ab}$ is independent of $u$, meaning $\Lie_t \Lie_\Xi g_{ab}=0$, where $t = \dbd{u}$ is the timelike Killing vector. Since $\Lie_t g_{ab}=0$, we have
\begin{equation}
\Lie_t\Lie_\Xi g_{ab}=\Lie_t\Lie_\Xi g_{ab}-\Lie_\Xi\Lie_t g_{ab}=\Lie_{[t,\Xi]}g_{ab}, 
\end{equation}
where $[t,\Xi]^a$ is the commutator. The requirement $\Lie_t \Lie_\Xi g_{ab}=0$ thereby implies $\Lie_{[t,\Xi]}g_{ab}=0$, meaning $[t,\Xi]^a$ can only be a linear combination of Killing vectors. $[t,\Xi]^a$ evaluates to $\partial_u\Xi^a$ in advanced Kerr-Newman coordinates, and so we conclude
\begin{equation}\label{eq:Xi Kerr}
\Xi^a = u(\alpha  t^a + \beta \varphi^a),
\end{equation}
where $\varphi=\dbd{\varphi_*}$ is the axial Killing vector and $\alpha,\beta$ are real constants. In particular, unlike in the flat-space calculation, here there cannot be any terms quadratic in $u$. The reason is that in flat spacetime, $\partial_u\Xi^a$ could be a linear combination of all Killing vectors of flat spacetime, including the generators of boosts, which are themselves linear in $u$. Similarly, Eq.~\eqref{eq:Xi Kerr} must also be modified in Schwarzschild to account for the spacetime's additional Killing vectors. We detail the Schwarzschild case in \ref{app:gauge vec mode decomp}.

Next, we note that $l^b\Lie_t g_{ab}=0=l^b\Lie_\varphi g_{ab}$ trivially. Therefore $t^a$ and $\varphi^a$ can be written in the form~\eqref{eq:residual IRG}. A short calculation shows that the Held coefficients in that form are
\begin{align}
t^\circ_l = 2t^\circ_n = 1, &\qquad t^\circ_m = 0,\label{eq: t Held}\\
\varphi^\circ_l = 2\varphi^\circ_n = -a\sin^2\theta, &\qquad \varphi^\circ_m = \frac{i\sin\theta}{\sqrt{2}}.\label{eq:phi Held}
\end{align}
From this and Eqs.~\eqref{eq:Xi Kerr}, \eqref{eq:residual IRG}, \eqref{eq:xi circ decomp}, we read off
\begin{subequations}\label{eq:Xi Held}
\begin{align}
\Xi^\circ_l &= 2\Xi^\circ_n = u(\alpha - a\beta\sin^2\theta),\\
\Xi^\circ_m &= \frac{u \beta i\sin\theta}{\sqrt{2}}.
\end{align}
\end{subequations}
We can now determine the coefficients $\alpha,\beta$, as well as $\langle \xi^{\circ}_{l,n,m} \rangle$, by substituting the vector~\eqref{eq:xi circ decomp}, into Eqs.~\eqref{HeldSDaggerXLiexigmmb:eq1}--\eqref{HeldSDaggerXLiexignm:eq6} and \eqref{HeldSDaggerXLiexignn:eq3} and picking out the stationary piece of each equation. The results are
\begin{subequations}\label{eq:stationary gauge}
\begin{align}
\langle a^\circ \rangle = &\ \alpha - \frac{3}{2}a\beta\sin^2\theta  +\half\left ( \chand{1}{0}^\dagger \langle \bar{\xi}^\circ_m \rangle + \chand{1}{0} \langle \xi^\circ_m \rangle \right),\label{eq:stationary gauge 1} \nls\\
\langle b^\circ \rangle = &  -\half \left( \chand{1}{0}^\dagger {}_{0} \mathcal{L}_0 + \chand{1}{0} \, \chand{0}{0}^\dagger + 1 \right) \langle \xi^\circ_l \rangle  + \langle \xi^\circ_n \rangle \nonumber \nls\\
&  -ia\cos\theta \left( {}_1 \mathcal{L}_0 \langle \xi^\circ_m \rangle - \chand{1}{0}^\dagger \langle \bar{\xi}^\circ_m \rangle \right)
+ \frac{ia\sin\theta}{\sqrt{2}} \left( \langle \bar{\xi}^\circ_m \rangle-
\langle \xi^\circ_m \rangle \right),  \label{eq:stationary gauge 2} \nls\\
\langle \bar{d}^\circ_2 \rangle =&\ \frac{1}{2} a \beta  \sin^2\theta + \chand{-1}{0}^\dagger \langle \xi^\circ_m \rangle , \label{eq:stationary gauge 3} \nls\\
\langle \bar d^\circ_1 \rangle = &\ 2 i a^2 \beta  \sin^2\theta\cos\theta - \chand{-1}{0}^\dagger \chand{0}{0}^\dagger \langle \xi^\circ_l \rangle - i\sqrt{2}a\sin\theta \langle \xi^\circ_m \rangle,\label{eq:stationary gauge 4} \nls\\
{}_{2} \mathcal{L}_{0} \langle \bar{d}^\circ_3 \rangle + \langle c^\circ \rangle =&\ \frac{i\beta \sin\theta}{\sqrt{2}}, \label{eq:stationary gauge 5} \nls\\
\langle E^\circ \rangle =&\ -\frac{iMa\sin\theta}{\sqrt{2}} (\alpha - a\beta\sin^2\theta-\dot a/a) + M \chand{0}{0}^\dagger \langle \xi^\circ_l \rangle \nonumber \nls\\
&+ 2iMa \cos\theta \langle \xi^\circ_m \rangle , \label{eq:stationary gauge 6} \nls\\
\langle F^\circ \rangle =&\ 3M (\alpha-a\beta\sin^2\theta) + \dot{M}. \nls \label{eq:stationary gauge 7}
\end{align}
\end{subequations}

Here we have written Held's operators in terms of the Chandrasekhar-type operators $\chand{s}{\omega=0}^\dagger, \chand{s}{\omega=0}$ defined in \eqref{eq:Chandop}. The equations can be straightforwardly solved by expanding them in spin-weighted harmonics ${}_s Y_{\lm}(\theta,\varphi_*)$ (see \ref{sec:spin-weighted harmonics}), 
noting that (i) Chandrasekhar's operators act as spin raising and lowering operators, and (ii) the presence of trigonometric functions introduces nearest-neighbour coupling between $\l$ modes. The spin weight of each quantity is $s=(p-q)/2$, where  $\{p,q\}$ are the GHP weights given in Table~\ref{table:GHPWeights}. For example, $\langle a^\circ \rangle =\sum_{\l=0}^\infty\sum_m \langle a^\circ_{\lm} \rangle {}_0 Y_{\lm}$.

We first obtain $\alpha$ and $\beta$ from the $\ell=0$ mode of Eq.~\eqref{eq:stationary gauge 1} and the $\ell=1,\emm=0$ mode of \eqref{eq:stationary gauge 5}, finding
\begin{equation}\label{eq:alpha beta}
\alpha = \sqrt{\frac{1}{4\pi}}\langle a^\circ_{00} \rangle +a\beta,\qquad\beta = -i\sqrt{\frac{3}{4\pi}} \langle c^\circ_{10}\rangle.
\end{equation}
This remarkably simple result fully determines $\Xi^a$ in Eq.~\eqref{eq:Xi Kerr} in terms of integrals over $T_{ab}$.

There are multiple ways of finding the stationary pieces of the transformation, $\langle\xi^{\circ}_a\rangle$,  from Eqs.~\eqref{eq:stationary gauge 1}--\eqref{eq:stationary gauge 6}. For example, Eq.~\eqref{eq:stationary gauge 3} determines the $\l>1$ modes of $\langle \xi^{\circ}_m \rangle$; Eqs.~\eqref{eq:stationary gauge 4} and \eqref{eq:stationary gauge 6} then determine the $\l>0$ modes of $\langle \xi^{\circ}_l \rangle$ and the $\l=1$ mode of $\langle \xi^{\circ}_m \rangle$, up to terms proportional to $\varphi^\circ_l$ and $\varphi^\circ_m$ in ~\eqref{eq:phi Held}; and Eq.~\eqref{eq:stationary gauge 2} then determines $\langle \xi^{\circ}_n \rangle$. The $\l=0$ modes are only determined to be proportional to the Killing terms in Eq.~\eqref{eq: t Held}.

Since there are more equations than unknowns, the equations also encode consistency conditions on the various Held scalars $a^\circ,\ldots, g^\circ_{\dot a}$. Each of the scalars is an integral of the stress-energy tensor, so these conditions may be reducible to stress-energy conservation.

\subsubsection{Oscillatory sector}

The equations in the oscillatory sector are substantially simpler than those in the stationary sector. We seek solutions for the oscillatory piece of $\xi^{\circ}_a$, denoted $Z^\circ_a$ in Eq.~\eqref{eq:xi circ decomp}. To find it, we transform the equations into the frequency domain, writing all quantities in the form 
\begin{equation}
X^\circ(u,\theta,\varphi_*) = \int _{-\infty}^\infty d \omega  X_{\omega}^\circ(\theta, \varphi_*) e^{-i\omega u}.
\end{equation}
From Eqs.~\eqref{HeldSDaggerXLiexignm:eq1}, \eqref{HeldSDaggerXLiexigmmb:eq1}, and \eqref{HeldSDaggerXLiexigmmb:eq2}, we then read off in sequence, 
\begin{subequations}\label{eq:Z soln}
\begin{align}
-i\omega Z^\circ_{m,\omega} &= \chand{2}{\omega} \bar{d}^\circ_{3,\omega} + c^\circ_\omega,\nls \\
-i\omega Z^\circ_{l,\omega} &= a^\circ_\omega - \half \chand{1}{\omega}^\dagger Z^\circ_{\bar{m},\omega} - \half \chand{1}{\omega} Z^\circ_{m,\omega},\nls \\
Z^\circ_{n,\omega} &= b^\circ_\omega +\half \left( \chand{1}{\omega}^\dagger \chand{0}{\omega} + \chand{1}{\omega} \chand{0}{\omega}^\dagger +1\right)Z^\circ_{l,\omega} + ia\cos\theta \left( \chand{1}{\omega} Z^\circ_{m,\omega} - \chand{1}{\omega}^\dagger Z^\circ_{\bar{m},\omega} \right) \nls \nonumber\\
&\quad + \frac{ia\sin\theta}{\sqrt{2}} Z^\circ_{m,\omega} - \frac{ia\sin\theta}{\sqrt{2}} Z^\circ_{\bar{m},\omega},\nls
\end{align}
\end{subequations}
where the Chandrasekhar type operators $\chand{s}{\omega}, \chand{s}{\omega}^\dagger$ are as in \eqref{eq:Chandop}. The remaining four equations [\eqref{HeldSDaggerXLiexigmm:eq1}, \eqref{HeldSDaggerXLiexigmm:eq2}, \eqref{HeldSDaggerXLiexignm:eq6}, and \eqref{HeldSDaggerXLiexignn:eq3} with stationary terms set to zero] become consistency conditions. Again, the equations can be straightforwardly solved by expanding them in spin-weighted harmonics ${}_s Y_{\lm}(\theta,\varphi_*)$, noting that the $a\omega\sin\theta$ terms in Chandrasekhar's operators introduce nearest-neighbour coupling.

\subsection{Summary: metric perturbation in a shadowless gauge}\label{sec:Kerr summary}

We have now found the gauge vector $\xi^a$, which we may decompose as 
\beq
\xi_a=\xi^\textrm{S}_a+\Xi_a, 
\eeq
where $\xi^\textrm{S}_a=\langle\xi_a\rangle+Z_a$. By adding $-\Lie_\xi g_{ab}$ to the GHZ solution~\eqref{eq:GHZ Kerr}, we eliminate the shadow in the region $r>r_\textrm{max}$ and arrive at the generalization of the no-string solution for a spatially compact source in Kerr:
\begin{equation}\label{eq:shadowless Kerr}
h^\textrm{N}_{ab} = \hat h^\textrm{N}_{ab}  + \dot {\tilde g}_{ab} + \tilde h^\textrm{M}_{ab}\Theta_\textrm{M},
\end{equation}
where
\begin{equation}
\hat h^\textrm{N}_{ab} = 2\Re(\mathcal{S}^\dagger_{ab}\Phi^+)\Theta^+_\textrm{M}  + 2\Re(\mathcal{S}^\dagger_{ab}\Phi^-)\Theta^-_\textrm{M}
\end{equation}
is the shadowless reconstructed field in the vacuum region,
\begin{equation}\label{eq:tilde dotg}
\dot {\tilde g}_{ab} = \dot g_{ab}\Theta^+ - \Lie_\Xi g_{ab}\Theta^-
\end{equation}
is ``what is left'' of the shadow field outside the source after the transformation to the shadowless gauge, and
\begin{equation}\label{eq:tilde hM}
\tilde h^\textrm{M}_{ab} = 2\Re(\mathcal{S}^\dagger_{ab}\Phi^\textrm{M}) + x^\textrm{M}_{ab}-\Lie_{\xi^\textrm{S}} g_{ab} - \Lie_\Xi g_{ab}
\end{equation}
is the field inside the source region. The Heaviside functions $\Theta^\pm_\textrm{M}$ are equal to 1 in the vacuum regions $r<r_\textrm{min}$ and $r>r_\textrm{max}$, respectively, and $\Theta_\textrm{M}$ is equal to 1 in the region $r_\textrm{min}<r<r_\textrm{max}$. In Eqs.~\eqref{eq:tilde dotg} and \eqref{eq:tilde hM}, we have extended $\Xi_a$ throughout the spacetime to avoid linear growth of the metric perturbation, and we have allowed the gauge vector $\xi^a_\textrm{S}$ to extend inside the  matter region $r_\textrm{min}<r<r_\textrm{max}$. To prevent a shadow from forming in the region $r<r_\textrm{min}$, we attenuate this vector to zero somewhere inside the matter region.

Although we obtained this solution via a transformation from the GHZ solution that contained a shadow, we note that ultimately, the construction is equivalent to (i) reconstructing the shadowless solution outside the source region, in the same manner one reconstructs the no-string metric perturbation for a point particle, (ii) applying GHZ reconstruction in the interior of the source, with junction conditions at $r=r_{\rm min}$ enforcing the Einstein equation is satisfied there, (iii) adding the gauge perturbations $\Lie_\Xi g_{ab}$ and $\Lie_{\xi_\mathrm{S}} g_{ab}$ in the source region to ensure the Einstein equation is satisfied at $r=r_{\rm max}$.\footnote{To understand how these perturbations can contribute to the Einstein equation, consider the difference between a perturbation $\Lie_{\Theta_M^+\xi} g_{ab}$, which is a vacuum perturbation, and $\Theta_M^+\Lie_{\xi} g_{ab}$, which is not.}

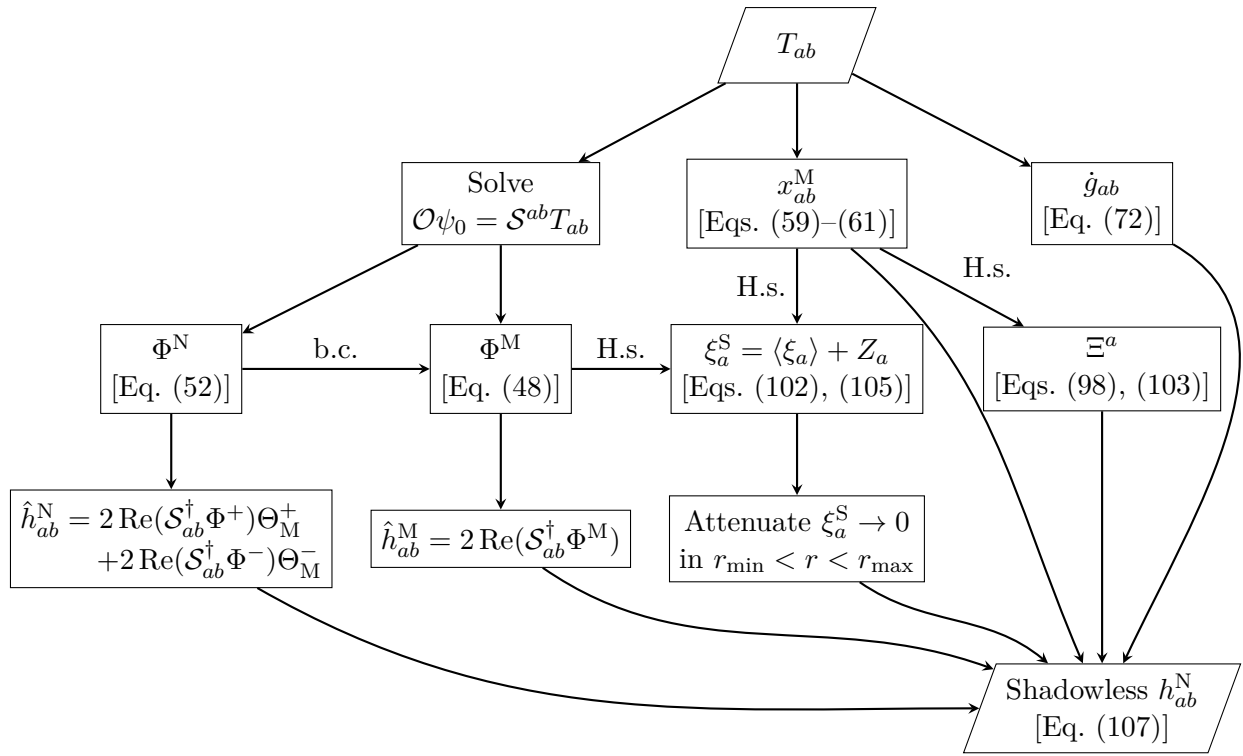
\begin{figure}
  \centering
  \begin{tikzpicture}
    \matrix[row sep=10mm, column sep=5mm]{
      &&\node[io] (Tab) {$T_{ab}$};&\\
      &\node[process] (psi0) {Solve\\$\O\psi_0 = \S^{ab}T_{ab}$};&\node[process] (xab) {$x_{ab}^{\rm M}$\\{[Eqs.~\eqref{eq:xmmbar Kerr}--\eqref{eq:xnn Kerr}]}};&\node[process] (gdot) {$\dot g_{ab}$\\{[Eq.~\eqref{eq:gdot}]}};\\
      \node[process] (PhiN) {$\Phi^{\rm N}$\\{[Eq.~\eqref{eq:-2Rpm}]} };& \node[process] (PhiM) {$\Phi^{\rm M}$\\{[Eq.~\eqref{eq:inversion modes}]}};&
      \node[process] (xiSo) {$\xi^{\textrm{S}}_a=\langle\xi_a\rangle+Z_a$\\{[Eqs.~\eqref{eq:stationary gauge}, \eqref{eq:Z soln}]}};&
      \node[process] (Xi) {$\Xi^a$ \\ {[Eqs.~\eqref{eq:Xi Kerr}, \eqref{eq:alpha beta}]}};\\
      \node[process, align=left] (hhatN) {$\hat h^\textrm{N}_{ab} = 2\Re(\mathcal{S}^\dagger_{ab}\Phi^+)\Theta^+_\textrm{M}$ \\ $\qquad \  + 2\Re(\mathcal{S}^\dagger_{ab}\Phi^-)\Theta^-_\textrm{M}$};&
      \node[process] (hhatM) {$\hat h^\textrm{M}_{ab}=2\Re(\mathcal{S}^\dagger_{ab}\Phi^\textrm{M})$};&
      \node[process] (xiS) {Attenuate $\xi^{\rm S}_a\to0$\\in $r_\textrm{min}<r<r_\textrm{max}$};
      &\\
      &&&\node[io] (hN) {Shadowless $h_{ab}^{\rm N}$\\{[Eq.~\eqref{eq:shadowless Kerr}]}};\\
    };
    \begin{scope}[every path/.style=line]
      \path (Tab) to (psi0);
      \path (Tab) to (xab);
      \path (Tab) to (gdot);
      \path (psi0) to (PhiN);
      \path (psi0) to (PhiM);
      \path (PhiN) to [edge label={b.c.}] (PhiM);
      \path (PhiN) to (hhatN);
      \path (PhiM) to (hhatM);
      \path (xab) to [edge label={H.s.}] (Xi);
      \path (xab) to [edge label'={H.s.}] (xiSo);
      \path (PhiM) to [edge label={H.s.}] (xiSo);
      \path (xiSo) to (xiS);
      \path (hhatN) to [out=-30, in=180] (hN);
      \path (hhatM) to [out=-35, in=160] (hN);
      \path (xiS) to [out=-35, in=140] (hN);
      \path (xab) to [out=-40, in=113] (hN);
      \path (Xi) to (hN);
      \path (gdot) to [out=-30, in=65] (hN);
    \end{scope}
  \end{tikzpicture}
  \caption{Summary of reconstruction procedure in shadowless gauge,
    starting from stress-energy tensor
    $T_{ab}$. Here ``H.s.''~refers to the Held scalars required as
    input to the gauge vectors $\xi^{\rm S}_a$ and $\Xi^a$.}
  \label{fig:flowchart-shadowless-gauge}
\end{figure}

To calculate the solution~\eqref{eq:shadowless Kerr} in practice, one carries out the following procedure (see Fig.~\ref{fig:flowchart-shadowless-gauge}):
\begin{enumerate}[leftmargin=*,labelindent=.75cm,labelsep=.5cm]
    \item[{\bf Step 1}] Solve the Teukolsky equation $\mathcal{O} \psi_0=\mathcal{S}^{ab}T_{ab}$ for $\psi_0$. At the level of a mode decomposition~\eqref{eq:psi0 mode decomp}, this yields the modes ${}_2 R_{\lm\omega}$.
    \item[{\bf Step 2}] From $\psi_0$, obtain the shadowless Hertz potential $\Phi^\textrm{N}$, Eq.~\eqref{eq:PhiN Kerr}, in the regions $r<r_\textrm{min}$ and $r>r_\textrm{max}$, and the Hertz potential $\Phi^\textrm{M}$ in the matter region. In the mode decomposition~\eqref{eq:Phi mode decomp}, $\Phi^\textrm{N}$ is given by Eq.~\eqref{eq:-2Rpm}, and $\Phi^\textrm{M}$ is given by the solution to Eq.~\eqref{eq:inversion modes} subject to the four junction conditions~\eqref{eq:rmin junction} at $r=r_\textrm{min}$.
    \item[{\bf Step 3}] From $\Phi^\textrm{N}$, calculate $\hat h^\textrm{N}_{ab}$ as in Eq.~\eqref{eq:hhatN Kerr} and from $\Phi^\textrm{M}$, calculate $\hat h^\textrm{M}_{ab}=2\Re(\mathcal{S}^\dagger_{ab}\Phi^\textrm{M})$.
    \item[{\bf Step 4}] Add the correction $\dot g_{ab}$ in the vacuum region $r>r_\textrm{max}$ and the corrector tensor $x_{ab}$ in the matter region  $r_\textrm{min}<r<r_\textrm{max}$. These corrections carry invariant content in each region. Here $\dot g_{ab}$ is in the IRG form~\eqref{eq:gdot}, where the mass and spin perturbations $\dot M$ and $\dot a$ are given by Eq.~\eqref{eq:Mdot adot} with Eq.~\eqref{eq:Q def}. The corrector tensor $x_{ab}$ is given by Eqs.~\eqref{eq:xmmbar Kerr}--\eqref{eq:xnn Kerr}. 
    \item[{\bf Step 5}] Complete the metric perturbation by adding the gauge perturbation $-\lie{{\xi^\textrm{S}}} g_{ab}$ in the matter region $r_\textrm{min}< r\leq r_\textrm{max}$, attenuating $\xi^\textrm{S}_a$ to zero somewhere in that region, and the gauge perturbation $-\Lie_\Xi g_{ab}$ in the entire region $r<r_\textrm{max}$. The vector $\Xi^a$ is given by Eq.~\eqref{eq:Xi Kerr} with Eq.~\eqref{eq:alpha beta}. Modulo its attenuation, the vector $\xi^\textrm{S}_a$ is $\langle\xi_a\rangle+Z_a$, where $Z_a$ is in the form~\eqref{eq:residual IRG} with $\xi^\circ_a\to Z^\circ_a$, and $\langle\xi_a\rangle$ can be obtained from Eq.~\eqref{eq:residual IRG} by substituting $\xi^\circ_a\to\Xi^\circ_a+\langle\xi^{\circ}_a\rangle$ on the right-hand side and then setting $u=0$ (such that $\langle\xi_a\rangle$ includes the contribution from $u$ derivatives of $\Xi^\circ_a$). The Held scalars $\Xi^\circ_a$, $\langle\xi^{\circ}_a\rangle$, and $Z^\circ_a$ are obtained from Eqs.~\eqref{eq:Xi Held}, \eqref{eq:stationary gauge} [as described below Eq.~\eqref{eq:alpha beta}], and Eq.~\eqref{eq:Z soln}.
\end{enumerate}
Step 5 requires as input the Fourier modes of the quantities $a^\circ$, $b^\circ$, $c^\circ$, and $d^\circ_3$ for all frequencies, as well as the zero-frequency modes of $d^\circ_0$, $d^\circ_1$, $d^\circ_2$, and $e^\circ$. Equations~\eqref{eq:ab to C1C3}, \eqref{eq:cHeld}, and \eqref{eq:eHeld} give $a^\circ$, $b^\circ$, $c^\circ$, and $e^\circ$. Equation~\eqref{eq:d from B} gives $d^\circ_i$ in terms of $B^\circ_i$, where $B^\circ_i$ is obtained from the four junction conditions~\eqref{eq:rmax junction} at $r=r_\textrm{max}$. The quantity $f^\circ$, given in Eq.~\eqref{eq:fHeld}, is not needed, but it may be used together with Eq.~\eqref{HeldSDaggerXLiexignn:eq3} as a consistency check. 

However, we note that Step 5 can largely be skipped if one is only calculating first-order self-force quantities because the quantities of interest~\cite{Barack:2018yvs} are invariant under the transformation generated by $\xi^\mathrm{S}_a$. In that case, Step 5 only involves finding $\Xi_a$, which only requires the modes $\langle a^\circ_{00}\rangle$ and $\langle c^\circ_{10}\rangle$. We comment on this further in later sections.

\section{Teukolsky puncture scheme}
\label{sec:puncture}

In the preceding section, we formulated metric reconstruction in a shadowless gauge for an extended source. We now outline how that method can be used to calculate the point-mass metric perturbation in a more regular gauge than the standard no-string and half-string radiation gauges.

\subsection{Singular and regular fields}

A core feature of self-force theory is its division of the physical metric into singular and regular pieces~\cite{Poisson:2011nh,Pound:2021qin}. The division is chosen such that (i) the regular piece is a solution to the vacuum Einstein equations, and (ii) the motion of the particle is geodesic in the effective metric $g_{ab}+\varepsilon h^\RDW_{ab}+\varepsilon^2 j^\RDW_{ab}+\ldots$. At linear order, this split has a simple form: $h_{ab}=h^\SDW_{ab}+h^\RDW_{ab}$, where the singular field $h^\SDW_{ab}$ is a certain particular solution to the linearized Einstein equation with a point particle source,
\beq
{\cal E}_{ab}(h^\SDW)=T_{ab},
\eeq
and $h^\RDW_{ab}$ is a certain homogeneous solution,
\beq
{\cal E}_{ab}(h^\RDW)=0.
\eeq
In terms of $h^\RDW_{ab}$, the self-acceleration (as measured relative to a geodesic of the background spacetime) is given by
\beq\label{eq:GSF}
v^b\nabla_b v^a = -\frac{\varepsilon}{2}(g^{ac}+v^a v^c)\left(2\nabla_b h^\RDW_{cd}-\nabla_c h^\RDW_{db}\right)v^b v^d.
\eeq

Based on the fact that $h^\RDW_{ab}$ is a vacuum solution, Ref.~\cite{Keidl:2006wk} proposed long ago that one might be able to calculate it by defining $\psi^\SDW_0 = {\cal T}^{ab}h^\SDW_{ab}$ and $\psi^\RDW_0 = \psi_0 - \psi^\SDW_0$ and then carrying out the CCK procedure starting from $\psi^\RDW_0$. It is not obvious how to implement this idea because $h^\RDW_{ab}$ is only quasilocally defined and satisfies noncausal boundary conditions~\cite{Poisson:2011nh}. However, using our results from the previous section, we {\em can} formulate a variant of it using the notion of a puncture and residual field.

We start with a local expansion of $h^\SDW_{ab}$ around the particle. For example, in Fermi normal coordinates centered on the particle, $h^\SDW_{ab}$ can be written as\footnote{This differs by an overall minus sign relative to formulas in the literature due to our mostly negative signature.}
\beq\label{eq:hS local}
h^\SDW_{ab} = -\frac{2m\delta_{ab}}{s} + O(s),
\eeq
where $\delta_{ab}$ is the Kronecker delta and $s$ is the geodesic distance to the particle. Here we have adopted the Lorenz gauge condition $\nabla^a \left(h_{ab}-\half g_{ab} g^{cd}h_{cd}\right)=0$, which is the common choice for the singular field. Expansions of the form~\eqref{eq:hS local}, and expansions in a similar covariant form, are known analytically through order $s^4$~\cite{Heffernan:2012su,Pound:2014xva} in the Lorenz gauge and through order $s^2$ in the highly regular gauge~\cite{Upton:2021oxf}. For our purpose we can take $h^\P_{ab}$ to be in either of these gauges.

We next define the puncture field $h^\P_{ab}$ to be the local expansion of $h^\SDW_{ab}$ truncated at some finite order $n$ and then attenuated to zero outside some finite region $\mathscr{P}$ around the particle. The remainder, or residual field $h^\RR_{ab} = h_{ab}-h^\P_{ab}$, satisfies the perturbed Einstein equation with an effective source $\Teff_{ab}$, 
\begin{align}\label{eq:EhR=TR}
\E_{ab} (h^\RR) = \Teff_{ab} \equiv T_{ab} - \mathcal E_{ab}(h^\P).
\end{align} 
If $h^\P_{ab}$ is an $n$th-order puncture (i.e., if it includes terms through order $s^n$), then $\Teff_{ab}$ is a $C^{n-2}$ field at $s=0$. Section~\ref{sec:regularity requirements} discusses the behaviour of $\Teff_{ab}$ and its consequences in more detail.

Unlike $h^\RDW_{ab}$, the residual field $h^\RR_{ab}$ is globally defined. Since it reduces to $h_{ab}$ outside the support of the puncture, it also satisfies retarded boundary conditions. This means one can calculate it simply by finding the retarded solution to Eq.~\eqref{eq:EhR=TR}. Historically, this has always been done by imposing the same gauge condition on $h^\RR_{ab}$ as on $h^\P_{ab}$ and then solving Eq.~\eqref{eq:EhR=TR} directly~\cite{Wardell:2015kea,Pound:2021qin}. This has meant in practice that $h^\RR_{ab}$ has always been calculated in the Lorenz gauge. However, in principle $h^\RR_{ab}$ can be in any gauge; it does not need to be in the same gauge as $h^\P_{ab}$. So, in particular, we can solve Eq.~\eqref{eq:EhR=TR} using the reconstruction procedure of the previous section. The resulting field $h^\RR_{ab}$ will contain gauge singularities away from the particle, but they will be weaker than the standard string (or no-string) singularity because the source $\Teff_{ab}$ is less singular than $T_{ab}$. Alternatively stated, by putting the most singular piece of the field, $h^\P_{ab}$, in a gauge that localizes the singularity to the particle, we leave a milder singularity for the reconstruction procedure to transport away from the particle.

We discuss the reconstruction procedure for $h^\RR_{ab}$ next, and in Sec.~\ref{sec:regularity requirements} we assess its singularity structure.

\subsection{Reconstruction of the residual field in the shadowless gauge}\label{sec:hR reconstruction}

We assume the particle's worldline is a bound geodesic that oscillates between a minimum and maximum radius. The support of the puncture, $\mathscr{P}$, can then be confined to a region with $r_{\rm min}<r<r_{\rm max}$, and our reconstruction procedure in Sec.~\ref{sec:Kerr summary} can be applied directly to obtain $h^\RR_{ab}$.

To see how the procedure applies, first note that $h^\RR_{ab}$ has an associated Weyl scalar 
\begin{equation}
    \psi_0^\RR \equiv {\cal T}^{ab}h^\RR_{ab} = \psi_0 - \psi_0^\P,
\end{equation}
which satisfies
\beq\label{eq:OpsiR}
{\cal O}\psi^\RR_0 = T_0^\textrm{eff}\equiv \S^{ab}\Teff_{ab} = \S^{ab}T_{ab} - \O\psi_0^\P 
\eeq
with retarded boundary conditions. Similarly, there is a Hertz potential $\Phi^\RR$ satisfying $\th^4 \widebar{\Phi}^\RR=-2\psi_0^\RR$, from which one can reconstruct $\hat{h}_{ab}^\RR = 2 \Re(\S_{ab}^\dagger\Phi^\RR)$. Finally there is a corrector field $x^\RR_{ab}$ given by Eqs.~\eqref{eq:xmmbar Kerr}, \eqref{eq:xnm Kerr}, and \eqref{eq:xnn Kerr} with $\Teff_{ab}$  substituted  for $T_{ab}$. Adding these pieces together yields the GHZ solution $2\Re(\S_{ab}^\dagger\Phi^\RR) + x^\RR_{ab}$ for $h^\RR_{ab}$. 

$h^\RR_{ab}$ can be brought to the shadowless gauge by solving \eqref{eq:xidet} as described in Sec.~\ref{subsec:Solve:eq:xidet}. $h^\RR_{ab}$ is then given by Eq.~\eqref{eq:shadowless Kerr} with a label $\RR$ placed on almost all quantities. (The label ``M'' no longer refers to a ``matter region'', since the source $\Teff_{ab}$ is only an effective one, but it does refer to the ``middle region'' $r_{\rm min}<r<r_{\rm max}$.) The lone quantity that does {\em not} require an $\RR$ label is $\dot g_{ab}$: the values of $\dot M$ and $\dot a$ remain \eqref{eq:Mdot adot} with Eq.~\eqref{eq:Q def}, where $T_{ab}$ does not carry an $\RR$. This is a consequence of the facts that $\E_{ab}(h^\P+h^\RR)=T_{ab}$ and that $h^\RR_{ab}=h_{ab}$ outside the region $\mathscr{P}$. Given these facts, we can follow the same derivation that led to Eq.~\eqref{eq:Mdot adot}.

\begin{figure}
  \centering
  \begin{tikzpicture}
    \matrix[row sep=10mm, column sep=-5mm]{
      &\node[io] (xp) {Worldline $x_{\rm p}(\tau)$};&\\
      \node[process] (hPLor) {Construct $h_{ab}^{\rm P}$\\in Lorenz or\\ highly regular gauge}; && \node[process] (T) {$T^{ab} = \mu \int v^a v^b \delta^4(x, x_p(\tau))d\tau$};\\
      \node[process] (psi0P) {$\psi_0^{\rm P} = \T^{ab} h_{ab}^{\rm P}$}; && \node[process] (psi0) {Solve\\$\O\psi_0 = \S^{ab}T_{ab}$};\\
      &\node[process,align=left] (psi0R) {$\psi_0^{\rm R} = \psi_0 - \psi_0^{\rm P}$,\\$T_{ab}^{\rm R}=T_{ab}-\E_{ab}(h^{\rm P})$};&\\
      &\node[process] (hRShadowless) {Reconstruct $h^{\rm R}_{ab}$\\ in shadowless gauge\\(Steps 2--5 in Sec.~\ref{sec:Kerr summary})};&\\
      &\node[io] (h) {$h_{ab} = h_{ab}^{\rm P} + h_{ab}^{\rm R}$};&\\
    };
    \begin{scope}[every path/.style=line]
      \path (xp) to (hPLor);
      \path (xp) to (T);
      \path (hPLor) to (psi0P);
      \path (T) to (psi0);
      \path (psi0P) to (psi0R);
      \path (psi0) to (psi0R);
      \path (psi0R) to (hRShadowless);
      \path (hRShadowless) to (h);
      \path (hPLor) to [out=-140, in=150] (h);
    \end{scope}
  \end{tikzpicture}
  \caption{Summary of puncture scheme for constructing
    $h_{ab}$ given a worldline $x_{\rm p}(\tau)$. $h_{ab} =
    h_{ab}^{\rm P} + h_{ab}^{\rm
      R}$ is a sum of two terms, a puncture $h^{\rm
      P}_{ab}$ in Lorenz or highly regular gauge, and a residual piece
    $h^{\rm R}_{ab}$ which is reconstructed in the shadowless gauge.}
  \label{fig:flowchart-puncture}
\end{figure}
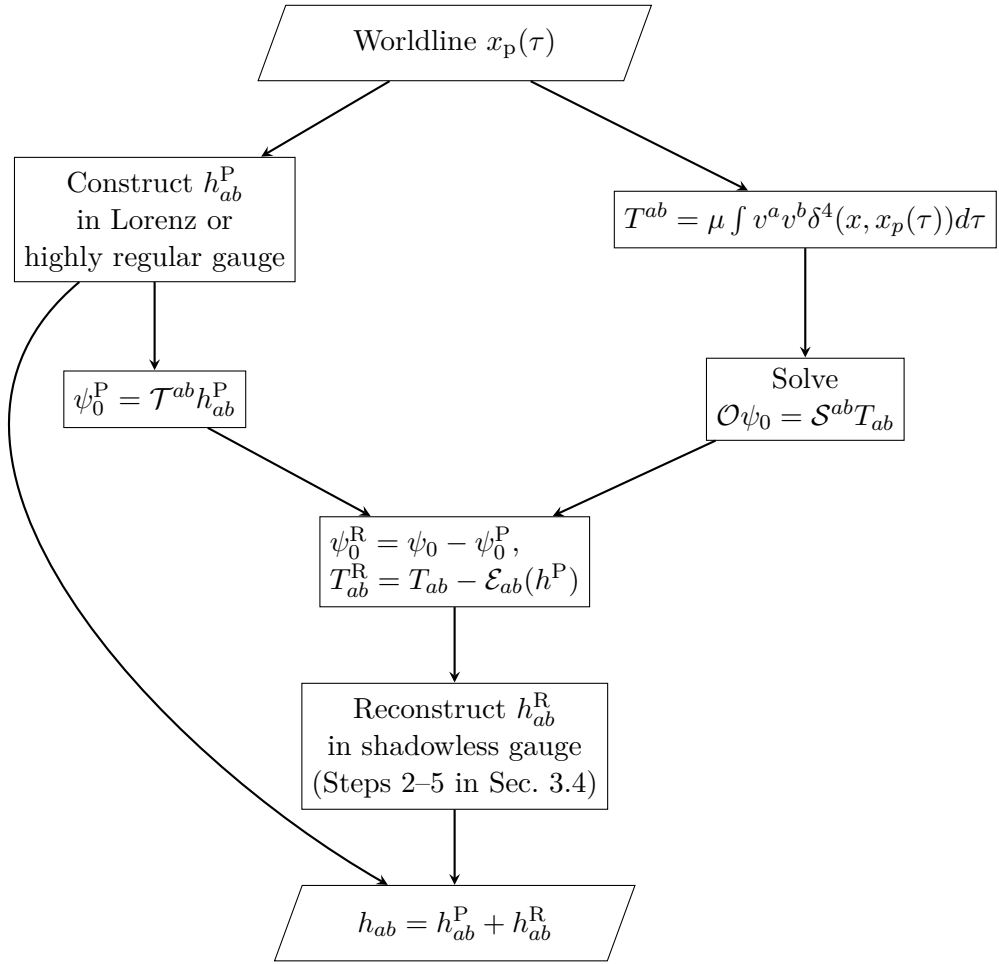

In summary, we have the following procedure, also illustrated in Fig.~\ref{fig:flowchart-puncture}:
\begin{enumerate}[leftmargin=*,labelindent=.5cm,labelsep=.5cm,widest={\bf Steps 2--6}]
\item[{\bf Step 0}] Compute the puncture field $h^{\rm P}_{ab}$ in the Lorenz or highly regular gauge.
\item[{\bf Step 1}] Compute the puncture scalar $\psi_0^\P = \T^{ab}h^\P_{ab}$ and the retarded solution to ${\cal O}\psi_0 = {\cal S}^{ab}T_{ab}$. From them, compute the residual Weyl scalar $\psi_0^\RR = \psi_0 - \psi_0^\P$. 
\item[{\bf Steps 2--5}] Starting from $\psi_0^\RR$ and $\Teff_{ab}$, follow Steps 2--5 of the metric reconstruction procedure in Sec.~\ref{sec:Kerr summary} to obtain $h^\RR_{ab}$ in the shadowless gauge.
\item[{\bf Step 6}] Add the puncture field to obtain the total metric perturbation $h_{ab}=h^{\rm R}_{ab}+h^{\rm P}_{ab}$. 
\end{enumerate}
An alternative Step 1 would be to calculate $\psi^\RR_0$ as the retarded solution to Eq.~\eqref{eq:OpsiR}, without ever calculating $\psi_0$. However, this would generally be more involved and numerically expensive than obtaining $\psi_0$ with a point-particle source. 

Before concluding this section, we emphasize one important point: in the vacuum regions $r<r_\mathrm{min}$ and $r>r_\mathrm{max}$, the calculation of the shadowless field $\hat h^\mathrm{N}_{ab}=2\Re(\S^\dagger_{ab}\Phi^-)\Theta^-_\mathrm{M} + 2\Re(\S^\dagger_{ab}\Phi^+)\Theta^+_\mathrm{M}$ is wholly identical to the calculation for a point particle. In each of the two disjoint regions, $\Phi^-$ (for $r<r_\mathrm{min}$) or $\Phi^+$ (for $r>r_\mathrm{max}$) is determined directly from the point-particle $\psi_0$ in that region. It follows that our total shadowless solution $\hat h^\mathrm{N}_{ab}+\dot{\tilde g}_{ab}$ in these regions is identical to the no-string solution, up to the choice of gauge for $\dot g_{ab}$ in the region $r>r_\mathrm{max}$ and the value of the gauge vector $\Xi^a$ in the region $r<r_\mathrm{min}$ (which is also tied to the gauge of $\dot g_{ab}$).

\subsection{Softened string and regularity requirements}\label{sec:regularity requirements}

To serve our purposes, the metric perturbation $h_{ab}=h^{\rm R}_{ab}+h^{\rm P}_{ab}$ should satisfy two regularity requirements: (i) the  self-force~\eqref{eq:GSF} should be calculable with the simple replacement $h^\RDW_{ab}\to h^\RR_{ab}$, and (ii) $G^{(2)}_{ab}(h,h)\sim h\partial\partial h + \partial h\partial h$ should be well defined as a distribution. In the remainder of this section, we assess whether our puncture method meets these requirements.

The field $h_{ab}$ that we obtain with our procedure has two types of singularities: a singularity in $h^\P_{ab}$ at the particle, with the form~\eqref{eq:hS local} (or a similar form in the highly regular gauge); and a softened string singularity in $h^{\rm R}_{ab}$ that extends away from the particle. In the shadowless gauge, the softened string singularity is confined to the middle region $r_{\rm min}<r<r_{\rm max}$. We can understand its form by analyzing the form of $\Teff_{ab}$ together with the equations~\eqref{eq:xmmb}, \eqref{eq:xnm}, and \eqref{eq:xnn} for $x^\RR_{ab}$ and the inversion relation $\th^4\Phi^\RR = -2\psi^\RR_0$ for $\Phi^\RR$.

Start by considering $\Teff_{ab}$. We can write it as $\Teff_{ab} = \E_{ab} (h^\SDW-h^\P)=\E_{ab}(h^{\SDW,n+1})+O(s^n)$, where $n$ is the highest power of $s$ included in $h^{\rm P}_{ab}$, and $h^{\SDW,n+1}_{ab}$ is the order-$s^{n+1}$ term in the local expansion of $h^{\SDW}_{ab}$. Concretely, in the Lorenz gauge~\cite{Pound:2014xva}
\beq
h^{\SDW,n+1}_{ab}\sim \frac{(\Delta x)^{3n+6}}{s_0^{2n+5}},
\eeq
where $x^a$ are any smooth coordinates, $\Delta x^a$ denotes the coordinate distance from a reference point $x_\mathrm{p}$ on the worldline, and $(\Delta x)^k$ denotes a polynomial of homogeneous order $k$, which we can write as $P_{ab a_1\cdots a_k}(x_\mathrm{p})\Delta x^{a_1}\cdots \Delta x^{a_k}$ for some $P_{aba_1\cdots a_k}$. The quantity $s_0$ is the leading term in a coordinate expansion of $s$, given by $\sqrt{-(g_{ab}+v_a v_b)\Delta x^a\Delta x^b}$ with $g_{ab}$ and $v_a$ evaluated at $x_\mathrm{p}$. $h^{\SDW,n+1}_{ab}$ is $C^n$ at $s=0$ and scales as $\sim s^{n+1}$, implying $\Teff_{ab}$ is $C^{n-2}$ there and scales as $\sim s^{n-1}$. Using the fact that a derivative acts as $\partial s_0 \propto (\Delta x)^1/s_0$, we can write
\beq\label{eq:Teff form}
\Teff_{ab}\sim \frac{(\Delta x)^{3n+8}}{s_0^{2n+9}}.
\eeq

We bring Eq.~\eqref{eq:Teff form} into a more explicit but simple form by adopting a local coordinate system $(\tau,x,y,z)$ that is comoving with the particle and orthonormal on the worldline, where $\tau$ is proper time along the worldline, $x=y=z=0$ there, and $g_{ab}={\rm diag}(1,-1,-1,-1)$ there. Choose the axes such that $l \propto (\partial_\tau+\partial_z)$ at the worldline, where the constant of proportionality is positive. $z<0$ then corresponds to $r<r_\mathrm{p}$, and $z>0$ to $r>r_\mathrm{p}$. It will also be useful to define the cylindrical coordinates $(z,\varrho,\phi)$ such that $x=\varrho\cos\phi$ and $y=\varrho\sin\phi$. $z$ will represent distance along the string, $\varrho$ will represent a distance from it, and $\phi$ will represent the angle around it. An example of such local coordinates is constructed explicitly in Ref.~\cite{Pound:2014xva}. In these coordinates the denominator in Eq.~\eqref{eq:Teff form} is an odd power of $s_0=\sqrt{z^2+\varrho^2}$. The numerator is a sum of terms of the form $P_{n_1 n_2 n_3}(\tau)z^{n_1}(\varrho\cos\phi)^{n_2}(\varrho\sin\phi)^{n_3}$ with $n_1+n_2+n_3=3n+8$, where $P_{n_1 n_2 n_3}$ is a smooth function of $\tau$. 

Finding $x^\RR_{ab}$ and $\S^\dagger_{ab}\Phi^\RR$ involves integrating such functions along integral curves of $l^a$. For simplicity, we can extend the local coordinates off the worldline and choose the normalization of $l^a$ in such a way that $l=l^\tau\partial_\tau+\partial_z$, making $z$ the parameter along the integral curves. Along one of those curves, we then have $P_{n_1 n_2 n_3}(\tau)=P_{n_1 n_2 n_3}(\tau_0)+O(s)$, where $\tau_0$ is the value of $\tau$ at which $z=0$ on the curve. We can discard the subleading term, meaning that $P_{n_1 n_2 n_3}$ can be moved outside the integral. 

If we now integrate a function of the form~\eqref{eq:Teff form} with respect to $z$, with regular boundary conditions at some $z<0$, then we propagate the singularity at $s=0$ along the entire curve $\varrho=0$ for $z>0$. More precisely, a $C^{k-1}$ function of $(x,y,z)$ scaling as $\sim s^k$ near $s=0$ is transported as a $C^k$ function of $(x,y)$ scaling as $\sim\varrho^{k+1}$ near $\varrho=0$. 
Hence, the integration increases the singularity's dimension by one but reduces its strength by one as well. Once we have introduced this string singularity through a first integration, we have simple rules for how further linear operations affect the strength of the singularity. Additional $z$ integrations do not alter the scaling with $\varrho$ or differentiability at $\varrho=0$, and $z$ derivatives likewise have no affect unless they act on an only once-integrated function. $\tau$ derivatives and multiplication by a smooth function have no effect, while an $x$ or $y$ derivative decreases the scaling and differentiability by one order. All of these statements are straightforwardly verified using the explicit functions of $(z,\varrho,\phi)$ described in the previous paragraph.

Now consider $x^\RR_{ab}$ and $\S^\dagger_{ab}\Phi^\RR$. By virtue of Eq.~\eqref{eq:xmmb} and the above rules, $x^\RR_{m\bar m}\sim\varrho^{n}$ at the string. Similarly, Eqs.~\eqref{eq:xnm} and \eqref{eq:xnn} imply that $x^\RR_{mn}\sim \varrho^{n-1}$ and $x^\RR_{nn}\sim \varrho^{n-2}$. To examine $\Phi^\RR$, first note that 
\beq
\psi^\RR_0 = \T^{ab}(h_{ab}-h^\P_{ab}) = \T^{ab}(h^\RDW_{ab}) + \T^{ab}(h^{\SDW,n+1}_{ab})+O(s^n). 
\eeq
Since $\psi_0$ is gauge invariant, we can use the Lorenz-gauge fields in these equalities.  $\T^{ab}(h^\RDW_{ab})$ is $C^\infty$,  while $\T^{ab}(h^{\SDW,n+1}_{ab})$ has the same singularity structure as $\E_{ab}(h^{\SDW,n+1})$. So $\psi^\RR_0$ has the same singular form~\eqref{eq:Teff form} as $\Teff_{ab}$, meaning it scales as $\sim s^{n-1}$ and is $C^{n-2}$ at the particle. The inversion relation $\th^4\Phi^\RR = -2\psi^\RR_0$ then propagates this singularity along the string, implying $\Phi^\RR\sim \varrho^{n}$ at the string. Consequently, the metric perturbation $\S^\dagger_{ab}\Phi^\RR$ is $\sim \varrho^{n-2}$ at the string, like $x^\RR_{ab}$. In all cases, a function scaling as $\sim \varrho^k$ is a $C^{k-1}$ function of $(x,y)$ at $\varrho=0$

We can therefore conclude that, at worst, $h^\RR_{ab}$ behaves as $\sim \varrho^{n-2}$ and is $C^{n-3}$ at the string. Given this estimate, the highest-order puncture available in the literature would lead to a $\sim \varrho^2$, $C^1$ residual field on the string. However, this very likely overestimates the strength of the singularity. Our flat-spacetime results indicate that the strongest singularities cancel between $x^\RR_{ab}$ and ${\cal S}^\dagger_{ab}\Phi^\RR$. Section~\ref{sec:S*Phi+x metric} and \ref{sec:mode sums} indicate that without a puncture, $x_{ab}$ and $\S^\dagger_{ab}\Phi$ each behave as $\sim\varrho^{-4}$. Here we formally write $\sim\varrho^{-2-k}$ for the $k$th derivative of the angular delta function because it can be written as $k+2$ derivatives of a function $\sim\ln\varrho$.\footnote{This counting could be rephrased in terms of the order of the distribution.} If we note that the unpunctured case corresponds to $n=-2$, making $\Teff_{ab}=T_{ab}$, then we see that this $\sim\varrho^{-4}$ behaviour agrees with the prediction from the general scaling arguments. But the total GHZ solution $h_{ab}$ in the flat-spacetime case diverges as only $\sim\varrho^{-2}$, and this is also known to be true in Kerr~\cite{Pound:2014xva}. This suggests that $h^\RR_{ab}$ could behave as $\sim \varrho^{n}$ for an $n$th-order puncture, rather than as $\sim\varrho^{n-2}$. A more detailed local analysis, following the approach in Sec.~III of Ref.~\cite{Pound:2013faa}, could confirm this regularity conjecture. We defer that analysis to a future paper.

Characterizing the regularity of $h_{ab}$ also entails assessing the behaviour of $h^\RR_{ab}$ at $s=0$. This is a simpler task because regardless of the string, an integral always increases the scaling with $s$ and a derivative always decreases it. Starting from $\Teff_{ab}\sim \psi^\RR_0\sim s^{n-1}$, referring to Eqs.~\eqref{eq:xmmb}--\eqref{eq:xnn}, and counting integrals and derivatives, we infer $x^\RR_{m\bar m}\sim x^\RR_{mn}\sim\S^\dagger_{ab}\Phi^\RR\sim s^{n+1}$,  and $x^\RR_{nn}\sim s^n$. However, the flat-space calculation again suggests that the solution, specifically $x^\RR_{nn}$, will be more regular than this. Equation~\eqref{eq:x inner outer} with \eqref{eq:Fab} show that every component of $x_{ab}$ behaves as $\sim s^{-1}$, two orders more regular than the source (formally treating the angular delta function as $\sim 1/s^2$ and the three-dimensional delta function as $\sim 1/s^3$). This agrees with the generic predictions above for $x_{m\bar m}$ and $x_{mn}$,  but it is one order more regular than predicted for $x_{nn}$. The additional order of regularity, which arises from cancellations on the right-hand side of Eq.~\eqref{eq:xnn flat}, suggests $x^\RR_{nn}\sim s^{n+1}$. Therefore we conjecture that $h^\RR_{ab}$ is generally two orders smoother at $s=0$ than $\Teff_{ab}$. Or in other words, for a puncture of order $n$, $h^\RR_{ab}$ is $C^{n}$ at $s=0$.

We can now estimate the order $n$ that is required to satisfy the two regularity criteria listed at the beginning of the section. For this we adopt the two conjectures that $h^\RR_{ab}$ contains a $\sim \varrho^{n}$ ($C^{n-1}$) string singularity and a $\sim s^{n+1}$ ($C^{n}$) singularity at $s=0$. The same analysis can also be performed with the worst-case estimates $\sim \varrho^{n-2}$ ($C^{n-3}$) and $\sim s^{n}$ ($C^{n-1}$).

To see what is required to use $h^\RR_{ab}$ to calculate the self-force~\eqref{eq:GSF}, we note that if $h^\RR_{ab}$ is $C^1$ at $s=0$, then $h^\RR_{ab}+h^\P_{ab}$ is related to the Lorenz-gauge (or highly regular gauge) solution by a differentiable gauge transformation. It then follows ~\cite{Pound:2013faa,Pound:2017psq} that we may use $h^\RR_{ab}$ in place of $h^\RDW_{ab}$ in Eq.~\eqref{eq:GSF}. Therefore any $n\geq 1$ meets our requirement for computing the self-force.

Now we turn to $G^{(2)}_{ab}(h,h)$. Since it is bilinear in its arguments, we may divide it into $G^{(2)}_{ab}(h^\P,h^\P)$, $G^{(2)}_{ab}(h^\P,h^\RR)$ (defining it such that it is symmetric in its arguments whenever the arguments differ), and $G^{(2)}_{ab}(h^\RR,h^\RR)$. $G^{(2)}_{ab}(h^\P,h^\P)$ is smooth away from the particle. At the particle it behaves as $\sim s^{-4}$ if $h^\P_{ab}$ is in the Lorenz gauge or as $\sim s^{-2}$ if $h^\P_{ab}$ is in a highly regular gauge. It is well understood how to obtain the physical solution to the second-order field equation~\eqref{eq:2nd Einstein} with these sources~\cite{Pound:2012dk,Upton:2021oxf}. The cross terms $G^{(2)}_{ab}(h^\P,h^\RR)$ behave as $\sim \varrho^{n-2}$ at the string (arising from two derivatives on $h^\RR_{ab}$) and as $\sim s^{-3}$ at the particle (arising from two derivatives on $h^\P_{ab}$). Using the area element $\varrho dzd\varrho d\phi$, we see that the string singularity is locally integrable if $n\geq2$. Since $G^{(2)}_{ab}(h^\P,h^\RR)$ is linear in $h^\P_{ab}$, its $\sim s^{-3}$ singularity at the particle arises from the action of a linear second-order differential operator (with $C^n$ coefficients) on the integrable function $h^\P_{ab}$, meaning it is also well defined as a distribution if $n\geq2$. Last, $G^{(2)}_{ab}(h^\RR,h^\RR)$ behaves as $\sim \varrho^{2n-4}$, implying that $G^{(2)}_{ab}(h^\RR,h^\RR)$ is locally integrable at the string for any puncture with $n\geq2$.

Combining the above estimates, we infer that our regularity requirements are satisfied if the puncture is of order $n\geq2$. Since $n=2$ punctures are commonly used in practice (e.g., in ~\cite{Pound:2019lzj,Warburton:2021kwk}), this does not represent an obstacle. If our regularity conjectures are incorrect, then the requirements for the puncture will be more stringent. However, the next section will provide additional evidence for the validity of our conjectures.

\section{Demonstration of the puncture scheme: return to flat spacetime}\label{sec:return to flat space}

As a first test of our puncture scheme, we return to the model problem of a static particle in flat spacetime. This will illustrate the procedure as well as confirm many of its key aspects. We closely follow the steps outlined in Sec.~\ref{sec:hR reconstruction}, and we adopt the same coordinates and tetrad as in Sec.~\ref{sec:static particle}.

\subsection{Step 0: construction of $h^\P_{ab}$}

The procedure begins by constructing a puncture. As mentioned above, typical puncture schemes obtain the puncture from the singular field in the Lorenz gauge, in which it has the easily identifiable, Coulomb-type structure~\eqref{eq:hS local} near the worldline. For the static particle in flat spacetime, in inertial coordinates that are static with respect to the particle, the leading-order term in Eq.~\eqref{eq:hS local} is the exact solution to $\E_{ab}(h)=T_{ab}$, and the regular field $h^\RDW_{ab}$ vanishes. We write this solution in covariant form as
\beq\label{eq:hL flat}
h^\L_{ab} = \frac{\mu (g_{ab}-2v_a v_b)}{4\pi s}, 
\eeq
defining  $v^a$ at points off the worldline by parallel transport. In the retarded coordinates of Sec.~\ref{sec:static particle}, $v=\partial_u$ and the proper spatial distance from the worldline is $s=\sqrt{r^2+r^2_\mathrm{p}-2rr_\mathrm{p}\cos\gamma}$, where $\gamma$ is the angle between $(\theta,\varphi)$ and $(\theta_\mathrm{p},\varphi_\mathrm{p})$. We can immediately write Eq.~\eqref{eq:hL flat} in terms of the tetrad~\eqref{eq:NP tet flat} by noting $v^a=n^a+\tfrac{1}{2}l^a$:
\beq\label{eq:hL flat tetrad}
h^\L_{ab} = -\frac{\mu}{2\pi s} (\tfrac{1}{4} l_a l_b +n_a n_b +m_{(a}\bar m_{b)}).
\eeq

We take the punctured region to be a shell of radial width $2c>0$ extended in time:
$\mathscr{P}=\mathbb{S}^2  \times \{ r \in (\rp-c,\rp+c) \} \times \mathbb R$, such that $r_\mathrm{min}=r_\mathrm{p}-c$ and $r_\mathrm{max}=r_\mathrm{p}+c$.
Inside $\mathscr{P}$ we cut out the field around the particle completely.
To do this, we introduce a radial window function $\W$ with compact support around $r=\rp$ such that the puncture field reads
\begin{equation}\label{eq:punc choice}
h^\P_{ab} = \W h^\L_{ab}.
\end{equation} 
In the Lorenz gauge, the residual metric is then simply the total Lorenz gauge metric outside of $\mathscr{P}$, or
\begin{equation}\label{eq:res def}
    h^\RR_{ab} = (1-\W)h^{\L}_{ab}.
\end{equation}
The residual field we obtain with our reconstruction procedure will necessarily be related to this field by a gauge transformation. 

To make our example concrete, we take the window function to be a box distribution of width $2c$ centered at the particle,
\begin{align}\label{eq:W box}
\W = \Theta(r-r_{\rm min}) - \Theta(r-r_{\rm max}) = \Theta_{\textrm M}.
\end{align}
With this choice, the Lorenz-gauge residual field~\eqref{eq:res def} identically vanishes in the puncture region. As a consequence, the residual field we obtain with our procedure will be pure gauge in this region.

In implementations of puncture schemes, the puncture is typically decomposed into a basis of modes. Using the standard decomposition of $1/s$ in terms of Legendre polynomials ${P}_\l$, we write
\begin{equation}\label{eq:hL static}
h^\L_{ab} =  -\frac{\mu}{2\pi} (\tfrac{1}{4} l_a l_b +n_a n_b +m_{(a}\bar m_{b)}) \sum_{\l\geq 0} \frac{r_<^\l}{r_>^{\l+1}} \, P_\l(\cos\gamma),
\end{equation}
which can be expressed in terms of spherical harmonics using ${P}_\l(\cos\gamma) = \frac{4\pi}{2\l+1} \sum_{m=-\l}^\l \bar{Y}_{\lm}^{\textrm{p}} Y_{\lm}$.
In our flat-spacetime toy problem, we have the luxury of obtaining each $\lm$ mode as an exact function of $r$. However, in curved spacetime the modes are obtained by directly performing the integral of the puncture against a spherical harmonic. Evaluating those integrals analytically requires expanding them in powers of the radial distance $\Delta r\equiv r-r_\mathrm{p}$~\cite{Wardell:2015ada}. We can see the effect of this expansion by carrying it out on Eq.~\eqref{eq:hL static}. The only $r$ dependence in the solution is
\beq\label{eq:Dr expansion}
\frac{r^\l_<}{r^{\l+1}_>} = \frac{1}{r_>}\left[1-\l_\pm (\Delta r/\rp) + \frac{1}{2}\l_\pm(\l_\pm+1)(\Delta r/\rp)^2+O(\Delta r/\rp)^3\right], 
\eeq
where the $\pm$ signs in $\l_\pm \equiv \pm\l$ apply for $r>r_\mathrm{p}$ ($+$) and $r<r_\mathrm{p}$ ($-$) respectively. The $n$th-order term in this expansion grows as $\sim \l^n$. Hence, if we truncate the expansion at any finite order, the sum over $\l$ diverges for all $\Delta r\neq 0$. Away from $\Delta r=0$, $h^\P_{ab}$ is perfectly smooth and the sum~\eqref{eq:hL static} converges exponentially with $\l$, but the expansion in $\Delta r$ has made it divergent at all points (a consequence of $\Delta r=0$ being a singular point of the sum). Because of this feature, we avoid such an expansion here. We comment on the practical implications of Eq.~\eqref{eq:Dr expansion} in the conclusion.

\subsection{Step 1: construction of $\psi^\RR_0$}

The  puncture field \eqref{eq:punc choice} gives rise to a spin-(+2) Weyl puncture
\begin{align}\label{eq:psi0P flat}
\psi_0^\P =\half \edth^2 h^\P_{ll}
=  \half \Theta_{\textrm M} \,\edth^2 h^\L_{ll} = \Theta_{\textrm M}\psi_0
\end{align}
inside $\mathscr{P}$. In this case, the differential operators are purely angular, meaning no derivatives act on the window function $\Theta_{\textrm M}$. That will not be the case generically.

We already obtained the retarded $\psi_0$ in Sec.~\ref{sec:CCK-Ori flat}. Due to the simple form of $\psi^\P_0$, the residual Weyl scalar is simply 
\beq
\psi^\RR_0 = \psi_0-\psi^\P = (1-\Theta_{\textrm M})\psi_0,
\eeq
equal to $\psi_0$ outside $\mathscr{P}$ and vanishing inside. Its modes can be read off Eq.~\eqref{eq:psi0lm sol}.

\subsection{Steps 2 and 3: $\Phi^\RR$ and the reconstructed field $\hat h^\RR_{ab}$}

We next calculate the shadowless Hertz potential 
$\Phi^\RR$ and reconstruct the field $\hat h^\RR_{ab}$.
Both in and outside the puncture region $\mathscr{P}$, the fields satisfy  $\partial^4_r\bar\Phi^\RR=-2\psi^\RR_0$ and $\hat h^\RR_{ab} = 2\Re\S^\dagger_{ab}\Phi^\RR$. However, in the shadowless solution these equations do not necessarily hold on the boundaries between regions.

Outside the puncture region, the relevant potential is the shadowless potential $\Phi^\mathrm{N}$:
\beq
\Phi^\RR=\Phi^\mathrm{N} = \Phi^+\Theta^+_{\textrm M} + \Phi^-\Theta^-_{\textrm M} \qquad\text{outside } \mathscr{P}.
\eeq
$\Phi^\pm$ are the same functions appearing in the first term in Eq.~\eqref{eq:Phi modes}. They satisfy the inversion relation $\partial_r^4\bar\Phi^\pm = -2\psi_0$, with $\Phi^+$ obtained by integrating from $r=\infty$ and $\Phi^-$ obtained by integrating from $r=0$. 

Inside the puncture region, $\psi^\RR_0=0$, making the Hertz potential a solution to $\partial^4_r\Phi^\RR=0$. We can hence write its modes as
\beq
\Phi^\RR_{\lm} = \Phi^\mathrm{M}_{\lm} = \sum_{j=0}^3 \delta^\circ_{j,\lm}\rho^{-j}\qquad\text{in }\mathscr{P},
\eeq
where we follow the notation of Sec.~\ref{sec:Kerr summary} in labelling the solution in this region with an M. 
The coefficients $\delta^\circ_j$ are determined by the junction conditions~\eqref{eq:rmin junction} across $r_\textrm{min}$, which in this case read simply $\partial_r^n\Phi^{\textrm M}=\partial^n_r\Phi^{\textrm N}$ at $r=r_\mathrm{min}$ for $n=0,\ldots,3$. These conditions enforce that $\Phi^\RR$ {\em does} satisfy $\partial_r^4\Phi^\RR=-2\psi^\RR_0$ at $r_\textrm{min}$. Evaluating the junction conditions and solving for $\delta^\circ_{j,\lm}$, we find
\begin{subequations}\label{eq:delta modes}
\begin{align}
\delta^\circ_{0,\lm} &= -\frac{\mu  r_\textrm{p} \C{-2}\bYp{}}{6(2+\l)(2\l+1)}  (1-c/r_\textrm{p})^{2+\l}, \\
\delta^\circ_{1,\lm} &=  -\frac{\mu \C{-2}\bYp{}}{2(1+\l)(2\l+1)}  (1-c/r_\textrm{p})^{1+\l} ,\\
\delta^\circ_{2,\lm} &=  -\frac{\mu \C{-2}\bYp{}} {2r_\textrm{p}\l(2\l+1)}  (1-c/r_\textrm{p})^{\l} ,\\
\delta^\circ_{3,\lm} &=  -\frac{\mu \C{-2}\bYp{}} {6r_\textrm{p}^2(\l-1)(2\l+1)}  (1-c/r_\textrm{p})^{\l-1}. 
\end{align}
\end{subequations}
Since $0<c/r_\textrm{p}<1$, the factor $(1-c/r_\mathrm{p})^\l$ decays exponentially with $\l$. This implies that, as expected, subtracting the puncture from the metric perturbation has eliminated the string singularity in the Hertz potential. In the general case in Kerr, subtracting a finite-order puncture will only soften the string; in the present model problem, the leading-order puncture is the exact singular field, so the the string has been entirely removed.

Moving to Step 3 in our procedure, we apply the reconstruction operator $\S^\dagger_{ab}$ to these potentials. Outside the punctured region, $\mathscr{P}$, the reconstructed metric is
\begin{equation}\label{eq:hR ouside}
    \hat h^\RR_{ab} = \hat h^\mathrm{N}_{ab} =
    \begin{cases}
    2 \Re (\S^\dagger_{ab}\Phi^+) , &\qquad r>r_\mathrm{max}, \\
    2\Re(\S^\dagger_{ab}\Phi^-),&\qquad r<r_\mathrm{max}, 
    \end{cases}
\end{equation}
and 
\begin{equation}\label{eq:hR inside}
    \hat h^\RR_{ab} = \hat h^\mathrm{M}_{ab} = 2 \Re \left(\S^\dagger_{ab}\sum_{j=0}^3 \delta_j^\circ \rho^{-j}\right)\qquad\ \ \  \text{inside }\mathscr{P}.
\end{equation}
$\hat h^\mathrm{M}_{ab}$ is given explicitly by the formula~\eqref{eq:hhatS Kerr} with the replacement $d^\circ_j\to \delta^\circ_j$ and the simplifications $\Omega^\circ=\tau^\circ=0$ and $\rho=-1/r$. $\hat h^\mathrm{N}_{ab}$ is identical to the no-string reconstructed field~\eqref{eqs:hhat}, but now restricted to the region outside $\mathscr{P}$.

\subsection{Step 4: invariant correction terms}

To complete the metric inside and outside of $\mathscr{P}$, we must add two quantities containing invariant information: the corrector tensor $x^\RR_{ab}$ in $\mathscr{P}$, and an analog of $\dot{g}_{ab}$ in the vacuum region $r>r_\mathrm{max}$.

Start with the analog of $\dot g_{ab}$, which is made up of the $\l=0,1$ modes. It must be written in the IRG to be compatible with our gauge corrections in the next section. The only nonvanishing $\l=0$ piece is simply
\begin{align}
\tilde h^{\l=0}_{nn} &= -\frac{2m}{r} \qquad\text{for } r>r_\mathrm{max}.
\end{align}
This is the flat-spacetime limit of Eq.~\eqref{eq:gdot IRG}; it differs from the $m/r$ term in Eq.~\eqref{eq:x ell=0} by a gauge refinement that eliminates the $m\bar m$ component. The $\l=1$ term is given by the terms proportional to the mass dipole moment $\mu_i$ in Eq.~\eqref{eq:x ell=1 flat}. $\mu_i$ here describes the location of the center of mass (in this case, the particle's position) relative to the origin in Cartesian coordinates.

Now consider the corrector tensor $x^\RR_{ab}$ in $\mathscr{P}$. It is obtained from the ODEs~\eqref{eq:xR_mmb eq flat}--\eqref{eq:xnn flat} with $x_{ab}$ replaced by $x^\RR_{ab}$ and $T_{ab}$ replaced by  $T^\RR_{ab}=T_{ab}-\E_{ab}(h^\P)$. The relevant NP components of $T^\RR_{ab}$ are given in retarded coordinates by
\begin{align}\label{eq:Teff frame}
  \Teff_{ll} &= 2\Teff_{ln} =  -h_{m\mb}^\L\pd_r^2 \Theta_{\textrm M} - 2\left(\pd_r h_{m\mb}^\L+ \frac{1}{r}h_{m\mb}^\L\right) \pd_r \Theta_{\textrm M}, \nls \\
 \Teff_{lm} &= 0. \nls 
\end{align}
For our box window function \eqref{eq:W box}, the effective source consists entirely of terms proportional to 
$\delta(r-\rp\pm c)$ or $\delta'(r-\rp\pm c)$. Integrating Eqs.~\eqref{eq:xR_mmb eq flat}--\eqref{eq:xnn flat} from the origin,
we obtain the solutions~\eqref{eq:xmmbar Kerr}--\eqref{eq:xnn Kerr}. The integrals are straightforwardly evaluated if we use the simplifications $\rho=-1/r$ and $\Omega^\circ=0$ and read off the operators ${\cal N}$, ${\cal U}$, and ${\cal V}$ from the right-hand sides of Eqs.~\eqref{eq:xnm flat} and \eqref{eq:xnn flat}. Evaluating them for $r_{\rm min}<r<r_{\rm max}$, we find 
\begin{align}\label{eq:xab Toy regular}
x^\RR_{m\mb} &= \frac{2\Delta r_- }{r}\alpha^\circ +\beta^\circ,\\
x_{nm}^\RR &= \frac{\Delta r_-^2(2r_{\rm min}+\Delta r_-)}{3r^2r_{\rm min}}\Dbar \alpha^\circ + \frac{r^3 - r_{\rm min}^3}{3r^2 r_{\rm min}} \Dbar\beta^\circ,\\
x_{nn}^\RR  &= -\frac{\Delta r_-^2}{3r^2r_{\rm min}}\left[2\Delta r_-\Dbar'\Dbar \alpha^\circ+(2r+r_{\rm min})\Dbar'\Dbar \beta^\circ\right] + \frac{2\Drm}{r}\alpha^\circ + \beta^\circ, 
\label{eq:xnn Toy regular}
\end{align}
where $\Delta r_-\equiv r-r_{\rm min}$, $\alpha^\circ \equiv -\frac{1}{2}\left.(r\partial_r h^\L_{m\bar m})\right|_{r=r_{\rm min}}$, and $\beta^\circ \equiv - \left.h^\L_{m\bar m}\right|_{r=r_{\rm min}}$. 

We observe that $x^\RR_{ab}$ goes to a nonzero value at $r=r_{\rm min}$, joining discontinuously to its zero value for $r<r_{\rm min}$. This jump is due to the presence of $\delta'(r-r_{\rm min})$ in $T^\RR_{ab}$. It contrasts with the behaviour of the point-particle corrector tensor, which joined continuously to zero at $r=r_\mathrm{p}$. However, we also observe that cancellations in the right-hand side of Eq.~\eqref{eq:xnn flat} prevent a Dirac $\delta$ from arising in $x^\RR_{nn}$. 

We stress that the solution's radial nonsmoothness is an artifact of our choice of window function. In practice a smooth window function can be chosen, which would lead to a smooth corrector tensor at the window's boundaries. Here we are more interested in how our scheme smooths the string singularity in the GHZ solution. This can be discerned from the angular functions $\alpha^\circ$ and $\beta^\circ$, which can be read off Eqs.~\eqref{eq:hL static} or \eqref{eq:hL flat tetrad}. Explicitly,
\begin{align}\label{eq:Acirc toy}
\alpha^\circ &= \frac{\mu}{8\pi\rp}\sum_{\l\geq 0}\l (1-c/\rp)^\l {P}_{\l}(\cos\gamma),\nonumber\\
&=\frac{\mu}{8\pi\rp}
\frac{y(\cos\gamma-y)}{(1-2y\cos\gamma+y^2)^{3/2}}, \qquad y \equiv 1-c/\rp\in(0,1).
\end{align}
When $c\neq 0$, $\alpha^\circ$ and $\beta^\circ$ are smooth functions of angles. As expected, there is no string singularity. We can see the string emerge in the $c\to 0$ limit, shrinking the puncture region to zero size: at the angular location of the particle, we have $\alpha^\circ(\theta_\textrm{p},\varphi_\textrm{p})=\mu y(1-y)^{-2}/(8\pi r_\textrm{p})\sim 1/c^2$.
 
With the addition of $\tilde h^{\l=0,1}_{ab}$ and $x^\RR_{ab}$, our residual field is brought to the form
\beq\label{eq:hR before xi}
\hat h^-_{ab}\Theta^-_{\rm M} + \left(\hat h^\RR_{ab}+x^\RR_{ab}\right)\Theta_{\rm M} + \left(\hat h^+_{ab}+\tilde h^{\l=0,1}_{ab}\right)\Theta^+_{\rm M},
\eeq
where $\hat h^\pm_{ab}\equiv 2\Re\S^\dagger_{ab}\Phi^\pm$. The residual field is smooth everywhere except at the window's boundaries. In the vacuum regions outside $\mathscr{P}$, it is identical to the total no-string point-particle solution. All together, it satisfies $\E_{ab}(h^\RR)=\Teff_{ab}$ at all points except $r=r_{\rm max}$.

\subsection{Step 5: gauge correction terms}

The final step in our calculation of $h^\RR_{ab}$ is to add the gauge perturbations $-\Lie_\Xi g_{ab}$ and $-\Lie_{\xi^\mathrm{S}}g_{ab}$ in the region $r\leq r_\mathrm{max}$. Here $\xi = \Xi+\xi^\mathrm{S}$ is the vector that would bring the solution in the region $r>r_\mathrm{max}$ into the shadow gauge. $\Xi$ is made up of terms proportional to $u$ or $u^2$, and (in our static context) $\xi^\mathrm{S}$ is the piece that is independent of $u$. These terms must be added for $r\leq r_{\rm max}$ in order to satisfy the field equations; without them, the residual field differs from a solution to $\E_{ab}(h^\RR)=\Teff_{ab}$ by a field of the form $\Theta^+_\mathrm{M}\Lie_\xi g_{ab}$, which is not a vacuum solution. We must also extend $\Xi$ throughout the region $r\leq r_{\rm max}$ in order to ensure that our time coordinate is continuous. However, $\xi^\mathrm{S}$ is to be attenuated to zero in the puncture region, and since we allow ourselves to introduce radial singularities in this simple demonstration, we can choose the attenuation $\xi^\mathrm{S}\to\xi^\mathrm{S}\Theta^+_\mathrm{M}$. This will bring the field~\eqref{eq:hR before xi} to its final form:
\begin{multline}
h^\RR_{ab} = \left(\hat h^-_{ab}-\Lie_\Xi g_{ab}\right)\Theta^-_{\rm M} + \left(\hat h^\RR_{ab}+x^\RR_{ab}-\Lie_\Xi g_{ab}\right)\Theta_{\rm M}\\
-2\xi^{\rm S}_{(a}r_{b)}\delta(\Delta r_+)
+ \left(\hat h^+_{ab}+\tilde h^{\l=0,1}_{ab}\right)\Theta^+_{\rm M},
\end{multline}
where $\Delta r_+\equiv r-r_\mathrm{max}$.

By Eq.~\eqref{eq:residual IRG}
the gauge vector we seek has the form  
\begin{equation}\label{eq:resgauge flat}
\xi_l = \xi^\circ_l,\quad\xi_n = \xi^\circ_n-r\partial_u\xi^\circ_l, \quad\xi_m = -r\xi^\circ_m  - \tilde{\edth} \xi^\circ_l,
\end{equation}
after making various flat-spacetime simplifications. 
We can find $\xi^\circ_a$ by expanding Eqs.~\eqref{HeldSDaggerXLiexigmmb:eq1}-- \eqref{HeldSDaggerXLiexignn:eq3} in spin-weighted spherical harmonics. For $\l\neq1$, the resulting equations are a specialization of the Schwarzschild results in \ref{app:gauge vec mode decomp}. As an internal consistency check, we have calculated the complete GHZ solution in the region $r>r_{\rm max}$, including the corrector tensor and $\Phi^\mathrm{S}$, independently found the gauge transformation to eliminate them, and verified that the results agree with those obtained from the formulas in \ref{app:gauge vec mode decomp}. However, since that calculation is not of intrinsic interest, we omit the details here. Instead, we focus on finding the vector $\Xi^a$; this is of interest because it directly affects the value of certain quasi-invariant quantities often calculated in self-force applications, as described in the next section.  

$\Xi^a$ is confined to the $\l=0,1$ piece of the gauge transformation. It can be calculated from Eqs.~\eqref{HeldSDaggerXLiexigmmb:eq1}, \eqref{HeldSDaggerXLiexigmmb:eq2}, and \eqref{HeldSDaggerXLiexignm:eq1}, which reduce to
\begin{subequations}
\label{eq:abc equations}
\begin{align}
    a^\circ &= \partial_u \xi^\circ_l + \frac{1}{2}\Dbar\xi^\circ_{\mb} + \frac{1}{2}\Dbar'\xi^\circ_m ,\\
    b^\circ &= -\frac{1}{2}\left( 2 \Dbar \Dbar' + 1  \right) \xi^\circ_l +\xi^\circ_n, \\
    c^\circ&=\partial_u \xi^\circ_m.
\end{align}
\end{subequations}
The Held scalars on the left-hand side are given by the integrals~\eqref{eq:coeffs static vals} with $T_{ab}$ replaced by $\Teff_{ab}$. These evaluate to 
\begin{align}
a^\circ &  =  \alpha^\circ +  \delta^\circ+ \half (\beta^\circ+ \gamma^\circ), \label{eq:ao pnct}\nls \\
b^\circ &= r_{\rm min}\alpha^\circ  + r_{\rm max} \delta^\circ, \label{eq:bo pnct} \nls \\
c^\circ  &=   -\frac{1}{3r_{\rm min}r_{\rm max}} \left[r_{\rm max}(\Dbar \alpha^\circ+\Dbar \beta^\circ)+r_{\rm min}( \Dbar \gamma^\circ +\Dbar \delta^\circ)\right] \label{eq:co pnct},
\end{align}
where 
$\gamma^\circ\equiv h_{m\mb}^\L\vert_{r=r_{\rm max}}$, $\delta^\circ\equiv\frac12( r \pd_r h_{m\mb}^\L)\vert_{r=r_{\rm max}}$. Substituting spin-weighted spherical harmonic expansions into Eq.~\eqref{eq:abc equations}, solving for the coefficients $\xi^\circ_{a,\lm}$, substituting the solutions into Eq.~\eqref{eq:resgauge flat}, and picking out the terms linear and quadratic in $u$, we find
\begin{subequations}
\label{eq:Xi puncture flat}
\begin{align}
\Xi_l &= u a^\circ_{00}Y_{00}+\frac{1}{2}\sum_m \left(2ua^\circ_{1m}+u^2 c^\circ_{1m}\right)Y_{1m},\\
\Xi_n &= \frac{1}{2}\Xi^{00}_lY_{00}-\frac{1}{2}\sum_m\left(\Xi^{1m}_l+2ur c^\circ_{1m}\right)Y_{1m},\\
\Xi_m &= -\frac{1}{2}\sum_m \left(2ua^\circ_{1m}+u^2 c^\circ_{1m}+2ur c^\circ_{1m}\right){}_1Y_{1m}.
\end{align}
\end{subequations}
The $\l=0$ terms can be written as $\Xi^a_{\l=0} = u a^\circ_{00}Y_{00} t^a$, in agreement with Eqs.~\eqref{eq:Xi Kerr} and \eqref{eq:alpha beta}.

\subsection{Calculation of a quasi-invariant quantity}

The final step of our procedure is to add $h^\P_{ab}$ to our residual field to obtain the total metric perturbation. However, the residual field itself is usually the object of interest in self-force applications, and we can use it as a strong check on our procedure.

In a typical self-force calculation, the physical outputs are quasi-invariants constructed from the regular field on the particle. These are quantities that are invariant within a class of gauges that manifestly preserve the metric's asymptotic flatness and set of discrete frequencies~\cite{Barack:2018yvs}. The most important example of such a quantity is the Detweiler redshift $h^\RR_{ab}(x_\mathrm{p})v^a v^b$~\cite{Detweiler:2008ft}, which has a central role in the first law of binary mechanics~\cite{LeTiec:2011ab}, is closely related to the particle's perturbed Hamiltonian~\cite{Fujita:2016igj} and to a binary's binding energy~\cite{LeTiec:2011dp}, and has facilitated numerous synergies between self-force theory and other models of binary systems~\cite{LeTiec:2014oez,Barack:2018yvs}.

Although usually formulated in black hole spacetimes, $h^\RR_{ab}v^a v^b$ has the same quasi-invariance in flat spacetime. Under a smooth gauge transformation generated by a vector field $\xi^a$, we have $h^\RR_{ab}\to h^\RR_{ab}+\Lie_\xi g_{ab}$, such that the transformation of $h^\RR_{ab}v^a v^b$ is
\beq\label{eq:Dhvv}
(\Lie_\xi g_{ab})v^a v^b = 2\partial_t\xi^t;
\eeq
here $t=u+r$ is the usual time coordinate, in terms of which $v=\frac{\partial}{\partial t}$. Simultaneously, we know from the general argument in Sec.~\ref{sec:stationary sector} that $\xi^a$ can only preserve manifest stationarity ($\Lie_t h_{ab}=0$) if $\partial_t\xi^a$ vanishes or is a Killing vector. Hence, within the class of manifestly stationary gauges, Eq.~\eqref{eq:Dhvv} vanishes or is equal to the $t$ component of a Killing vector. But if it is the latter it does not vanish in the limit $r\to\infty$, violating manifest asymptotic flatness. Therefore $h^\RR_{ab}v^a v^b$ is invariant within the class of manifestly stationary and asymptotically flat gauges. 

In the present context, this implies that $h^\RR_{ab}v^a v^b=0$ in all such gauges. To see this, note that in the Lorenz gauge, $h^\RR_{ab}v^a v^b=0$ trivially because $h^\RR_{ab}$ itself vanishes. Our shadowless gauge falls within the same class, and so our residual field at the particle should satisfy $h^\RR_{ab}v^a v^b=(\hat h^\RR_{ab}+x^\RR_{ab}-\Lie_\Xi g_{ab})v^a v^b=0$. Note that $\Xi^a$ is precisely the type of vector that preserves manifest stationarity but contributes a nonzero amount to $\Lie_\Xi g_{ab}v^a v^b=0$; it brings the asymptotically nonflat shadow gauge to the asymptotically flat shadowless gauge. The remainder of this section will confirm that $h^\RR_{ab}v^a v^b=0$ in the shadowless gauge and thereby verify that our $\Xi^a$ is correct.

We first consider the $\l>1$ contributions to $h_{ab}^\RR v^a v^b$. Since $\Xi^a$ only contributes $\l=0,1$ modes, we require only $\hat{h}^{\textrm R}_{ab}v^a v^b=\hat{h}^{\textrm R}_{nn}$ and $x^\RR_{ab}v^a v^b= x^\RR_{nn}$ (recalling $v^a=n^a+l^a/2$ and $\hat{h}^\RR_{ab}l^b=x^\RR_{ab}l^b=0$). To compute these we insert the coefficients $\delta_{i,\lm}^\circ$ from Eq.~\eqref{eq:delta modes} into \eqref{Sdnn} and add the modes of $x^\RR_{nn}$ given in \eqref{eq:xnn Toy regular} with coefficients $\alpha^\circ_{\lm}=-\frac12 (r \pd_r h_{m\mb,\lm})\vert_{r=r_{\rm min}}$ and $\beta^\circ_{\lm}=- h_{m\mb,\lm}\vert_{r=r_{\rm min}}$. We find that, mode-by-mode, $\hat h^\RR_{nn}+x^\RR_{nn} = 0$ for all $x\in \mathscr{P}$, and therefore at the particle
\begin{align}
    h^\RR_{ab}v^a v^b = 0,  \qquad \l>1.
\end{align}
Equivalently,
\beq\label{eq:hhatvv+xvv}
(\hat h^\RR_{ab}+x^\RR_{ab})v^av^b = (x^{\RR}_{nn})^{\l=0,1}.
\eeq

We next consider the contributions from the $\l=0$ and 1 modes. For this we need
$(\Lie_\Xi g_{ab})v^a v^b = 2\partial_t\Xi^t$. From Eq.~\eqref{eq:Xi puncture flat}, we have $\Xi^t = \Xi_n + \frac{1}{2}\Xi_l = ua^\circ_{00}Y_{00} -ur\sum_m c^\circ_{1m}Y_{1m}$, and so
\beq\label{eq:LieXigvv}
(\Lie_\Xi g_{ab})v^a v^b = 2a^\circ_{00}Y_{00} -2r_\mathrm{p}\sum_m c^\circ_{1m}Y_{1m}(\theta_{\rm p},\varphi_{\rm p})
\eeq
at the particle. To see how this combines with $(x^{\RR}_{nn})^{\l=0,1}$, we need to examine the Held scalars $\alpha^\circ$, $\beta^\circ$, $\gamma^\circ$, and $\delta^\circ$ appearing in $a^\circ$ and $c^\circ$ in Eq.~\eqref{eq:abc equations} and in $x^\RR_{nn}$ in Eq.~\eqref{eq:xnn Toy regular}. From their definitions in terms of $h^\L_{m\bar m}$, and the mode decomposition of $h^\L_{m\bar m}$ in Eq.~\eqref{eq:hL static}, we derive the relationships $\alpha^\circ_{\l=0}=0$, $\delta^\circ_{\l=0}=-\frac{1}{2}\gamma^\circ_{\l=0}$, $\alpha^\circ_{\l=1}=\frac{1}{2}\beta^\circ_{\l=1}$, and $\delta^\circ_{\l=1}=-\gamma^\circ_{\l=1}$. Substituting  these into Eqs.~\eqref{eq:abc equations} and \eqref{eq:xnn Toy regular}, we find that the terms in Eq.~\eqref{eq:LieXigvv} are related to $x^\RR_{nn}$ on the particle by
\beq
2a^\circ_{\l=0} = (x^{\RR}_{nn})^{\l=0} =\beta^\circ_{\l=0}\qquad\text{and} 
\qquad-2r_p\sum_m c^\circ_{1m} Y_{1m}=(x^\circ_{nn})^{\l=1}=\frac{r_\mathrm{p}}{r_\mathrm{min}}\beta^\circ_{\l=1}.
\eeq
So on the particle,
\beq\label{eq:hRvv ell=0,1}
(x^\RR_{nn})^{\l=0,1}-(\Lie_\Xi g_{ab})v^av^b=0; 
\eeq
the contribution from $\Xi^a$ precisely cancels the contribution from the low modes of $x^\RR_{ab}$. 

Combining Eq.~\eqref{eq:hRvv ell=0,1} with Eq.~\eqref{eq:hhatvv+xvv}, we obtain our final result: 
\beq
h^\RR_{ab}v^av^b=(\hat h^\RR_{ab}+x^\RR_{ab}-\Lie_\Xi g_{ab})v^av^b = 0.
\eeq
This provides the first significant validation of our puncture scheme in general and of our method of determining the gauge completion $-\Lie_\Xi g_{ab}$ in particular.

\section{Discussion}\label{sec:discussion}

Our aims in this paper were twofold: (i) to illuminate the relationship between the GHZ procedure and prior reconstruction methods used in self-force calculations, and (ii) to use the GHZ formalism to develop a Teukolsky puncture scheme that yields the metric perturbation in a gauge that is more regular than the no-string radiation gauge currently used. 

On the first point, we have clarified how the GHZ procedure completes earlier methods. Specifically, the corrector tensor completes Ori's description of nonvacuum reconstruction, thus determining the complete half-string solution (or half-shadow solution for an extended source). In the case of the no-string solution, which is currently used in practice, the GHZ procedure determines the coefficient of the Dirac delta function supported on the sphere $r=r_\mathrm{p}(t)$ that intersects the particle's position at each instant. Although that coefficient is not explicitly required to calculate the self-force and related quantities, it is needed to have a complete solution to the linearized Einstein equation. Moreover, it provides valuable information about whether the time coordinate is continuous across the sphere. We have shown how to use that information to obtain a simple formula for the ``gauge completion'' that ensures continuity of the time coordinate. The gauge completion is the perturbation $-\Lie_\Xi g_{ab}$ that must be added inside the sphere, where $\Xi^a$ is given by Eq.~\eqref{eq:Xi Kerr} with Eqs.~\eqref{eq:alpha beta}, \eqref{eq:corrector constants}, and \eqref{eq:cHeld}.

In order to develop our puncture scheme,  we have also extended and fleshed out the GHZ formalism for generic spatially compact sources. Our primary result in that context, as summarized in Sec.~\ref{sec:Kerr summary} and Fig.~\ref{fig:flowchart-shadowless-gauge}, is a concrete, practical procedure for constructing the metric perturbation in a ``shadowless'' gauge in which the corrector tensor is confined to the interior of the source. In the point-particle limit, this reduces to the no-string gauge.

That general procedure forms the basis of our puncture scheme, which is summarized in Sec.~\ref{sec:hR reconstruction} and Fig.~\ref{fig:flowchart-puncture}, and which we demonstrated in Sec.~\ref{sec:return to flat space}. The key idea of the scheme is to put the puncture in a well-behaved gauge and then reconstruct the residual field using our shadowless reconstruction procedure. By putting the most singular piece of the field in a more regular gauge, this procedure avoids the strong singularities that extend off the particle in the half-string and no-string gauges. 

Our primary motivation for developing this method was to obtain a metric perturbation that is sufficiently regular to use as input for second-order self-force calculations. But we expect it to have benefits even for first-order applications. In addition to providing a simple prescription for the gauge completion, it may also eliminate the large-cancellation problem that arises in no-string calculations. Current self-force calculations in the no-string gauge rely on CCK vacuum reconstruction in the frequency domain, which fails inside the orbital libration region. To overcome this failure, they rely on the method of extended homogeneous solutions to analytically extend the vacuum solutions from outside the libration region. For orbits with moderate-to-high eccentricity, this method involves large numerical cancellations requiring $30+$ digits of precision~\cite{vandeMeent:2017bcc}. Since the GHZ formalism allows non-pointlike sources, it has no fundamental need to use extended homogeneous solutions in this way. Extended solutions are typically used in any case, outside the no-string context, to overcome the Gibbs phenomenon associated with Fourier series representations of nonsmooth fields~\cite{Barack:2008ms}, but since our puncture scheme works with the residual field rather than the physical field, it involves smoother functions that would suffer less severe Gibbs phenomena. When combined with series acceleration methods, this may eliminate the need for extended solutions.

Another potential advantage of our method at first order is that it directly yields the regular field on the particle, rather than requiring a mode-by-mode subtraction of the singular field from the retarded field. While such subtraction is often simpler than a puncture scheme, its current implementation in Kerr requires a computationally expensive projection from spin-weighted spheroidal harmonics onto scalar spherical harmonics~\cite{vandeMeent:2017bcc}, which our method avoids.

Our scheme also simplifies considerably when only first-order information is required. This is because any quantity of interest will be invariant under the transformation generated by $\xi^\mathrm{S}_a$ in Step 5 of the procedure. Hence, first-order calculations do not need to calculate this vector.

This is borne out by the explicit demonstration of our puncture scheme
in Sec.~\ref{sec:return to flat space}. There we showed that our scheme yields the correct result for the type of quantity often calculated in self-force applications, and we never required $\xi^\mathrm{S}_a$. On the other hand, these types of quantities are {\em not} invariant under transformations generated by the gauge-completion vector $\Xi^a$, and our demonstration has validated our method of obtaining $\Xi^a$.

Our demonstration also revealed a potential pitfall. Puncture schemes historically have expanded each $\lm$ mode of the puncture in powers of radial distance $\Delta r$ from the particle. Such an expansion has been required in practice to obtain analytical expressions for each mode. As Eq.~\eqref{eq:Dr expansion} makes clear, a puncture that has been expanded in this way diverges at large $\l$ for all values of $\Delta r$. There is a subtle reason such divergences have not posed a problem in the past: if we solve $\E_{ab}(h^\RR)=T_{ab}-\E_{ab}(h^\P)$ for the $\lm$ modes of $h^\RR_{ab}$ and impose the same gauge condition on $h^\RR_{ab}$ as on $h^\P_{ab}$, then the total field $h^{\P,\lm}_{ab}+h^{\RR,\lm}_{ab}$ will be identical to the retarded field mode $h^{\lm}_{ab}$ in that gauge. In a gauge such as the Lorenz gauge, where $h_{ab}$ is smooth away from the particle, this implies that $h^{\RR,\lm}_{ab}$ necessarily cancels the  large-$\l$ behaviour in $h^{\P,\lm}_{ab}$. We cannot expect these cancellations to occur in our puncture scheme, where the residual field is computed in a different gauge than the puncture. This means that we cannot afford a $\Delta r$ expansion of the puncture's $\lm$ modes.  Fortunately, they may not be needed. Our procedure utilizes $h^\P_{ab}$ in three ways: to compute $\psi^\P_0$, which is needed to find $\psi^\RR_0=\psi_0-\psi^\P_0$; to compute $T^\RR_{ab}$, which is needed to find $x^\RR_{ab}$; and ultimately to obtain the total field $h^\RR_{ab}+h^\P_{ab}$ (in scenarios where the total field is needed). The only one of these steps that requires $\lm$ modes is the calculation of $\psi^\RR_0$, which requires modes because in practice $\psi_0$ is computed at the level of modes. Hence, we only need to calculate $\psi^{\P,\lm}_0$. At worst, this will involve computing numerical integrals of $\psi^{\P}_0$ against spin-weighted spheroidal harmonics, unless an analytical method can be found that avoids a $\Delta r$ expansion.

We conclude by noting that our puncture scheme represents only one possible path to obtaining the metric perturbation in a regular gauge. Other reconstruction methods are also under development~\cite{Dolan:2019hcw,Wardell:2020naz,Dolan:2021ijg,Loutrel:2020wbw,Ripley:2020xby}. Although these are not yet able to reconstruct a nonvacuum metric perturbation, they may soon offer a viable alternative to our method.

\begin{acknowledgements}
We thank Leor Barack, Sam Dolan, Chris Kavanagh, Maarten van de Meent, and Barry Wardell for helpful discussions. AP gratefully acknowledges the support of a Royal Society University Research Fellowship. AP and AS acknowledge the support of a Royal Society Research Fellows Enhancement Award. SH\ is grateful to the Max-Planck Society for supporting the collaboration between MPI-MiS and Leipzig U., grant Proj.~Bez.\ M.FE.A.MATN0003. VT is grateful to International Max Planck Research School for support through the studentship.
 
\end{acknowledgements}

\appendix 

\section{GHP and Held formalisms}\label{app:GHP and Held}
\subsection{GHP Formalism}\label{sec:GHP}
Throughout the paper we utilize the GHP formalism \cite{Geroch:1973am}, which we now briefly review. 

The GHP formalism, like the NP formalism, adopts a double-null tetrad $(l^a,n^a,m^a,\bar{m}^a)$, in which
\begin{equation}\label{eq:NP met}
g_{ab} = 2l_{(a}n_{b)}-2m_{(a}\bar m_{b)}.
\end{equation}
We choose the normalization $n_al^a=1$ and $m_a\bar m^a=-1$ corresponding to mostly negative signature for $g_{ab}$. The GHP formalism refines the NP formalism by putting equations in a form that is invariant under the structure group of spin and boost transformations of the tetrad. This group consists of two real-valued parameters and is isomorphic to the group of multiplication by a complex number, $\lambda$.  To express the weight of a GHP quantity $f$ we  use the notation $f \GHPwt \{p,q\}$, meaning that under a spin and boost transformation
  \begin{equation}
      f \xrightarrow{} \lambda^p \bar{\lambda}^q f \ \ \Longleftrightarrow \ \  f \GHPwt \{p,q\}.
  \end{equation}
  By writing equations purely in terms of quantities with a definite (homogeneous) transformation law under spin and boost transformations, these are readily transformed from any one choice of frame to another (similarly to writing geometric equations only in terms of $\nabla_a, R_{abcd},$ etc.). 
  
  In particular, the tetrad legs are assigned the weights
\begin{align}\label{eq:leg weights}
l \GHPwt \{1,1\},&\qquad n \GHPwt \{-1,-1\} ,\no
m \GHPwt \{1,-1\},&\qquad \mb \GHPwt \{-1,1\},
\end{align}
under this rule.
The weights of the spin coefficients $\rho,\rho',\tau,\tau',\sigma,\sigma, \kappa, \kappa'$ follow directly from the weights assigned to the tetrad above by counting the numbers of the legs appearing in their definition (see \cite{Geroch:1973am}). In particular, $\rho\GHPwt\{1,1\}$ and $\tau\GHPwt\{1,-1\}$.  The GHP formalism also includes priming  $f' \circeq \{-p,-q\}$ and complex conjugation  $\bar{f}\circeq \{q,p\}$ operations, which are already implicit in the naming of the above spin coefficients. 

The remaining spin coefficients do not have definite weight and do not appear in the GHP equations, but appear in the GHP derivative operators 
\begin{align}\label{eq: GHP ops}
\th &= l^a\nabla_a - p \epsilon-q\bar\epsilon, \qquad \ \ \ \  \th' = n^a \nabla_a+p\epsilon'+q\bar{\epsilon}', \non
\edth&=m^a\nabla_a - p\beta+q\bar{\beta}', \qquad \edth'=\bar{m}^a\nabla_a+p\beta'-q\bar{\beta},
\end{align}
which are GHP covariant and have weights given by the corresponding weights of legs \eqref{eq:leg weights}; e.g. $\th f \GHPwt \{p+1,q+1\}$.

The main utility of the GHP formalism is in the context of algebraically 
special spacetimes because the tetrad legs can be chosen to be aligned with geometrically preferred directions, leading to many simplifications.
In any tetrad aligned with both principal null directions of a type D metric we have $\kappa = \kappa' = \sigma = \sigma' = 0$ by the Goldberg-Sachs theorem, so we shall always use such a tetrad in this paper.
Whenever possible in the course of our analysis, we also make simplifications using the GHP vacuum Ricci equations \cite{Geroch:1973am}. In any vacuum type D metric and tetrad aligned with both principal null directions, these read 
\begin{align}\label{eq:GHP eqs}
\th \rho &= \rho^2, 
\qquad \th'\rho =\edth'\tau+\rho\bar\rho'-\tau\bar\tau-\Psi_2, \qquad  \edth\rho=(\rho-\bar\rho)\tau, \nonumber \\ 
 \edth \tau &=\tau^2, 
 \qquad \th \tau = (\tau-\bar\tau') \rho.
\end{align}
These equations are supplemented in Kerr by 
\begin{equation}\label{eq:type D ids}
\frac{\tau}{\bar\tau'}=\frac{\tau'}{\bar\tau}=-\frac{\rho}{\bar\rho}=-\frac{\rho'}{\bar\rho'} = -\frac{\Psi_2^{1/3}}{\bar{\Psi}_2^{1/3}}, \qquad  \th\tau'=2\rho\tau', \qquad \edth'\rho= 2 \rho\tau',
\end{equation}
which follow from the existence of a Killing tensor, the commutation relations, and the Bianchi identities.  
  The commutation relations also give rise to the further identities
\begin{align}
\th'\rho&=\rho\rho'+\tau'(\tau-\tabp)-\half \Psi_2- \frac{\rho}{2\rhb}\bar{\Psi}_2,\\
\edth'\tau&=\tau\tau'+\rho(\rho'-\rhb')+\half \Psi_2- \frac{\rho}{2\rhb}\bar{\Psi}_2.
\end{align}
Finally, we note two identities that reduce the number of terms required in many manipulations. These are $\edth\tau'=\edth'\tau$ and $\th\rho'=\th'\rho$.

An example of an aligned tetrad is the Kinnersly frame, where the non-zero
GHP coefficients are given by
\begin{align}\label{eq:GHP coeffs in Kinn}
\rho &= - \frac{1}{\bar{\Gamma}},  \qquad \rho' = \frac{\Delta}{2\Sigma \bar{\Gamma}}, \\
\tau &= \frac{ia \sin\theta}{\sqrt{2}\Sigma}, \qquad \tau'= \frac{ia\rho^2\sin\theta}{\sqrt{2}},\\
\Psi_2 &= -\frac{M}{\bar{\Gamma}^3}
\end{align}
and the remaining non-zero NP coefficients required for the GHP directional derivative operators are
\begin{align}\label{eq:NP coeffs in Kinn}
\beta = -\frac{\bar\rho\cot\theta}{2\sqrt{2}},\qquad \beta' = \bar\beta - \frac{ia\rho^2\sin\theta}{\sqrt{2}}, \qquad\epsilon' = -\frac{\rho\Delta + r-M}{2\Sigma}.
\end{align}

\subsection{Held Formalism}\label{sec:Held}
Held \cite{Held:1974,Held:1975} introduced a specialized version of the GHP formalism particularly adapted to non-accelerating vacuum Type-D spacetimes such as Kerr that is extremely useful when making covariant expansions in the quantity $\rho$ (essentially the inverse radius). The formalism involves a new set of operators, called Held derivatives,
\begin{subequations}
  \label{eq:Hops}
  \begin{align}
    \Nbar &=\th' - \tab \Dbar - \tau \Dbar' + \tau \tab \left( \frac{p}{\bar \rho} + \frac{q}{\rho} \right) + \frac{1}{2} \left( \frac{q \bar{\Psi}_2}{\bar{\rho}} + \frac{p \Psi_2}{\rho} \right)   \GHPwt \{ -1,-1 \}, \\
    \Dbar &=\frac{1}{\bar \rho} \edth + \frac{q\tau}{\rho}  \GHPwt \{0,-2 \}, \\
    \Dbar' &= \frac{1}{\rho} \edth' + \frac{p\tab}{\bar \rho}  \GHPwt \{-2,0\}.
  \end{align}
\end{subequations}

Intrinsic to the formalism are the Held scalars, by which we mean quantities annihilated by $\th$. The point is that, if $h^\circ$ is a Held scalar, then so are its Held derivatives but not the derivatives by 
$\th', \edth, \edth'$. Indeed, it follows from  \eqref{eq:type D ids}
that~\cite{Held:1974}
\begin{equation}\label{eq:th Dbar commute}
\left[\th,\Dbar'\right]f = \left[\th,\Dbar\right]f=0,
\end{equation}
for any $f\GHPwt\{p,q\}$. Additionally, $[\th,\Nbar]f = 0$ if $f$ has ``no $\rho$-dependence'', i.e. if it is a Held scalar $f = h^\circ$  \cite{Held:1974}. An example of a Held scalar is 
\begin{equation}
\label{eq:OmegaH}
 \Omega^\circ= \frac{1}{\rhb}-\frac{1}{\rho} \GHPwt \{-1,-1\}.
\end{equation} 
When acting on Held scalars, the Held operators in outgoing Kerr-Newman coordinates and Kinnersley's frame  give 
\begin{equation}
\label{eq:Hop}
\begin{split}
\Dbar & = -\frac{1}{\sqrt{2}} \left( \dbd{\theta} + i \csc \theta \dbd{\varphi} + ia \sin \theta \dbd{u} - \half (p-q) \cot \theta \right)  , \\
\Dbar'   &= -\frac{1}{\sqrt{2}} \left( \dbd{\theta} - i \csc \theta \dbd{\varphi} - ia \sin \theta \dbd{u} + \half (p-q) \cot \theta \right)  , \\
\tilde{\th}'  & = \dbd{u},
\end{split}
\end{equation}
where the last equation holds only when acting on a Held scalar $h^\circ$.
The operators $\Dbar$ and $\Dbar'$ are spin-raising and lowering operators when acting on spin-weighted spherical harmonics $\Y{s}$;  see Eq. \eqref{eq:Ys}. In the Kinnersley frame and outgoing Kerr-Newman coordinates, Held's operators are equal up to trivial 
normalization and notational change to Chandrasekhar's operators  $\chand{-s}{\omega}^\dagger, \chand{+s}{\omega}$,
\begin{equation}
\label{eq:Chandop}
\begin{split}
\tilde{\edth} & \equiv \chand{-s}{\omega}^\dagger  = -\frac{1}{\sqrt{2}} \left( \dbd{\theta} 
+ i \csc \theta \dbd{\varphi} + a\omega \sin \theta - s \cot \theta \right), \\
\tilde{\edth}'   &\equiv \chand{+s}{\omega}   = -\frac{1}{\sqrt{2}} \left( \dbd{\theta} - i \csc \theta \dbd{\varphi} - a\omega \sin \theta  + s \cot \theta \right), \\
\tilde{\th}'  &= -i\omega,
\end{split}
\end{equation}
when acting on a Held scalar with $u$-dependence $e^{-i\omega u}$ (here $s=\half (p-q)$).
In addition to the derivatives \eqref{eq:Hops}, Held introduced the coefficients $\tah \GHPwt \{-1,-3\},\rhoph  \GHPwt \{-2,-2\}$ and $\Psih  \GHPwt \{-3,-3\} $. In a tetrad in which $l^a$ and $n^a$ are aligned with principal null directions, the vacuum Ricci and Bianchi identities give the following relations between GHP and Held coefficients:
\begin{subequations}
  \label{eq:Held coeffs}
  \begin{align}
    \tau &= \tau^\circ \rho\bar\rho, \\
    \tau' &= -\tab^\circ \rho^2, \\
    \rho' &= \rho^{\prime \circ} \bar \rho - \half \Psi^\circ \rho^2 - \left(\Dbar \tab^\circ + \half \Psi^\circ\right) \rho \bar \rho - \tau^\circ \tab^\circ \rho^2 \bar \rho,\\
    \Psi_2 &= \Psi^\circ \rho^3.
  \end{align}
\end{subequations}
In the Kinnersley tetrad the Held scalars reduce to
\begin{subequations}\label{eq:Held coeffs in Kinn}
\begin{align}
\tau^\circ &= -\frac{ia\sin\theta}{\sqrt{2}},\qquad \rho^{\prime\circ}=-1/2, \qquad
\Psi^\circ = M, \qquad \Omega^\circ=-2ia\cos\theta.
\end{align}
\end{subequations}

\subsection{Tetrad components of a gauge perturbation}\label{app:Lie}

We frequently require expressions for a gauge perturbation $\lie{\xi} g_{ab} = 2\nabla_{(a} \xi_{b)}$ in Held form. Here we write these expressions in a generic type D spacetime.

The components along $l^a$ are
\begin{subequations}\label{Liexigla}
\begin{align}
\left( \lie{\xi} g \right)_{ll} &=  2 \th \xi_l,\label{Liexigll} \nls\\
\left( \lie{\xi} g \right)_{ln} &= \th \xi_n + \th' \xi_l + \left( \tau' + \bar{\tau} \right) \xi_m + \left( \tau + \bar{\tau}' \right) \xi_{\bar{m}}\nonumber \nls\\
&= \Big[ \tilde{\th}' + \bar{\tau}^\circ \rho \bar{\rho} \tilde{\edth} + \tau^\circ \rho \bar{\rho} \tilde{\edth}' - \tau^\circ \bar{\tau}^\circ \rho \bar{\rho} \left( \rho + \bar{\rho} \right) - \half \left( \Psi^\circ \rho^2 + \bar{\Psi}^\circ \bar{\rho}^2 \right) \Big] \xi_l\nonumber \nls\\
&\quad+ \th \xi_n - \left( \bar{\tau}^\circ \rho^2 \bar{\rho} \Omega^\circ \right) \xi_m + \left( \tau^\circ \rho \bar{\rho}^2 \Omega^\circ \right) \xi_{\bar{m}},\label{Liexigln} \nls\\
\left( \lie{\xi} g \right)_{lm} &= \left( \edth + \bar{\tau}' \right) \xi_l + \left( \th + \bar{\rho} \right) \xi_m\nonumber \nls\\
&= \bar{\rho} \left( \tilde{\edth} - 2 \tau^\circ \bar{\rho} \right) \xi_l + \left( \th + \bar{\rho} \right) \xi_m. \nls \label{Liexiglm}
\end{align}
\end{subequations}
The remaining components along $n^a$ are
\begin{subequations}\label{Liexigna}
\begin{align}
\left( \lie{\xi} g \right)_{nn} &= 2 \th' \xi_n\nonumber \nls\\
&= 2 \Big[ \tilde{\th}' + \bar{\tau}^\circ \rho \bar{\rho} \tilde{\edth} + \tau^\circ \rho \bar{\rho} \tilde{\edth}' + \tau^\circ \bar{\tau}^\circ \rho \bar{\rho} \left( \rho + \bar{\rho} \right) + \half \left( \Psi^\circ \rho^2 + \bar{\Psi}^\circ \bar{\rho}^2 \right) \Big] \xi_n,\label{Liexignn} \nls\\
\left( \lie{\xi} g \right)_{nm} &= \left( \edth + \tau \right) \xi_n + \left( \th' + \rho' \right) \xi_m\nonumber\\
&= \bar{\rho} \left( \tilde{\edth} + \tau^\circ (\rho + \bar{\rho}) \right) \xi_n\nonumber \nls\\
&\quad+ \Big( \tilde{\th}' + \bar{\tau}^\circ \rho \bar{\rho} \tilde{\edth} + \tau^\circ \rho \bar{\rho} \tilde{\edth}' + \rho \bar{\rho}^2 \tau^\circ \bar{\tau}^\circ - 2 \rho^2 \bar{\rho} \tau^\circ \bar{\tau}^\circ + \half \bar{\Psi}^\circ \bar{\rho}^2 + {\rho'}^\circ \bar{\rho}\nonumber \nls\\
&\quad- \Psi^\circ \rho^2 + \half \rho \bar{\rho} \left( \rho^{\prime \circ} + \bar{\rho}^{\prime \circ} \right) \Omega^\circ - \half  \rho \bar{\rho} \bar{\Psi}_2^\circ \Big) \xi_m,\label{Liexignm} \nls
\end{align}
\end{subequations}
and the remaining components along $m^a$ are
\begin{subequations}\label{Liexigma}
\begin{align}
\left( \lie{\xi} g \right)_{mm} &= 2 \edth \xi_m = 2 \bar{\rho} \left( \tilde{\edth} + \bar{\rho} \tau^\circ \right) \xi_m,\label{Liexigmm} \nls\\
\left( \lie{\xi} g \right)_{m\bar{m}} &= \left( \rho'+ \bar{\rho}' \right) \xi_l + \left( \rho + \bar{\rho} \right) \xi_n + \edth' \xi_m + \edth \xi_{\bar{m}}\nonumber \nls\\
&= \left[ {\rho'}^\circ \bar{\rho} + \bar{\rho}^{\prime \circ} \rho - \half \left( \Psi^\circ \rho^2 + \bar{\Psi}^\circ \bar{\rho}^2 \right) - \half \left( \Psi_2^\circ + \bar{\Psi}_2^\circ \right) \rho \bar{\rho} - \tau^\circ \bar{\tau}^\circ \rho \bar{\rho} ( \rho + \bar{\rho} ) \right] \xi_l\nonumber \nls\\
&\quad+ \left( \rho + \bar{\rho} \right) \xi_n + \rho \left( \tilde{\edth}' - \bar{\tau}^\circ \rho \right) \xi_m + \bar{\rho} \left( \tilde{\edth} - \tau^\circ \bar{\rho} \right) \xi_{\bar{m}}.\label{Liexigmmb} \nls
\end{align}
\end{subequations}

\section{Linear differential operators}
\label{sec:operators}
This appendix defines the linear differential operators ${\cal E}$, ${\cal O}$, $\mathcal{S}$, ${\cal T}$, their adjoints, and ${\cal F}$. The adjoint of a GHP covariant operator ${\cal D}$ is defined by the relation
\begin{equation}
\varphi{\cal D}\psi = ({\cal D}^\dagger\varphi)\psi + \nabla_a v^a
\end{equation}
for arbitrary smooth tensor/GHP quantities $\varphi$ and $\psi$ of appropriate rank/weight so that the above bilinear is a scalar of trivial GHP weight, and for some vector field $v^a$ constructed from $\varphi$ and $\psi$ and their derivatives.

The linearized Einstein operator, which maps a metric perturbation to its linearized Einstein tensor, is
\begin{equation}\label{eq:E}
  {\cal E}_{ab}(h) =  
  \half \left[ -\nabla^c \nabla_c h_{ab} - \nabla_a \nabla_b h^c_{\phantom{c}c} + 2 \nabla^c \nabla_{(a}h_{b)c} + g_{ab} \left( \nabla^c \nabla_c h^d_{\phantom{d}d} - \nabla^c \nabla^d h_{cd}\right) \right].
\end{equation}
It is self-adjoint.

The wave operator $\O$ appearing in the Teukolsky equation is
\begin{equation}\label{eq:O}
  \O  = 2 \left[ (\th - 4\rho - \bar \rho)(\th'-\rho') - (\edth - 4\tau - \bar \tau')(\edth' - \tau') -3\Psi_2 \right],
 \end{equation}
and its adjoint is
\begin{align}
  \label{eq:Odag}
  \O^\dagger = 2\left[ (\th'-\rho')(\th + 3 \rho) - (\edth' - \tau')(\edth +3\tau) -3\Psi_2 \right].
\end{align}

The operator $\mathcal{S}^{ab}$, which acts on a stress-energy to return the source in the Teukolsky equation, is
\begin{subequations}
  \begin{align}\label{eq:S}
    \mathcal{S}^{ab}T_{ab} ={}& (\edth - \bar \tau' - 4\tau)\left[(\th - 2\bar \rho) T_{lm} - (\edth - \bar \tau') T_{ll} \right] \nonumber \\
            & + (\th - \bar \rho - 4\rho)\left[(\edth - 2\bar \tau') T_{lm} - (\th - \bar \rho) T_{mm} \right],
  \end{align}
\end{subequations}
and its adjoint is
\begin{align}\label{eq:Sdagger}
\mathcal{S}^\dagger_{ab} &= - l_a l_b(\edth-\tau)(\edth+3\tau)  - m_a m_b (\th - \rho)(\th + 3\rho)  \nonumber\\
&\quad 
+ l_{(a} m_{b)}[(\th - \rho + \bar \rho)(\edth + 3\tau) + (\edth-\tau+\bar \tau')(\th +3\rho) ].
\end{align}

The operator ${\cal T}^{ab}$, which acts on a metric perturbation to return its Weyl scalar $\psi_0$, is
\begin{align}
    \label{eq:T}{\cal T}^{ab} h_{ab} ={}& \half(\edth-\bar \tau')(\edth-\bar \tau') h_{ll} + \half (\th - \bar \rho)(\th - \bar \rho) h_{mm} \nonumber \\
            &-\half\left[(\th- \bar \rho)(\edth-2\bar \tau') + (\edth -\bar \tau')(\th -2\bar \rho)\right] h_{(lm)},
\end{align}
and its adjoint is
\begin{align}\label{eq:Tdag}
  {\cal T}^\dagger_{ab} ={}
  & \half l_a l_b(\edth-\tau)(\edth-\tau) + \half m_a m_b (\th-\rho)(\th-\rho) \nonumber\\
  & - \half l_{(a} m_{b)}\{(\edth + \tab' -\tau)(\th - \rho) + (\th - \rho + \bar \rho)(\edth-\tau) \}. 
\end{align}

Finally, the operator ${\cal F}_{ab}$ that returns the corrector tensor in Eq.~\eqref{eq:x inner outer} is
\begin{align}\label{eq:Fab}
{\cal F}_{ab} &= \frac{\mu\Dr}{3r^2\rp^2}l_a l_b\left[(\Dr)^2\Dbar'\Dbar - 3r \rp\right]  +\frac{\mu(\Dr)^2(\Dr+3\rp)}{3r^2\rp^2} l_{(a}\left (m_{b)}\Dbar'+\bar m_{b)}\Dbar\right) \nonumber\\
&\quad - \frac{2\mu\Dr} {rr_\textrm{p}} m_{(a}\mb_{b)}.
\end{align}

\section{Spin-weighted spherical harmonics}
\label{sec:spin-weighted harmonics}

We define the spin-weighted harmonics with the conventions of \cite{PenroseRindler1}:
\begin{align}\label{eq:Ys}
    \Y{s}\equiv
       \C{|s|} \begin{cases} 
           (\sqrt{2})^s \left( \prod^{s-1}_{j=0} \chand{-j}{0}^\dagger \right) \Y{},  & 0 < s \leq \l,  \\
           Y_{\ell \emm},  & s=0,\\
            (-\sqrt{2})^{(-\ess)} \left( \prod^{-(s+1)}_{j=0} \chand{j}{0} \right) \,\Y{}, & -\l \leq s < 0,
        \end{cases}
        \quad \text{with } \C{s}=\sqrt{\frac{(\l-s)!}{(\l+s)!}}.
\end{align}
The angular derivatives $\chand{s}{0}$ and $\chand{-s}{0}^\dagger$ are given in Eq.~\eqref{eq:Chandop}. In our calculations in flat spacetime, where we do not perform a Fourier decomposition, we instead use $\Dbar$ and $\Dbar'$ when working with spherical harmonics. In terms of those operators, we can use the replacements $\left( \prod^{s-1}_{j=0} \chand{-j}{0}^\dagger \right) \to \Dbar^s$ and $\left( \prod^{-(s+1)}_{j=0} \chand{j}{0} \right)\to \Dbar^{-s}$ in the above definition.

The harmonics satisfy $\Y{0} = \Y{}$, $\Y{s}=0$ for $\l<|s|$,
the identities
\begin{subequations}\label{SpheroidalHarmonicsIdentities}
\begin{align}
\chand{-s}{0}^\dagger \swsy{\ess} &= \sqrt{\frac{(\ell - \ess)(\ell + \ess + 1)}{2}} \swsy{\ess + 1},\\
\chand{s}{0} \swsy{\ess} &= - \sqrt{\frac{(\ell + \ess)(\ell - \ess + 1)}{2}} \swsy{\ess - 1},\\
\chand{\ess+1}{0} \chand{-\ess}{0}^\dagger \swsy{\ess} &= - \frac{(\ell - \ess)(\ell + \ess + 1)}{2} \swsy{\ess},\\
\bY{\ess} &= (-1)^{\emm + \ess} \preindex{Y}{-\ess}_{\ell, -\emm},\label{eq:bar sYlm}
\end{align}
\end{subequations}
and the completeness relation 
\begin{equation}\label{eq:completeness}
\dOm = \sum_{\ell\geq|s|,m}\bY{s}(\theta_\textrm{p},\varphi_\textrm{p})\,\Y{s}(\thA).
\end{equation}

We also frequently appeal to the standard addition theorem for scalar harmonics,
\begin{equation}\label{eq:addition theorem}
\sum_{m=-\l}^{\l}\bY{}(\theta_\textrm{p},\varphi_\textrm{p})\,\Y{}(\thA) =\frac{2\l+1}{4\pi} P_{\l}(\cos\gamma),
\end{equation}
where $\gamma$ is the angle between $(\theta_\textrm{p},\varphi_\textrm{p})$ and $(\thA)$, and $\cos\gamma = \cos\theta\cos\theta_\textrm{p} + \cos(\varphi - \varphi_\textrm{p}) \sin\theta\sin\theta_\textrm{p}$. Addition theorems for higher spin weights are straightforwardly obtained by applying $\Dbar$ or $\Dbar'$ to this relation.

If a spin-weight-$s$ quantity $f$ is expanded as $f = \sum_{\lm}f^{\lm}\, \Y{\ess}$, then Eq.~\eqref{eq:bar sYlm} implies that the complex conjugate of $f$ can be expanded in spin-weight-$(-\ess)$ harmonics as $\bar f = \sum_{\lm}\bar f^{\lm}\, \Y{-\ess}$, where 
\begin{equation}\label{eq:bar flm}
\bar f^{\lm} = (-1)^{\emm+\ess}\overline{f^{\l,-m}}.
\end{equation}

\section{Evaluation of singular mode sums}
\label{sec:mode sums}

In Sec.~\ref{sec:CCK-Ori flat}, we found a string contribution to the Hertz potential, given by
\begin{equation}\label{eq:PhiS distribution1}
\Phi^\textrm{S} = \sum_{\l=2}^\infty\sum_{m}\Phi^\textrm{S}_{\lm}{}_{-2}Y_{\l m}
\end{equation}
with coefficients~\eqref{eq:PhiS modes}. Treated as an ordinary function, this sum diverges at all points. However, it should be interpreted as a distributional solution, not as an ordinary function, and the sum should be evaluated in the sense of distributions. The action of $\Phi^\textrm{S}$ on a spin-weight $+2$ test function $\phi(\theta,\varphi)$ is therefore
\begin{equation}\label{eq:<PhiS,phi>}
\langle \Phi^\textrm{S},\phi\rangle =\lim_{N\to\infty} \int\phi\sum_{\l=2}^N\sum_{m}\Phi^\textrm{S}_{\lm}{}_{-2}Y_{\l m} d \Omega. 
\end{equation}
This evaluates to the sum
\begin{equation}\label{eq:<PhiS,phi> sum}
\langle \Phi^\textrm{S},\phi\rangle = \sum_{\l=2}^\infty\sum_{m}\Phi^\textrm{S}_{\lm}\overline{\bar{\phi}_{\l m}},
\end{equation}
where $\bar{\phi}_{\l m}$ denotes the $\lm$ mode of $\bar\phi$ (rather than the conjugate of the $\lm$ mode of $\phi$). Equation~\eqref{eq:<PhiS,phi> sum} converges exponentially with $\l$ because $\phi_{\lm}$ decays exponentially for any smooth $\phi$.

We can use this fact to find $\Phi^\textrm{S}$ in closed form at all points off the string. We first write $\Phi^\textrm{S}$ as a different limit,
\begin{equation}\label{eq:PhiS distribution2}
\Phi^\textrm{S} = \lim_{a\to1^-}\sum_{\l=2}^\infty \sum_{m}a^\l \Phi^\textrm{S}_{\lm}{}_{-2}Y_{\l m},
\end{equation}
where $0<a<1$. The sum is exponentially convergent, yielding a smooth function for each value of $a$. Writing the action on a test function as in Eq.~\eqref{eq:<PhiS,phi>}, we find
\begin{equation}
\langle \Phi^\textrm{S},\phi\rangle = \lim_{a\to1^-}\sum_{\l=2}^\infty\sum_{m}a^\l \Phi^\textrm{S}_{\lm}\overline{\bar{\phi}_{\l m}} = \sum_{\l=2}^\infty\sum_{m}\Phi^\textrm{S}_{\lm}\overline{\bar{\phi}_{\l m}},
\end{equation}
in agreement with Eq.~\eqref{eq:<PhiS,phi>}. It follows that Eq.~\eqref{eq:PhiS distribution2} is the same distribution as \eqref{eq:PhiS distribution1}.  But unlike Eq.~\eqref{eq:PhiS distribution1}, Eq.~\eqref{eq:PhiS distribution2} can be evaluated as an ordinary smooth function away from the string. The integral of that smooth function against a test function necessarily agrees with the action of Eq.~\eqref{eq:PhiS distribution1} for all test functions whose support excludes the string. 

To carry out this strategy, we rewrite Eq.~\eqref{eq:PhiS distribution1} using Eqs.~\eqref{eq:addition theorem} and \eqref{eq:Ys} as
\begin{equation}\label{eq:PhiS summed}
\Phi^\textrm{S} = \frac{4m}{r^2_p}\Dbar^\prime{}^{2}\left\{\left[r_\textrm{p}^2\Delta r+r_\textrm{p}(\Delta r)^2\right]\Phi^\textrm{S}_1 + \frac{1}{6}(\Delta r)^3\Phi^\textrm{S}_2\right\},
\end{equation}
where
\begin{align}
\Phi^\textrm{S}_1 &= \lim_{a\to 1^-}\sum_{\l=2}^\infty \frac{a^\l (2\l+1)}{(\l+2)(\l+1)\l(\l-1)}P_\l(\cos\gamma),\\
\Phi^\textrm{S}_2 &= \lim_{a\to 1^-}\sum_{\l=2}^\infty \frac{a^\l (2\l+1)}{(\l+2)(\l-1)}P_\l(\cos\gamma).
\end{align}
These sums can be evaluated using the generating function
\begin{equation}\label{eq:generating function}
f_{(a)}(\gamma)\equiv \sum_{\l=2}a^\l P_\l(\cos\gamma) = \frac{1}{\sqrt{1+a^2-2a\cos\gamma}} - 1 - a\cos\gamma
\end{equation}
by expressing quantities of the form $F(\l)a^\l$ in terms of derivatives and integrals of $a^\l$; for example, $\frac{a^\l}{\l+1} = a^{-1}\int^a_0 d a_1 a_1^\l$. Explicitly, we write
\begin{align}
\Phi^\textrm{S}_1 &= \lim_{a\to 1^-} a^{-2}\int^a_0 d a_1 \int^{a_1}_0 d a_2 \left(2 a_2\int^{a_2}_0 d a_3\frac{d}{d a_3}+\int^{a_2}_0 d a_3\right)\int^{a_3}_0 d a_4 a_4^{-2}f_{(a_4)}(\gamma),\\
\Phi^\textrm{S}_2 &= \lim_{a\to 1^-} a^{-2}\int^a_0 d a_1 a^2_1 \int^{a_1}_0 d a_2 a_2^{-2}\left(2 a_2\frac{d}{da_2} + 1\right)f_{(a_2)}(\gamma),
\end{align}
which evaluate to
\begin{align}
\Phi^\textrm{S}_1 &= \frac{1}{4}+\ln \left(\sin (\gamma/2)\right) + \cos \gamma \Bigg[\frac{1}{12}+\frac{1}{2}\ln
   \left(\frac{1-\cos\gamma+2 \sin \left(\frac{\gamma }{2}\right)}{1-\cos\gamma}\right) \nonumber\\
   &\qquad\qquad\qquad\qquad\qquad\quad -\frac{1}{2} \textrm{arctanh}\left(\sin(\gamma/2)\right)+\frac{1}{2} \log (2 \csc\gamma)\Bigg],\\
\Phi^\textrm{S}_2 &= -\frac{1}{2} + \cos\gamma\Bigg[-\frac{4}{3} + \frac{4}{3}\ln
   \left(\frac{1-\cos\gamma+2 \sin \left(\frac{\gamma}{2}\right)}{1-\cos\gamma}\right)+\frac{1}{3}\log \left(\frac{-1+\cos\gamma +2 \sin \left(\frac{\gamma}{2}\right)}{1+\cos\gamma}\right)\nonumber\\
   &\qquad\qquad\qquad - \textrm{arctanh}\left(\sin(\gamma/2)\right) +\ln(2 \csc\gamma)\Bigg].
\end{align}

For small $\gamma$, these behave as $\Phi^\textrm{S}_1\sim \gamma^2\ln\gamma$ and $\Phi^\textrm{S}_2\sim \ln\gamma$. Since Eq.~\eqref{eq:PhiS summed} involves two derivatives of these quantities, we can infer $\Phi^\textrm{S}\sim 1/\gamma^2$. However, the derivatives should be treated distributionally, meaning that angular delta functions may arise in addition to (or instead of) the power-law divergence.

\section{Dipole mode in flat spacetime}
\label{sec:dipole}
In this appendix we analyze the $\l=1$ mode of the solution described in Sec.~\ref{sec:S*Phi+x metric}. We can put the solution in more intuitive form by introducing Cartesian 3-vectors, with indices $i,j$ raised and lowered with the Kronecker delta $\delta_{ij}$. Specifically, we introduce the unit vectors $\Omega^i=(\cos\varphi\sin\theta,\sin\varphi\sin\theta,\cos\theta)$, which points toward the field point, and $\Omega^i_p=\Omega^i(\theta_\textrm{p},\varphi_\textrm{p})$, which points toward the particle.

In terms of these quantities, from Eqs.~\eqref{eq:x inner outer}, \eqref{eq:Fab}, \eqref{eq:completeness}, and \eqref{eq:addition theorem}, the nonzero components of $x^{\l=1}_{ab}$ are
\begin{subequations}\label{eq:x ell=1 flat}
\begin{align}
x^{\l=1}_{nn} &= \left(-\frac{2\mu_i\Omega^i}{r^2}+ 2ra_i\Omega^i\right)\Theta^+,\\
x^{\l=1}_{n\bar m} &= -r\bar m^A\left(\frac{2\mu_i\Omega^i_A}{r^2} - \frac{3p_i\Omega^i_{A}}{r} + r a_i\Omega^i_A\right)\Theta^+,\\
x^{\l=1}_{m\bar m} &= \left( - \frac{6p_i\Omega^i}{r} + 6s_i\Omega^i\right)\Theta^+ .
\end{align}
\end{subequations}
Here we have used $\cos\gamma=\Omega^i_p\Omega_i$ in Eq.~\eqref{eq:addition theorem}, introduced uppercase Latin indices $A,B$ for vectors on the sphere spanned by $\theta^A=(\theta,\varphi)$, and defined $\Omega^i_A=\partial_A\Omega^i$. Uppercase indices are lowered and raised with the metric $\Omega_{AB}$ of the unit sphere; our notation here follows Ref.~\cite{Poisson:2011nh}, for example. We have also introduced the 3-vectors $a^i = \frac{m\Omega^i_p}{r_\textrm{p}^2}$, $s^i = \frac{m\Omega^i_p}{r_\textrm{p}}$,  $p^i = m\Omega^i_p$, and $\mu^i = m r_\textrm{p} \Omega^i_p$.

The 3-vector $\mu_i$ is a mass dipole moment, representing the displacement (times the mass $m$) of the center of mass relative to the origin. In Kerr, such a quantity would be pure gauge because it would represent an order-$\varepsilon$ displacement away from the ``center'' of the black hole. But in flat spacetime, it is invariant under perturbative gauge transformations, as it is an order-1 displacement that would require an order-1 translation to remove.

The $a_i$, $p_i$, and $s_i$ terms in $x^{\l=1}_{ab}$ are pure gauge in the region $r>r_\textrm{p}$. In particular, the $a_i$ terms represent a uniform acceleration of the coordinate system; see Eq.~(10.37) of Ref.~\cite{Poisson:2011nh}. We can eliminate these terms with a gauge transformation generated by
\begin{equation}\label{eq:Xi ell=1}
\Xi_{\l=1} = \alpha_i\Omega^i \dbd{u} + \beta_i\Omega^i \dbd{r} + r^{-1}\gamma^i\Omega_i^A \dbd{x^A},
\end{equation}
where $\Omega_i^A=\Omega^{AB}\delta_{ij}\Omega^j_B$.  $\alpha_i$, $\beta_i$, and $\gamma_i$ are functions of $(u,r)$ given by
\begin{subequations}
\begin{align}
\alpha_i &= \frac{1}{2}a_i u^2,\\
\beta_i &= -\frac{1}{2}a_i u^2 - a_i u r -3p_i,\\
\gamma_i &= -\frac{1}{2}a_i u^2 - a_i u r-3s_i r.
\end{align}
\end{subequations}
One can straightforwardly check that $\partial_u\Xi_{\l=1}$ is the generator of a boost in the direction of the particle, $\partial_u\Xi_{\l=1} = a^i \left(t\frac{\partial}{\partial x^i}+x_i \frac{\partial}{\partial t}\right)$ in inertial coordinates $(t=u+r,x^i)$. Although the $p_i$ terms in $x^{\l=1}_{ab}$ do not need to be gauged away to obtain manifest asymptotic flatness, doing so puts the metric perturbation in the canonical $\sim 1/r^2$ form of a stationary dipolar perturbation.

It is easily checked that the $a_i$ terms in this transformation are unique: any transformation that gauges away the $a_i$ terms in $x_{ab}$ must have precisely this $u$ dependence. Just as for $\l=0$, if we apply this transformation only for $r>r_\textrm{p}$, then we introduce a (quadratic-in-$u$) growth in the coefficient of the Dirac delta function in the no-string solution. More generally, it is impossible to transition the transformation to zero in any way without introducing such growth in the transition region.

On the other hand, the $p_i$ and $s_i$ terms in the transformation can be deformed in any desirable manner while preserving the $\sim 1/r^2$ form of the perturbation at large $r$. For concreteness, we apply $\Xi^a$ as written in  Eq.~\eqref{eq:Xi ell=1} for all $r$, leading to the no-string solution~\eqref{eq:h no string + l=1,2}, but we keep in mind that only the linear-and-quadratic-in-$u$ terms actually need to be applied globally in this way.

\section{Dependent equations in transformation to the shadowless gauge}
\label{app:dependent equations}

This appendix lists the equations that follow from Eq.~\eqref{eq:xidet}.
These equations are divided into independent and dependent ones and are constructed by first taking individual NP components and then expanding the result as a finite power series in $\rho$ whose coefficients are Held scalars. We thereby get equations for each power of $\rho$, but these are not necessarily independent from each other:
\begin{itemize}
\item The $m\bar{m}$ component of Eq.~\eqref{eq:xidet} gives:
\begin{equation}
\textrm{$m\bar{m}$ component of \eqref{eq:xidet}} = \left( \frac{\rho}{\bar{\rho}} + \frac{\bar{\rho}}{\rho} \right) \eqref{HeldSDaggerXLiexigmmb:eq1} + \left( \rho + \bar{\rho} \right)  \eqref{HeldSDaggerXLiexigmmb:eq2}.
\end{equation}
Then we divide by the highest power of $\bar \rho$ and use 
$1/\bar \rho = 1/\rho + \Omega^\circ$, to find that
\eqref{HeldSDaggerXLiexigmmb:eq1} and 
\eqref{HeldSDaggerXLiexigmmb:eq2} hold. 
\item The $mm$ component of Eq.~\eqref{eq:xidet} gives:
\begin{equation}
\textrm{$mm$ component of \eqref{eq:xidet}} = 2 \eqref{HeldSDaggerXLiexigmm:eq1} + 2 \bar{\rho} \eqref{HeldSDaggerXLiexigmm:eq2}.
\end{equation}
Then we divide by the highest power of $\bar \rho$ and use 
$1/\bar \rho = 1/\rho + \Omega^\circ$ to find that
\eqref{HeldSDaggerXLiexigmm:eq1} and 
\eqref{HeldSDaggerXLiexigmm:eq2} hold. 
\item The $nm$ component of Eq.~\eqref{eq:xidet} gives:
\begin{equation}
\textrm{$nm$ component of \eqref{eq:xidet}} = \frac{1}{\bar{\rho}} \eqref{HeldSDaggerXLiexignm:eq1} + \bar{\rho} \eqref{HeldSDaggerXLiexignm:eq3} + \rho \bar{\rho} \eqref{HeldSDaggerXLiexignm:eq4} + \rho \bar{\rho}^2 \Omega^\circ \eqref{HeldSDaggerXLiexignm:eq5} + \rho^2 \bar{\rho} \Omega^\circ \eqref{HeldSDaggerXLiexignm:eq6}.
\end{equation}
Then we divide by the highest power of $\bar \rho$ and use 
$1/\bar \rho = 1/\rho + \Omega^\circ$ to find that
\eqref{HeldSDaggerXLiexignm:eq1}, \eqref{HeldSDaggerXLiexignm:eq6} and
the following equations hold. 
\begin{multline}
\label{HeldSDaggerXLiexignm:eq3}
- \tilde{\edth}' \bar{d}^\circ_1 - \Omega^\circ \tilde{\edth}' \bar{d}^\circ_2 + 2 \bar{\tau}^\circ \bar{d}^\circ_2 - \Omega^{\circ 2} \tilde{\edth}' \bar{d}^\circ_3 + \Omega^\circ \tilde{\edth} a^\circ + 4 \tau^\circ a^\circ + \tilde{\edth} b^\circ - \Omega^{\circ 2} c^\circ = \nls\\
- \half \Omega^\circ \tilde{\edth} \tilde{\th}' \xi^\circ_l + 3 \tau^\circ \tilde{\th}' \xi^\circ_l - \rho^{\prime \circ} \tilde{\edth} \xi^\circ_l + \tilde{\edth} \xi^\circ_n + \bar{\tau}^\circ \tilde{\edth} \xi^\circ_m + \tau^\circ \tilde{\edth}' \xi^\circ_m - \half \left( \Psi^\circ - \bar{\Psi}^\circ \right) \xi^\circ_m \nls\\
+ \half \left( \rho^{\prime \circ} + \bar{\rho}^{\prime \circ} \right) \Omega^\circ \xi^\circ_m, \nls
\end{multline}
\begin{multline}
\label{HeldSDaggerXLiexignm:eq4}
- 2 \tilde{\edth}' \bar{d}^\circ_0 - 2 \Omega^\circ \tilde{\edth}' \bar{d}^\circ_1 + 2 \bar{\tau}^\circ \bar{d}^\circ_1 - 2 \Omega^{\circ 2} \tilde{\edth}' \bar{d}^\circ_2 + 2 \Omega^\circ \bar{\tau}^\circ \bar{d}^\circ_2 - 2 \Omega^{\circ 3} \tilde{\edth}' \bar{d}^\circ_3 + 2 \tau^\circ b^\circ + 2 e^\circ = \nls\\
- 2 \bar{\tau}^\circ \tilde{\edth}^2 \xi^\circ_l - \tau^\circ \left( \tilde{\edth} \tilde{\edth}' + \tilde{\edth}' \tilde{\edth} \right) \xi^\circ_l + 2 \Psi^\circ \tilde{\edth} \xi^\circ_l + \left( \rho^{\prime \circ} + \bar{\rho}^{\prime \circ} \right) \tau^\circ \xi^\circ_l + 2 \tau^\circ \xi^\circ_n + 2 \Omega^\circ \bar{\tau}^\circ \tilde{\edth} \xi^\circ_m \nls\\
+ \Omega^\circ \tau^\circ \tilde{\edth}' \xi^\circ_m + 2 \tau^\circ \bar{\tau}^\circ \xi^\circ_m - 2 \Omega^\circ \Psi^\circ \xi^\circ_m - \Omega^\circ \tau^\circ \tilde{\edth} \xi^\circ_{\bar{m}} - 2 \tau^{\circ 2} \xi^\circ_{\bar{m}}  + 2 g^\circ_{\dot a}, \nls
\end{multline}
\begin{multline}
\label{HeldSDaggerXLiexignm:eq5}
\Omega^\circ \tau^\circ a^\circ - \tau^\circ b^\circ = \half \tau^\circ \left( \tilde{\edth} \tilde{\edth}' + \tilde{\edth}' \tilde{\edth} \right) \xi^\circ_l + \Omega^\circ \tau^\circ \tilde{\th}' \xi^\circ_l - \half \left( \rho^{\prime \circ} + \bar{\rho}^{\prime \circ} \right) \tau^\circ \xi^\circ_l - \tau^\circ \xi^\circ_n \nls\\
+ \tau^\circ \bar{\tau}^\circ \xi^\circ_m + \Omega^\circ \tau^\circ \tilde{\edth} \xi^\circ_{\bar{m}} + {\tau^\circ}^2 \xi^\circ_{\bar{m}}. \nls
\end{multline}

\item The $nn$ component of Eq.~\eqref{eq:xidet} gives:
\begin{multline}
\textrm{$nn$ component of \eqref{eq:xidet}} = \left( \frac{1}{\rho} + \frac{1}{\bar{\rho}} \right) \eqref{HeldSDaggerXLiexignn:eq1} + \eqref{HeldSDaggerXLiexignn:eq2} + \left( \rho + \bar{\rho} \right) \eqref{HeldSDaggerXLiexignn:eq3} + \rho \bar{\rho} \eqref{HeldSDaggerXLiexignn:eq4}\\
+ \rho \bar{\rho} \left( \rho + \bar{\rho} \right) \eqref{HeldSDaggerXLiexignn:eq5} + \rho^2 \bar{\rho}^2 \eqref{HeldSDaggerXLiexignn:eq6}.
\end{multline}
Then we divide by the highest power of $\bar \rho$ and use 
$1/\bar \rho = 1/\rho + \Omega^\circ$ to find that \eqref{HeldSDaggerXLiexignn:eq3} and
the following equations hold.
\begin{equation}
\label{HeldSDaggerXLiexignn:eq1}
\Re \left\{ - \tilde{\edth}^{\prime 2} \bar{d}^\circ_3 + \tilde{\th}' a^\circ - \tilde{\edth}' c^\circ \right\} = \tilde{\th}^{\prime 2} \xi^\circ_l, \nls
\end{equation}
\begin{multline}
\label{HeldSDaggerXLiexignn:eq2}
2 \Re \left\{ - \tilde{\edth}^{\prime 2} \bar{d}^\circ_2 + 4 \bar{\tau}^\circ \tilde{\edth}' \bar{d}^\circ_3 + \dfrac{5}{2} \Omega^\circ \tilde{\edth}^{\prime 2} \bar{d}^\circ_3 + \tilde{\edth}' \tilde{\edth} a^\circ - 2 \rho^{\prime \circ} a^\circ - \dfrac{5}{2} \Omega^\circ \tilde{\edth}' c^\circ + 4 \bar{\tau}^\circ c^\circ \right\} = \nls\\
- \left( \rho^{\prime \circ} + \bar{\rho}^{\prime \circ} \right) \tilde{\th}' \xi^\circ_l + 2 \tilde{\th}' \xi^\circ_n, \nls
\end{multline}
\begin{multline}
\label{HeldSDaggerXLiexignn:eq4}
2 \Re \Big\{ - \tilde{\edth}^{\prime 2} \bar{d}^\circ_0 - \dfrac{3}{2} \Omega^\circ \tilde{\edth}^{\prime 2} \bar{d}^\circ_1 - 2 \Omega^{\circ 2} \tilde{\edth}^{\prime 2} \bar{d}^\circ_2 + 3 \Omega^\circ \bar{\tau}^\circ \tilde{\edth}' \bar{d}^\circ_2 + 2 \bar{\tau}^{\circ 2} \bar{d}^\circ_2 - \dfrac{5}{2} \Omega^{\circ 3} \tilde{\edth}^{\prime 2} \bar{d}^\circ_3 + 8 \Omega^{\circ 2} \bar{\tau}^\circ \tilde{\edth}' \bar{d}^\circ_3 \nls\\
+ \half \Omega^{\circ 2} \left( \tilde{\edth}' \tilde{\edth} + \tilde{\edth} \tilde{\edth}' \right) a^\circ + 6 \tau^\circ \bar{\tau}^\circ a^\circ + \frac{3}{2} \Omega^\circ \left( \Psi^\circ - \bar{\Psi}^\circ \right) a^\circ + 2 \bar{\tau}^\circ \tilde{\edth} b^\circ + \Psi^\circ b^\circ - \dfrac{3}{2} \Omega^{\circ 3} \tilde{\edth}' c^\circ \nls\\
+ 2 \Omega^{\circ 2} \bar{\tau}^\circ c^\circ + \tilde{\edth}' e^\circ \Big\} = - \Omega^\circ \bar{\tau}^\circ \tilde{\edth} \tilde{\th}' \xi^\circ_l + \Omega^\circ \tau^\circ \tilde{\edth}' \tilde{\th}' \xi^\circ_l + 8 \tau^\circ \bar{\tau}^\circ \tilde{\th}' \xi^\circ_l + 2 \Omega^\circ \left( \Psi^\circ - \bar{\Psi}^\circ \right) \tilde{\th}' \xi^\circ_l \nls\\
- 2 \rho^{\prime \circ} \bar{\tau}^\circ \tilde{\edth} \xi^\circ_l - 2 \bar{\rho}^{\prime \circ} \tau^\circ \tilde{\edth}' \xi^\circ_l + \rho^{\prime \circ} \Psi^\circ \xi^\circ_l + \bar{\rho}^{\prime \circ} \bar{\Psi}^\circ \xi^\circ_l + 2 \bar{\tau}^\circ \tilde{\edth} \xi^\circ_n + 2 \tau^\circ \tilde{\edth}' \xi^\circ_n + \left( \Psi^\circ + \bar{\Psi}^\circ \right) \xi^\circ_n \nls\\
+ \Omega^{\circ 2} \bar{\tau}^\circ \tilde{\th}' \xi^\circ_m + 2 \Omega^\circ \rho^{\prime \circ} \bar{\tau}^\circ \xi^\circ_m + \Omega^{\circ 2} \tau^\circ \tilde{\th}' \xi^\circ_{\bar{m}} - 2 \Omega^\circ \bar{\rho}^{\prime \circ} \tau^\circ \xi^\circ_{\bar{m}}, \nls
\end{multline}
\begin{multline}
\label{HeldSDaggerXLiexignn:eq5}
2\Re \Big\{ - \half \Omega^\circ \tilde{\edth}^{\prime 2} \bar{d}^\circ_0 - \bar{\tau}^\circ \tilde{\edth}' \bar{d}^\circ_0 - \half \Omega^{\circ 2} \tilde{\edth}^{\prime 2} \bar{d}^\circ_1 + \bar{\tau}^{\circ 2} \bar{d}^\circ_1 - \half \Omega^{\circ 3} \tilde{\edth}^{\prime 2} \bar{d}^\circ_2 + \Omega^{\circ 2} \bar{\tau}^\circ \tilde{\edth}' \bar{d}^\circ_2 \nls\\
+ \Omega^\circ \bar{\tau}^{\circ 2} \bar{d}^\circ_2 - \half \Omega^{\circ 4} \tilde{\edth}^{\prime 2} \bar{d}^\circ_3 + 2 \Omega^{\circ 3} \bar{\tau}^\circ \tilde{\edth}' \bar{d}^\circ_3 + \half \Omega^{\circ 2} \Psi^\circ a^\circ + \tau^\circ \bar{\tau}^\circ b^\circ + \half \Omega^\circ \Psi^\circ b^\circ + \half \Omega^\circ \tilde{\edth}' e^\circ \nls\\
+ \bar{\tau}^\circ e^\circ \Big\} = - \bar{\tau}^{\circ 2} \tilde{\edth}^2 \xi^\circ_l - \tau^\circ \bar{\tau}^\circ \left( \tilde{\edth} \tilde{\edth}' + \tilde{\edth}' \tilde{\edth} \right) \xi^\circ_l - \tau^{\circ 2} \tilde{\edth}^{\prime 2} \xi^\circ_l + \dfrac{1}{4} \Omega^{\circ 2} \left( \Psi^\circ + \bar{\Psi}^\circ \right) \tilde{\th}' \xi^\circ_l \nls\\
+ \Psi^\circ \bar{\tau}^\circ \tilde{\edth} \xi^\circ_l + \bar{\Psi}^\circ \tau^\circ \tilde{\edth}' \xi^\circ_l + \left( \rho^{\prime \circ} + \bar{\rho}^{\prime \circ} \right) \tau^\circ \bar{\tau}^\circ \xi^\circ_l + \half \Omega^\circ \rho^{\prime \circ} \Psi^\circ \xi^\circ_l - \half \Omega^\circ \bar{\rho}^{\prime \circ} \bar{\Psi}^\circ \xi^\circ_l + 2 \tau^\circ \bar{\tau}^\circ \xi^\circ_n \nls\\
+ \half \Omega^\circ \left( \Psi^\circ - \bar{\Psi}^\circ \right) \xi^\circ_n + \Omega^\circ \bar{\tau}^{\circ 2} \tilde{\edth} \xi^\circ_m + \Omega^\circ \tau^\circ \bar{\tau}^\circ \tilde{\edth}' \xi^\circ_m - \half \Omega^\circ \bar{\tau}^\circ \left( \Psi^\circ - \bar{\Psi}^\circ \right) \xi^\circ_m - \Omega^\circ \tau^{\circ 2} \tilde{\edth}' \xi^\circ_{\bar{m}} \nls\\
- \Omega^\circ \tau^\circ \bar{\tau}^\circ \tilde{\edth} \xi^\circ_{\bar{m}} - \half \Omega^\circ \tau^\circ \left( \Psi^\circ - \bar{\Psi}^\circ \right) \xi^\circ_{\bar{m}}
+ 2 \Re \left\{ \half \Omega^\circ \tilde{\edth}' g^\circ_{\dot a} + \bar{\tau}^\circ g^\circ_{\dot a} \right\}, \nls
\end{multline}
\begin{multline}
\label{HeldSDaggerXLiexignn:eq6}
2 \Re \Big\{ - \half \Omega^{\circ 2} \tilde{\edth}^{\prime 2} \bar{d}^\circ_0 - \Omega^\circ \bar{\tau}^\circ \tilde{\edth}' \bar{d}^\circ_0 - \half \Omega^{\circ 3} \tilde{\edth}^{\prime 2} \bar{d}^\circ_1 + \Omega^\circ \bar{\tau}^{\circ 2} \bar{d}^\circ_1 - \half \Omega^{\circ 4} \tilde{\edth}^{\prime 2} \bar{d}^\circ_2 + \Omega^{\circ 3} \bar{\tau}^\circ \tilde{\edth}' \bar{d}^\circ_2 \nls\\
+ \Omega^{\circ 2} \bar{\tau}^{\circ 2} \bar{d}^\circ_2 - \half \Omega^{\circ 5} \tilde{\edth}^{\prime 2} \bar{d}^\circ_3 + 2 \Omega^{\circ 4} \bar{\tau}^\circ \tilde{\edth}' \bar{d}^\circ_3 + \Omega^{\circ 2} \tau^\circ \bar{\tau}^\circ a^\circ - \half \Omega^{\circ 3} \bar{\Psi}^\circ a^\circ + \half \Omega^{\circ 2} \Psi^\circ b^\circ \nls\\
+ \half \Omega^{\circ 2} \tilde{\edth}' e^\circ + \Omega^\circ \bar{\tau}^\circ e^\circ \Big\} = - \Omega^\circ \bar{\tau}^{\circ 2} \tilde{\edth}^2 \xi^\circ_l + \Omega^\circ \tau^{\circ 2} \tilde{\edth}^{\prime 2} \xi^\circ_l + 2 \Omega^{\circ 2} \tau^\circ \bar{\tau}^\circ \tilde{\th}' \xi^\circ_l \nls\\
+ \dfrac{1}{4} \Omega^{\circ 3} \left( \Psi^\circ + \bar{\Psi}^\circ \right) \tilde{\th}' \xi^\circ_l + \Omega^\circ \Psi^\circ \bar{\tau}^\circ \tilde{\edth} \xi^\circ_l - \Omega^\circ \bar{\Psi}^\circ \tau^\circ \tilde{\edth}' \xi^\circ_l + \half \Omega^{\circ 2} \rho^{\prime \circ} \Psi^\circ \xi^\circ_l + \half \Omega^{\circ 2}  \bar{\rho}^{\prime \circ} \bar{\Psi}^\circ \xi^\circ_l \nls\\
+ \half \Omega^{\circ 2} \left( \Psi^\circ + \bar{\Psi}^\circ \right) \xi^\circ_n + \Omega^{\circ 2} \bar{\tau}^{\circ 2} \tilde{\edth} \xi^\circ_m + \Omega^{\circ 2} \tau^\circ \bar{\tau}^\circ \tilde{\edth}' \xi^\circ_m + 2 \Omega^\circ \tau^\circ \bar{\tau}^\circ \xi^\circ_m - \half \Omega^{\circ 2} \bar{\tau}^\circ \left( \Psi^\circ + \bar{\Psi}^\circ \right) \xi^\circ_m \nls\\
+ \Omega^{\circ 2} \tau^{\circ 2} \tilde{\edth}' \xi^\circ_{\bar{m}} + \Omega^{\circ 2} \tau^\circ \bar{\tau}^\circ \tilde{\edth} \xi^\circ_{\bar{m}} - 2 \Omega^\circ \tau^\circ \bar{\tau}^\circ \xi^\circ_{\bar{m}} - \half \Omega^{\circ 2} \tau^\circ \left( \Psi^\circ + \bar{\Psi}^\circ \right) \xi^\circ_{\bar{m}}. \nls
\end{multline}
\end{itemize}

Each of the explicitly written equations can actually be expressed as a combination of the independent equations \eqref{HeldSDaggerXLiexigmmb:eq1}, \eqref{HeldSDaggerXLiexigmmb:eq2}, \eqref{HeldSDaggerXLiexigmm:eq1}, \eqref{HeldSDaggerXLiexigmm:eq2}, \eqref{HeldSDaggerXLiexignm:eq1}, \eqref{HeldSDaggerXLiexignm:eq6}, and \eqref{HeldSDaggerXLiexignn:eq3}:
\begin{subequations}\label{TheMasterEqCombinations}
\begin{align}
\eqref{HeldSDaggerXLiexignm:eq3} &= - \Omega^{\circ 2} \text{\eqref{HeldSDaggerXLiexignm:eq1}} - \tilde{\edth}' \text{\eqref{HeldSDaggerXLiexigmm:eq2}} - \Omega^\circ \tilde{\edth}' \text{\eqref{HeldSDaggerXLiexigmm:eq1}} + \tilde{\edth} \text{\eqref{HeldSDaggerXLiexigmmb:eq2}} + \Omega^\circ \tilde{\edth} \text{\eqref{HeldSDaggerXLiexigmmb:eq1}} + 4 \tau^\circ \text{\eqref{HeldSDaggerXLiexigmmb:eq1}}, \nls\\
\eqref{HeldSDaggerXLiexignm:eq4} &= 2 \text{\eqref{HeldSDaggerXLiexignm:eq6}} + 2 \bar{\tau}^\circ \text{\eqref{HeldSDaggerXLiexigmm:eq2}} + 2 \Omega^\circ \bar{\tau}^\circ \text{\eqref{HeldSDaggerXLiexigmm:eq1}} + 2 \tau^\circ \text{\eqref{HeldSDaggerXLiexigmmb:eq2}}, \nls\\
\eqref{HeldSDaggerXLiexignm:eq5} &= - \tau^\circ \text{\eqref{HeldSDaggerXLiexigmmb:eq2}} + \Omega^\circ \tau^\circ \text{\eqref{HeldSDaggerXLiexigmmb:eq1}}, \nls\\
\eqref{HeldSDaggerXLiexignn:eq1} &= \Re \left\{ - \tilde{\edth}' \text{\eqref{HeldSDaggerXLiexignm:eq1}} + \tilde{\th}' \text{\eqref{HeldSDaggerXLiexigmmb:eq1}} \right\}, \nls\\
\eqref{HeldSDaggerXLiexignn:eq2} &= 2 \Re \left\{ 4 \bar{\tau}^\circ \text{\eqref{HeldSDaggerXLiexignm:eq1}} - \dfrac{5}{2} \Omega^\circ \tilde{\edth}' \text{\eqref{HeldSDaggerXLiexignm:eq1}} - \tilde{\edth}^{\prime 2} \text{\eqref{HeldSDaggerXLiexigmm:eq1}} + \tilde{\th}' \text{\eqref{HeldSDaggerXLiexigmmb:eq2}} + \tilde{\edth} \tilde{\edth}' \text{\eqref{HeldSDaggerXLiexigmmb:eq1}} - 2 \rho^{\prime \circ} \text{\eqref{HeldSDaggerXLiexigmmb:eq1}} \right\}, \nls\\
\eqref{HeldSDaggerXLiexignn:eq4} &= 2 \Re \Big\{ \tilde{\edth}' \text{\eqref{HeldSDaggerXLiexignm:eq6}} - \dfrac{3}{2} \Omega^{\circ 3} \tilde{\edth}' \text{\eqref{HeldSDaggerXLiexignm:eq1}} + 2 \Omega^{\circ 2} \bar{\tau}^\circ \text{\eqref{HeldSDaggerXLiexignm:eq1}} + \half \Omega^\circ \tilde{\edth}^{\prime 2} \text{\eqref{HeldSDaggerXLiexigmm:eq2}} - 2 \bar{\tau}^\circ \tilde{\edth}' \text{\eqref{HeldSDaggerXLiexigmm:eq2}}\nonumber \nls\\
&\quad  - \Omega^{\circ 2} \tilde{\edth}^{\prime 2} \text{\eqref{HeldSDaggerXLiexigmm:eq1}}- \Omega^\circ \bar{\tau}^\circ \tilde{\edth}' \text{\eqref{HeldSDaggerXLiexigmm:eq1}} + 2 \bar{\tau}^{\circ 2} \text{\eqref{HeldSDaggerXLiexigmm:eq1}} + 2 \bar{\tau}^\circ \tilde{\edth} \text{\eqref{HeldSDaggerXLiexigmmb:eq2}} + \Psi^\circ \text{\eqref{HeldSDaggerXLiexigmmb:eq2}} + \Omega^{\circ 2} \tilde{\edth} \tilde{\edth}' \text{\eqref{HeldSDaggerXLiexigmmb:eq1}} \nonumber \nls\\
&\quad + 6 \tau^\circ \bar{\tau}^\circ \text{\eqref{HeldSDaggerXLiexigmmb:eq1}} + 3 \Omega^\circ \Psi^\circ \text{\eqref{HeldSDaggerXLiexigmmb:eq1}} \Big\}, \nls\\
\eqref{HeldSDaggerXLiexignn:eq5} &= 2 \Re \Big\{ \half \Omega^\circ \tilde{\edth}' \text{\eqref{HeldSDaggerXLiexignm:eq6}} + \bar{\tau}^\circ \text{\eqref{HeldSDaggerXLiexignm:eq6}} + \bar{\tau}^{\circ 2} \text{\eqref{HeldSDaggerXLiexigmm:eq2}} + \Omega^\circ \bar{\tau}^{\circ 2} \text{\eqref{HeldSDaggerXLiexigmm:eq1}} + \tau^\circ \bar{\tau}^\circ \text{\eqref{HeldSDaggerXLiexigmmb:eq2}} + \half \Omega^\circ \Psi^\circ \text{\eqref{HeldSDaggerXLiexigmmb:eq2}}\nonumber \nls\\
&\quad + \half \Omega^{\circ 2} \Psi^\circ \text{\eqref{HeldSDaggerXLiexigmmb:eq1}} \Big\}, \nls\\
\eqref{HeldSDaggerXLiexignn:eq6} &= 2 \Re \Big\{ \half \Omega^{\circ 2} \tilde{\edth}' \text{\eqref{HeldSDaggerXLiexignm:eq6}} + \Omega^\circ \bar{\tau}^\circ \text{\eqref{HeldSDaggerXLiexignm:eq6}} + \Omega^\circ \bar{\tau}^{\circ 2} \text{\eqref{HeldSDaggerXLiexigmm:eq2}} + \Omega^{\circ 2} \bar{\tau}^{\circ 2} \text{\eqref{HeldSDaggerXLiexigmm:eq1}} + \half \Omega^{\circ 2} \Psi^\circ \text{\eqref{HeldSDaggerXLiexigmmb:eq2}}\nonumber \nls\\
&\quad+ \Omega^{\circ 2} \tau^\circ \bar{\tau}^\circ \text{\eqref{HeldSDaggerXLiexigmmb:eq1}} + \half \Omega^{\circ 3} \Psi^\circ \text{\eqref{HeldSDaggerXLiexigmmb:eq1}} \Big\}. \nls
\end{align}
\end{subequations}
The equations \eqref{TheMasterEqCombinations} thus do not give us new information but could be used for a consistency check. 

\subsection{Mode decomposition for Schwarzschild spacetime}\label{app:gauge vec mode decomp}

In this subsection, we work out the nontrivial, independent equations in Sec.~\ref{subsec:Solve:eq:xidet}, 
which are used to find $\xi^\circ_m, \xi^\circ_n, \xi^\circ_m$, and hence the gauge vector field $\xi^a$, 
in the special case of Schwarzschild spacetime.  
Generally, we proceed as in Kerr, except for two main differences: (i) we use that in Schwarzschild, we have $\Omega^\circ = \tau^\circ = 0$, which results in many simplifications. (ii) On the other hand, Schwarzschild spacetime possesses two extra Killing fields, which results in a more complicated form of the growing piece $\Xi^a$ in $\xi^a$.

As in the Kerr case, we work in the Kinnersley frame and $u$, $r$, $\theta$ and $\varphi$ are the retarded Kerr-Newman coordinates which reduce to retarded Eddington-Finkelstein coordinates. 
Taking into account the simplifications for Schwarzschild, $\xi^a$ as in \eqref{eq:residual IRG} 
is given by
\begin{equation}\label{eq:residual xi Schwrz}
\xi^a = \left( \xi^\circ_n - r \partial_u \xi^\circ_l - \frac{M}{r} \xi^\circ_l \right) l^a + \xi^\circ_l n^a + \left( r \xi^\circ_m + \tilde{\edth} \xi^\circ_l \right) \bar{m}^a + \left( r \bar{\xi}^\circ_m + \tilde{\edth}' \xi^\circ_l \right) m^a.
\end{equation}
We now work out what becomes of the non-trivial, independent equations for $\xi^\circ_m$, $\xi^\circ_n$, and $\xi^\circ_{m}$ in Sec.~\ref{subsec:Solve:eq:xidet}, namely \eqref{HeldSDaggerXLiexigmmb:eq1}, \eqref{HeldSDaggerXLiexigmmb:eq2}, \eqref{HeldSDaggerXLiexigmm:eq1}, \eqref{HeldSDaggerXLiexigmm:eq2}, \eqref{HeldSDaggerXLiexignm:eq1}, \eqref{HeldSDaggerXLiexignm:eq6} and \eqref{HeldSDaggerXLiexignn:eq3}. It is convenient to decompose all these equations into spin-weighted spherical harmonics $\swsy{\ess}(\theta, \varphi)$, for which we adopt notations and conventions presented in \ref{sec:spin-weighted harmonics}.  The relations needed here are \eqref{SpheroidalHarmonicsIdentities}. Additionally, as in the Kerr case, we decompose each component into a growing (in $u$) part, a DC part, and an AC part as in \eqref{eq:xi circ decomp}.

For the growing in $u$ part, we derive similarly as in Kerr that
\begin{equation}\label{eq:Xi schwarzschild}
\Xi^a = u(\alpha  t^a + \beta z^a + \gamma y^a + \delta x^a),
\end{equation}
where $t^a=\dbd{t}$, and $x^a, y^a, z^a$ generates an infinitesimal rotation around the $x, y, z$ axis. The real constants $\alpha,\beta,\gamma,\delta$ are to be determined. Again, unlike in the flat spacetime calculation, there cannot be any terms quadratic in $u$ in $\Xi^a$.  The Killing vectors
$t^a, x^a, y^a, z^a$, when written in the form~\eqref{eq:residual xi Schwrz}, carry the Held coefficients 
\begin{subequations}
\begin{align}
t^\circ_l = 2 t^\circ_n = 1 , &\qquad t^\circ_m = 0,\label{eq: t Held schwarzschild}\\
z^\circ_l = z^\circ_n = 0 , &\qquad z^\circ_m = \frac{i\sin\theta}{\sqrt{2}},\label{eq:z Held schwarzschild}\\
y^\circ_l = y^\circ_n = 0 , &\qquad y^\circ_m = - \frac{\cos\varphi - i \cos\theta \sin\varphi}{\sqrt{2}},\label{eq:y Held schwarzschild}\\
x^\circ_l = x^\circ_n = 0 , &\qquad x^\circ_m = - \frac{\sin\varphi + i \cos\theta \cos\varphi}{\sqrt{2}},\label{eq:x Held schwarzschild}
\end{align}
\end{subequations}
Alternatively, in terms of spin-weighted spherical harmonics,
\begin{subequations}
\begin{align}
t^\circ_l = 2 t^\circ_n = \sqrt{4\pi} \swsY{0}{0}{0} , &\qquad t^\circ_m = 0,\label{eq: t Held schwarzschild modes}\\
z^\circ_l = 2 z^\circ_n = 0 , &\qquad z^\circ_m = 2i \sqrt{\frac{\pi}{3}} \swsY{1}{1}{0},\label{eq:z Held schwarzschild modes}\\
y^\circ_l = 2 y^\circ_n = 0 , &\qquad y^\circ_m = \sqrt{\frac{2\pi}{3}} (\swsY{1}{1}{+1} + \swsY{1}{1}{-1}),\label{eq:y Held schwarzschild modes}\\
x^\circ_l = 2 x^\circ_n = 0 , &\qquad x^\circ_m = -i \sqrt{\frac{2\pi}{3}} (\swsY{1}{1}{+1} - \swsY{1}{1}{-1}),\label{eq:x Held schwarzschild modes}
\end{align}
\end{subequations}
From this and Eqs.~\eqref{eq:Xi schwarzschild}, 
\eqref{eq:residual xi Schwrz} we can determine
\begin{subequations}
\begin{align}
\Xi^\circ_l &= 2\Xi^\circ_n = \alpha \sqrt{4\pi} \swsY{0}{0}{0} u,\\
\Xi^\circ_m &= 2i \beta \sqrt{\frac{\pi}{3}} \swsY{1}{1}{0} u + \gamma \sqrt{\frac{2\pi}{3}} (\swsY{1}{1}{+1} + \swsY{1}{1}{-1}) u - i \delta \sqrt{\frac{2\pi}{3}} (\swsY{1}{1}{+1} - \swsY{1}{1}{-1}) u.
\end{align}
\end{subequations}
Next, we determine the coefficients $\alpha,\beta,\gamma,\delta$, as well as $\langle \xi^{\circ}_{l,n,m}\rangle$ and $Z^{\circ}_{l,n,m}$, by substituting the vector~\eqref{eq:xi circ decomp}, into Eqs.~\eqref{HeldSDaggerXLiexigmmb:eq1}--\eqref{HeldSDaggerXLiexignm:eq6} and \eqref{HeldSDaggerXLiexignn:eq3} and picking out the AC and DC parts of each equation. Beginning with the DC part, the results are
\begin{subequations}
\begin{equation}\label{eq:stationary gauge Schwarzschild 1}
\langle a^\circ_{\ell\,\emm} \rangle = \sqrt{4\pi} \alpha \delta_{l, 0} \delta_{m, 0} - \frac{\C{-1}}{2\sqrt{2}}\left( \langle \xi^\circ_{m, \ell\, \emm} \rangle + (-1)^m \langle \bar{\xi}^\circ_{m, \ell\, -\emm} \rangle \right) \sforall \ell \geq 0,
\end{equation}
\begin{equation}\label{eq:stationary gauge Schwarzschild 2}
\langle b^\circ_{\ell\, \emm} \rangle = \half (\C{-1}^2 - 1) \langle \xi^\circ_{l, \ell\, \emm} \rangle + \langle \xi^\circ_{n, \ell\, \emm} \rangle \sforall \ell \geq 0,
\end{equation}
\begin{equation}\label{eq:stationary gauge Schwarzschild 3}
\langle \bar{d}^\circ_{2, \ell\, -\emm} \rangle = (-1)^\emm  \frac{\C{-2}}{\sqrt{2}\C{-1}} \langle \xi^\circ_{m, \ell\, \emm} \rangle \sforall \ell \geq 2,
\end{equation}
\begin{equation}\label{eq:stationary gauge Schwarzschild 4}
\langle \bar{d}^\circ_{1, \ell\, -\emm} \rangle = (-1)^{(\emm + 1)}  \half\C{-2} \langle \xi^\circ_{l, \ell\, \emm} \rangle \sforall \ell \geq 2,
\end{equation}
\begin{multline}\label{eq:stationary gauge Schwarzschild 5}
(-1)^{\emm + 1} \frac{\C{-2}}{\sqrt{2}\C{-1}} \langle \bar{d}^\circ_{3, \ell\, -\emm} \rangle + \langle c^\circ_{\ell\, \emm} \rangle = 2i \beta \sqrt{\frac{\pi}{3}} \delta_{l,1} \delta_{m,0} + \sqrt{\frac{2\pi}{3}} (\gamma - i\delta) \delta_{l,1} \delta_{m,1}\\
+ \sqrt{\frac{2\pi}{3}} (\gamma + i\delta) \delta_{l,1} \delta_{m,-1} \sforall \ell \geq 1,
\end{multline}
\begin{multline}\label{eq:stationary gauge Schwarzschild 6}
(-1)^\emm \frac{\C{-2}}{\sqrt{2}\C{-1}} \langle \bar{d}^\circ_{0, \ell\, -\emm} \rangle + \langle e^\circ_{\ell\, \emm} \rangle = M \frac{\C{-2}}{\sqrt{2}\C{-1}} \langle \xi^\circ_{l, \ell\, \emm} \rangle \\
+ i \sqrt{\frac{4\pi}{3}} M \dot a \delta_{\ell, 1} \delta_{\emm, 0}  \sforall \ell \geq 1,
\end{multline}
\begin{multline}\label{eq:stationary gauge Schwarzschild 7}
- \frac{1}{4} \C{-2} \left( \langle d^\circ_{1, \ell\, \emm} \rangle + (-1)^\emm \langle \bar{d}^\circ_{1, \ell\, -\emm} \rangle \right) + M \langle a^\circ_{\ell\, \emm} \rangle + 2 \langle f^\circ_{\ell\, \emm} \rangle\\
= 3 i M \sqrt{4\pi} \alpha \delta_{\ell, 0} \delta_{\emm, 0} + \sqrt{4 \pi} \dot M \delta_{\ell, 0} \delta_{\emm, 0}  \sforall \ell \geq 0.
\end{multline}
\end{subequations}
For example, we can determine $\alpha$ from the $\ell = 0$ mode of \eqref{eq:stationary gauge Schwarzschild 1}, whereas $\beta$, $\gamma$ and $\delta$ are determined from the $\ell = 1$ mode of \eqref{eq:stationary gauge Schwarzschild 5}, leading us to
\begin{equation}\label{eq:Xi coeffs mode sols}
\begin{split}
\alpha = &\  \frac{1}{\sqrt{4\pi}} \langle a^\circ_{0\, 0} \rangle,\\
\beta =&\ -i \sqrt{\frac{3}{4\pi}} \langle c^\circ_{1\, 0} \rangle,\\
\gamma = &\  \sqrt{\frac{3}{8\pi}} \left(\langle c^\circ_{1\, 1} \rangle + \langle c^\circ_{1\, -1} \rangle\right),\\
\delta = &\  i \sqrt{\frac{3}{8\pi}} \left(\langle c^\circ_{1\, 1} \rangle - \langle c^\circ_{1\, -1} \rangle\right).\\
\end{split}
\end{equation}
There are multiple ways of finding the stationary pieces of the transformation, $\langle \xi^{\circ}_{l,n,m}\rangle$ from these equations. We can first determine $\langle \xi^{\circ}_{l, \ell\, \emm} \rangle$ and $\langle \xi^{\circ}_{m, \ell\, \emm} \rangle$ from \eqref{eq:stationary gauge Schwarzschild 4} and \eqref{eq:stationary gauge Schwarzschild 3}. Then, get $\langle \xi^{\circ}_{n, \ell\, \emm} \rangle$ from \eqref{eq:stationary gauge Schwarzschild 2} by substituting $\langle \xi^{\circ}_{l, \ell\, \emm} \rangle$ into it.

We now turn to the AC equations 
\begin{subequations}
\begin{equation}\label{eq:Z soln Schwarzschild 1}
Z^\circ_{m, \omega \, \ell \, \emm} = \frac{i}{\omega} c^\circ_{\omega\, \ell\, \emm} + (-1)^{\emm + 1} \frac{i \C{-2}}{\sqrt{2}\C{-1} \omega} \bar{d}^\circ_{3,-\omega\, \ell\, -\emm} \sforall \ell \geq 1,
\end{equation}
\begin{equation}\label{eq:Z soln Schwarzschild 2}
Z^\circ_{l, \omega\, \ell\, \emm} = \frac{i}{\omega} a^\circ_{\omega\, \ell\, \emm} + \frac{i \C{-1}}{2\sqrt{2} \omega}\left( Z^\circ_{m, \omega\, \ell\, \emm} + (-1)^m \bar{Z}^\circ_{m, -\omega\, \ell\, -\emm} \right) \sforall \ell \geq 0,
\end{equation}
\begin{equation}\label{eq:Z soln Schwarzschild 3}
Z^\circ_{n, \omega\, \ell\, \emm} = b^\circ_{\omega\, \ell\, \emm} - \half (\C{-1}^2 - 1) Z^\circ_{l, \omega\, \ell\, \emm}\qquad\qquad\ \ \ \sforall \ell \geq 0.
\end{equation}
\end{subequations}
To solve these equations, we first obtain $Z^\circ_{m, \omega\, \ell\, \emm}$ from \eqref{eq:Z soln Schwarzschild 1}. Then, by substituting $Z^\circ_{m, \omega\, \ell\, \emm}$ into \eqref{eq:Z soln Schwarzschild 2} we get $Z^\circ_{l, \omega\, \ell\, \emm}$, which is then substituted 
into \eqref{eq:Z soln Schwarzschild 3} to yield the final component $Z^\circ_{n, \omega\, \ell\, \emm}$. We note that dividing by $\omega$, as in 
$\frac{i}{\omega} c^\circ_{\omega\, \ell\, \emm}$  for example, is not problematic as $\omega \to 0$ as $c^\circ_{\omega\, \ell\, \emm}$ represents 
the AC part of $c^\circ$.

\section{GHZ procedure for temporally noncompact sources}\label{sec:noncompact sources}

As formulated, the GHZ procedure applies to divergence free sources $T_{ab}$ such that the causal future of its support does not intersect a neighborhood of $\sH^-$ and $\sI^-$.
This assumption enters the construction of GHZ in the following ways:
(i) we need to find the retarded solution to the sourced (by ${\mathcal S}^{ab}T_{ab}$) Teukolsky equations for $\psi_0$, and this retarded solution makes rigorous sense a priori only for sources which vanish 
sufficently far in the past along any given null direction. (ii) While the corrector tensor $x_{ab}^-$ is obtained by forward integrations along the outgoing null curves tangent to $l^a$ starting from $\sH^-$ and is defined for sources $T_{ab}$ with or without its support excluding a neighborhood of $\sH^-$ and $\sI^-$, the proof that $2 \Re(\mathcal S_{ab}^\dagger \Phi^{-}) + x^{-}_{ab}$ is a solution of the Einstein equations with the given source relies on the support assumptions about $T_{ab}$. (Similarly, for the construction of the quantities $x^+_{ab}, \Phi^+$, we require integrations from $\sI^+$ backwards along the integral curves of $l^a$, and in this case we require sufficient decay (peeling) towards $\sI^+$ which we can expect for a source with suitable fall-off or support, but not in general.) 

For a point particle on a bound orbit which exists forever, $T_{ab}$ clearly does not fulfill the above support property. Similarly, in the puncture method, the effective source $T^\textrm{R}_{ab}$ is confined to $r_\textrm{min}<r<r_{\max}$, but it is not confined to a finite time interval. The resolution of this problem is to consider a cutoff source $T^T_{ab}$ labelled by some time $T$ far in the past such that $T^T_{ab} \to T_{ab}$ as $T \to -\infty$ pointwise in 
the exterior region $r > r_+$, and then to run the GHZ algorithm on this cutoff source. However, one potential obstacle to this type of argument is that a naive cutoff does not yield a conserved source, as required in order for there to exist any solution to the linearized Einstein equations in the first place. Thus, we must take some care when constructing the cutoff source. First, we pick some arbitrary (e.g., smooth) real-valued cutoff function $\chi_T(t)$ (here $t$ is any time function on the exterior of Kerr) which is $=1$ for $t \ge -T$ and $=0$ for $t < -T-1$. Then we consider:
\begin{equation}
\begin{split}
T_{mm}^T &= \chi_T  T_{mm}, \\
T_{nm}^T &= \chi_T T_{nm},\\
T_{ln}^T &= \chi_T T_{ln}.
\end{split}
\end{equation}
The remaining NP components of $T_{ab}^T$ are then determined by complex conjugation and the integration scheme described below in \ref{app:E}, which is obtained imposing the divergence-free condition and using the special properties of the type D background. 
Given that the spatial support of the original $T_{ab}$ is confined to 
$r_\textrm{min}<r<r_{\max}$, the nature of this integration scheme implies that the support of $T_{ab}^T$ is confined to the causal future  of that 
portion of support (e.g., the worldline) having $t>-T-1$ (or more generally of the set where $T_{mm}^T, T_{nm}^T, T_{ln}^T$ are nonzero), and that $\nabla^b T_{ab}^T=0$. In particular, 
the causal future of the support of $T_{ab}^T$ does not intersect a neighborhood of $\sH^-$ and $\sI^-$, 
meaning that we can apply the GHZ method. In fact the integration scheme implies that, on its support, 
$T_{ab}^T$ can differ from $T_{ab}$ only on that set of points 
which can be reached from the region where $\chi_T$ is not identically equal to $1$ by future directed integral curves along $n^a$ or $l^a$.
In particular, in the limit when $T \to -\infty$, $T_{ab}^T$ converges to 
$T_{ab}$ uniformly on any compact subset of $r>r_+$ (i.e., excluding $\sH^-$ and $\sI^-$). 

The GHZ scheme can be applied by construction to the cutoff source $T_{ab}^T$ and  yields perturbations
\begin{equation}
h^{\pm,T}_{ab} = 2 \Re {\mathcal S}^\dagger_{ab} \Phi^{\pm,T}  + x^{\pm, T}_{ab},
\end{equation}
where `$T$' refers to the cutoff source, and where $\pm$ refers to the solution obtained by, respectively, integrating the transport equations for the correctors $x^{\pm,T}_{ab}$ or Hertz potentials $\Phi^\pm$ inwards from $\sI^+$ respectively outwards from $\sH^-$. Due to the stated convergence properties of $T_{ab}^T$, it can be shown that $x^{\pm,T}_{ab} \to x_{ab}^\pm$ pointwise at least away from $\sH^-$ and $\sI^-$. For the Hertz potentials $\Phi^{\pm,T}$, which are obtained from integrations of the Weyl scalar $\psi_0^T$ obtained from the source $T_{ab}^T$ as the retarded solution to Teukolsy's equation, this is less clear to us but rather plausible. Indeed, convergence of $\psi_0^T \to \psi_0$ is expected  at least away from $\sH^-$ and $\sI^-$, and this statement is expected to descend to the reconstructed part, $2 \Re {\mathcal S}^\dagger_{ab} \Phi^{\pm,T}$. 

Thus the upshot of our discussion is that we can in principle (up to a mathematical proof of the above convergence statements) apply the GHZ method to the original source $T_{ab}$, with the understanding that the retarded solution $\psi_0$ sourced by ${\mathcal S}^{ab}T_{ab}$ is obtained by applying the retarded propagator to a suitably cut off source after which the cutoff is removed. At the level of modes, this procedure is expected to be equivalent (up to gauge) to the standard procedure of imposing the boundary conditions~\eqref{eq:Routdown} on each mode of the point-particle's discrete frequency spectrum, as outlined in~\cite{Pound:2021qin}, for example.

\subsection{Stress tensors of compact support}\label{app:E}

In any type-D background, the divergence of a symmetric rank 2 tensor $T_{ab}$ is given by 
\begin{equation}\label{divergence}
\begin{split}
\nabla^b T_{ab} =&+ \Big[ n_a (\th' - \rho' - \bar{\rho}') \Big] T_{ll}\\
&+ \Big[ n_a (\th - \rho - \bar{\rho}) + l_a (\th' - \rho' - \bar{\rho}') - \bar{m}_a (\tau + \bar{\tau}') - m_a (\tau' + \bar{\tau}) \Big] T_{ln}\\
&- \Big[ \bar{m}_a (\th' - \rho' - 2 \bar{\rho}') + n_a (\edth' - \tau' - 2 \bar{\tau}) \Big] T_{lm}\\
&- \Big[ m_a (\th' - 2 \rho' - \bar{\rho}') +  n_a (\edth - 2 \tau - \bar{\tau}') \Big] T_{l\bar{m}}\\
& + \Big[ l_a (\th - \rho - \bar{\rho}) \Big] T_{nn}\\
&- \Big[ \bar{m}_a (\th - 2 \rho - \bar{\rho}) + l_a (\edth' - 2 \tau' - \bar{\tau}) \Big] T_{nm}\\
&- \Big[ m_a (\th - \rho - 2 \bar{\rho}) + l_a (\edth - \tau - 2 \bar{\tau}') \Big] T_{n\bar{m}}\\
&+ \Big[ \bar{m}_a (\edth' - \tau' - \bar{\tau}) \Big] T_{mm}\\
&+ \Big[ \bar{m}_a (\edth - \tau - \bar{\tau}') + m_a (\edth' - \tau' - \bar{\tau}) - n_a (\rho + \bar{\rho}) - l_a (\rho' + \bar{\rho}') \Big] T_{m\bar{m}}\\
&+ \Big[ m_a (\edth - \tau - \bar{\tau}') \Big] T_{\bar{m}\bar{m}}
\end{split}
\end{equation}
in GHP form. 

In the context of the Kerr background, this formula can be used to construct a large class of tensors such that  $\textrm{supp} (T_{ab}) \cap \{r>r_+\}$ does not include an open neighborhood 
of ${\mathscr I}^-$ and an open neighborhood of ${\mathscr H}^-$. The construction is as follows. We first take $T_{mm},T_{nm},T_{ln},T_{m\mb}$
to be arbitrary smooth functions of compact support inside some compact set ${\mathscr S} \subset \{r>r_+\}$, so that their support in particular does not include an open neighborhood of ${\mathscr I}^-$ and an open neighborhood of ${\mathscr H}^-$.

The remaining NP components are then {\em uniquely determined} by the requirements that (i) $\nabla^a T_{ab} = 0$, (ii) $T_{ab}$ is real and symmetric, (iii) $\textrm{supp} (T_{ab}) \cap \{r>r_+\}$ does not include an open neighborhood of ${\mathscr I}^-$ and an open neighborhood of ${\mathscr H}^-$. For the proof of this claim, we note that in view of (ii) it is necessary to determine the NP components $T_{lm}, T_{ll}, T_{nn}$, with all other NP components then given by symmetry and complex conjugation. These are 4 real components which are found using the 4 conditions of (i) by integrating successively the following ODEs, with boundary conditions given by (iii).

1) By contracting the divergence  \eqref{divergence} into $n^a$, it is found that $T_{nn}$ must satisfy:
\begin{equation}
(\th-\rho-\rhb)T_{nn}=2 \Re \left\{ (\edth' - \bar{\tau} - 2 \tau')  T_{nm}+ \rho' T_{m\bar{m}} - \tfrac{1}{2}(\th' - \rho' - \bar{\rho}') T_{ln} \right\} .
\end{equation}
The right side is known by assumption. This is an ordinary differential equation along the orbits of $l^a$ for $T_{nn}$ which in view of (iii)
must be integrated with trivial initial conditions at ${\mathscr H}^-$. Thus, $T_{nn}$ satisfies (iii) and is unique.

2) By contracting the divergence  \eqref{divergence} into $m^a$, it is found that $T_{l\mb}$ must satisfy:
\begin{equation}
(\th' - \rho' - 2 \bar{\rho}') T_{lm} = - (\tau + \tab') T_{ln} - (\th - 2 \rho - \rhb) T_{nm} + (\edth - \tau - \tab') T_{m\mb} + (\edth' - \tau' - \bar{\tau}) T_{mm}.
\end{equation}
The right side is known by assumption. This is an ordinary differential equation along the orbits of $n^a$ for $T_{lm}$ which in view of (iii)
must be integrated with trivial initial conditions at ${\mathscr I}^-$. Thus, $T_{lm}$ satisfies (iii) and is unique.

3) By contracting the divergence  \eqref{divergence} into $l^a$, it is found that $T_{ll}$ must satisfy:
\begin{equation}
(\th' - \rho' - \bar{\rho}') T_{ll} = 2 \Re \left\{ (\edth' - 2 \bar{\tau} - \tau')T_{lm} + \rho T_{m\mb} - \tfrac{1}{2}(\th - \rho - \bar{\rho}) T_{ln} \right\} .
\end{equation}
The right side is known by assumption and 2). This is an ordinary differential equation along the orbits of $n^a$ for $T_{ll}$ which in view of (iii)
must be integrated with trivial initial conditions at ${\mathscr I}^-$. Thus, $T_{ll}$ satisfies (iii) and is unique.

The proof shows that if the support of $T_{mm},T_{nm},T_{ln},T_{m\mb}$ is originally chosen to be contained in some compact set 
${\mathscr S} \subset \{r>r_+\}$, then the support of the full $T_{ab}$ is contained in the set of all points that can be reached 
from ${\mathscr S}$ by a future-directed orbit of $l^a$ or $n^a$: $T_{nn}$
is obtained by integration along $l^a$ of a quantity supported in ${\mathscr S}$, and then $T_{lm},T_{ll}$ are obtained by intgration along $n^a$ of quantities not involving $T_{nn}$, which hence are also supported in ${\mathscr S}$.

Everything we said goes through for distributional $T_{mm},T_{nm},T_{ln},T_{\mb \mb}$, too.

\section{Held integration for $x_{ab}$}\label{app:integrations}
Here we utilize Held's geometric integration formalism \cite{Held:1974,Held:1975} to derive integral expressions for the corrector field $x_{ab}$ in type-D spacetimes. We then relate these results to the general solution for $x^\textrm{S}_{ab}$ as given in \eqref{eq:xS Kerr} to derive  expressions for the coefficients $a^\circ, b^\circ, c^\circ, e^\circ$ and $f^\circ$ in terms of explicit radial integrals of the NP components $T_{ab}$. We take $T_{ab}$ to have spatially compact support away from the past horizon in the radial interval $r_{\min} \leq r \leq r_{\max}$ but otherwise leave it arbitrary.  We integrate the source from the past horizon $ r_+<r_\textrm{min}$, where it vanishes, outwards along the orbits of the outgoing principal null vector $l^a$ to $r=\infty$. We focus on the string piece $x^\textrm{S}_{ab}$,  which has the general form \eqref{eq:xS Kerr} and ignore the contribution inside the source $x^{\mathrm M}_{ab}$, given already in integral form in Eqs.\eqref{eq:xmmbar Kerr},~\eqref{eq:xnm Kerr}, and \eqref{eq:xnn Kerr}.

In practice we exchange integration variables from $r$ to $\rho$ using the relations
$dr=\rho^{-2}d\rho$, $\rhb=\rho/(1+\rho\Oh)$ and $\rho_\textrm{max/min}=\rho(r_\textrm{max/min})$. Note that, as $\rho$ is complex unless $a \neq 0$, the real integration contour corresponding to 
varying $r$ along the real semi axis $r\ge r_{\rm min}$ at fixed $(u,\theta,\varphi_*)$, corresponds to a complex contour for $\rho$. The integrals below are all understood in 
terms of this particular complex contour. To derive the coefficients  $a^\circ, b^\circ, \ldots, f^\circ$ , we express both the general solution \eqref{eq:xS Kerr} and our integral forms of the solution as a Laurent polynomial in $\rho$ whose coefficints are Held scalars. Then the coefficients of $\rho$ can be easily equated in both forms, leading to algebraic relations for constants  $a^\circ, b^\circ, \ldots, f^\circ$ expressing them as source integrals. To put things into such a canonical form we simply  multiply by an appropriate factor which eliminates all positive powers of $\rhb$, and then we use the 
relation $1/\bar \rho = 1/\rho + \Oh$. 

We start with the seed equation governing the trace component $x_{m\mb}$, Eq.~ \eqref{eq:xmmb}.  Integrating \eqref{eq:xmmb} with our assumptions on $\mathrm{supp}(T_{ab})$ gives the result \eqref{eq:xmmbar Kerr}, which we recapitulate: 
\begin{align}\label{eq:xmmb ints recap}
x_{m\mb} = \frac{\rhb}{\rho}  \int_{\rho_\textrm{min}^-}^\rho
d\rho_2(1+\rho_2\Omega^\circ)  \int_{\rho_\textrm{min}^-}^{\rho_2} d\rho_1 \rho_1^{-4} T_{ll},
\end{align}
where
$\rho^-_\textrm{min} = \rho(r_\textrm{min})-0^+$.
In the source-free region $r>r_\textrm{max}$, the solution \eqref{eq:xmmb ints recap} can be rewritten as
\begin{align}\label{eq:xmmb Cs}
x^\textrm{S}_{m\mb} = \frac{C^\circ_1}{2}(\rho+\rhb) + C^\circ_3\frac{\rhb}{\rho},
\end{align}
where
\begin{subequations}
\begin{align}\label{eq:C1 C3 defs}
C^\circ_1 &= \int^{\rho_\textrm{max}^+}_{\rho_\textrm{min}^-} d\rho_1  \frac{T_{ll}}{\rho_1^{4}},  \\
C^\circ_3 &= \int^{\rho_\textrm{max}^+}_{\rho_\textrm{min}^-}
d\rho_2 (1+\rho_2\Omega^\circ) \int_{\rho_\textrm{min}^-}^{\rho_2} 
 d\rho_1  \frac{T_{ll}}{\rho_1^{4}} - C^\circ_1\rho_\textrm{max} (1+ \tfrac12 \rho_\textrm{max} \Omega^\circ),
\end{align}
\end{subequations}
where
$\rho^+_\textrm{max} = \rho(r_\textrm{max})+0^+$.
To relate $C^\circ_1$ and $C^\circ_3$ in \eqref{eq:xmmb Cs} to the coefficients $a^\circ$ and $b^\circ$ in the general solution,
\begin{equation}
  x^\textrm{S}_{m\bar{m}} = a^\circ \left( \frac{\rho}{\bar{\rho}} + \frac{\bar{\rho}}{\rho} \right) + b^\circ \left( \rho + \bar{\rho} \right),\label{eq:xmmb gen sol recap}\\  
\end{equation}
we multiply both sides of Eqs.~\eqref{eq:xmmb Cs} and \eqref{eq:xmmb gen sol recap} by the homogeneous solution $\rho/\rhb = 1+\rho\Omega^\circ$ and replace all remaining instances of $\rhb$ with $\rho$ and $\Oh$ to put the solution in canonical form,
\begin{align}
\frac{\rho}{\rhb}x^\textrm{S}_{m\mb} &= C^\circ_3  +  \frac{C^\circ_1}{2} \rho(2+\rho\Omega^\circ) \\
&= a^\circ(2+2\rho\Omega^\circ+\rho^2\Omega^\circ)+b^\circ \rho(2+\rho\Oh),
\end{align}
to find that
\begin{equation}\label{eq:ab to C1C3}
a^\circ = \half C^\circ_3, \qquad b^\circ = \half(C^\circ_1 -  \Oh C^\circ_3).
\end{equation}

Moving now to $x_{nm}$, recall that
\begin{align}\label{eq:xnm ints}
x_{nm} = 2\rho(\rhb+\rho)
\int_{\rho_\textrm{min}^-}^\rho  \frac{d\rho_2}{ \rho_2^{4} } \left(\frac{1+\rho_2  \Omega^\circ}{2+\rho_2\Omega^\circ}\right)^2\int_{\rho_\textrm{min}^-}^{\rho_2} \frac{d\rho_1} {\rho_1^{2}} \frac{2+\rho_1\Omega^\circ}{1+\rho_1\Omega^\circ}(T_{lm}+ \mathcal{N} x_{m\mb}),
\end{align}
where $\mathcal N$ is the differential operator appearing on the RHS of \eqref{eq:xnm}. In the vacuum region $r>r_\textrm{max}$, we have
\begin{equation}
x^\textrm{S}_{nm} = 2 \rho(\rho+\rhb) \left( D^\circ_3 - \frac{D^\circ_1}{6\rho^3} (1+\rho\Oh)(2+\rho\Oh)+ I_1(\rho) \right),
\end{equation}
where the coefficients $D^\circ_1$ and $D^\circ_3$ are given by the integrals
\begin{align}
D^\circ_1 & = \int_{\rho_\textrm{min}^-}^{\rho_\textrm{max}^+ } d \rho  \frac{2+\rho\Omega^\circ}{\rho^2(1+\rho\Omega^\circ)}
 (T_{lm} + \mathcal N x_{m\mb}), \\ 
D^\circ_3 &= \int_{\rho_\textrm{min}^-}^{\rho_\textrm{max}^+ }
d \rho_2\left(\frac{1 + \rho_{2} \Omega^\circ }{\rho_{2}^2 (2 +\rho_{2} \Omega^\circ)}\right)^2 \int_{\rho_\textrm{min}^-}^{\rho_2} \frac{d\rho_1}{\rho_1^2} \frac{2+\rho_1\Omega^\circ}{1+\rho_1\Omega^\circ}(T_{lm} + \mathcal N x_{m\mb} ) 
\nonumber \\
&-\frac{1+ 2\rho_\textrm{max}\Omega^\circ}{6\rho_\textrm{max}^3(2+\rho_\textrm{max}\Omega^\circ)}
D^\circ_1 
\end{align}
in the interior of the source, and
\begin{align}
I_1(\rho) &= \int_{\rho_\textrm{max}^+}^\rho  \frac{d\rho_2}{ \rho_2^{4} } \left(\frac{1+\rho_2  \Omega^\circ}{2+\rho_2\Omega^\circ}\right)^2\int_{\rho_\textrm{max}^+}^{\rho_2} \frac{d\rho_1} {\rho_1^{2}} \frac{2+\rho_1\Omega^\circ}{1+\rho_1\Omega^\circ}(\mathcal{N} x_{m\mb}) 
\end{align}
results from integration of $x_{m\mb}$ in the vacuum region $r>r_\textrm{max}$.
The integral $I_1$ is needed to determine $c^\circ$ and $e^\circ$. It has the general structure
$I_1 = \frac{(\rho-\rho_\textrm{max})^2 \rhb^2}{(\rho+\rhb)\rho^4} p_3(\rho)$,
where $p_3(\rho)$ is a third-order polynomial in $\rho$. 
The $\rhb^2$ contribution from the numerator of $I_1$ is the highest power of $\rhb$ in $x_{nm}$, and must be factored out to put the solution in canonical Laurent form. 
To compute the coefficients $c^\circ$ and $e^\circ$ we  also put the general solution \eqref{eq:xnm gen sol} in Laurent form.

After putting both the solutions in canonical form, equating coefficients of $\rho^{-3}$ and $\rho^2$ gives
\begin{align}
c^\circ &=- \frac{1}{6} \Big\{  
\frac{2\rhom^2}{1+\rhom \Omega^\circ}(\Dbar b^\circ + \Omega^\circ \Dbar a^\circ) + 2 D^\circ_1 \nonumber \\
&\ \ \ + \frac{ \tah\rhom^2(2+\rhom\Omega^\circ)^2}{(1+\rhom \Omega^\circ)^4}
\left[\rhom^2 \Omega^\circ b^\circ + \left(2+\rhom\Omega^\circ(4+\rhom\Omega^\circ)\right) a^\circ \right]
 \Big\}, \label{eq:cHeld} \\
 e^\circ &= 2D^\circ_3  -\frac{\tah (2+\rhom\Oh) }{6\rhom(1+\rhom\Omega^\circ)^4} \left(4+10\rhom\Oh+9\rhom^2\Oh{}^2+4\rhom^3\Oh{}^3 \right)a^\circ\nonumber \\
 &\ \ \ -\frac{\tah(6+16\rhom\Oh+13\rhom^2\Oh{}^2+4\rhom^3\Oh{}^3)}{6(1+\rhom\Omega^\circ)^4} b^\circ   -\frac{\Dbar b^\circ+\Oh\Dbar a^\circ }{3\rhom(1+\rhom \Oh)} \label{eq:eHeld}.
\end{align}

Lastly, for $x_{nn}$, we take on the task of computing $f^\circ$ from the final integral
\begin{equation}
x_{nn} =2 (\rho+\rhb)\int_{\rho_\textrm{min}^-}^\rho
\frac{d\rho_1}{\rho_1^4} \left(\frac{1+\rho_1\Omega^\circ}{2+\rho_1\Omega^\circ}\right)^2 \left(T_{ln} + \Re (\mathcal U x_{m\mb}) + \Re(\mathcal V x_{nm} )\right),
\end{equation}
where $\mathcal U$ and $\mathcal V$ are the differential operators on the RHS of \eqref{eq:xnn} acting on $x_{m\mb}$ and $x_{nm}$, respectively.
We first rewrite the solution as 
\begin{equation}\label{eq:xnn part}
x_{nn} =2 (\rho+\rhb) ( E^\circ_1 + I_2(\rho) ),
\end{equation}
where
\begin{equation}
E^\circ_1 =  \int_{\rho_\textrm{min}^-}^{\rho_\textrm{max}^+}
\frac{d\rho_1}{\rho_1^4} \left(\frac{1+\rho_1\Omega^\circ}{2+\rho_1\Omega^\circ}\right)^2 \left(T_{ln}+ \Re \mathcal U x_{m\mb} + \Re(\mathcal V x_{nm} )\right)
\end{equation}
 is a constant involving the stress-energy $T_{ln}$ and
\begin{equation}
I_2(\rho) = \int_{\rho_\textrm{max}^+}^\rho
\frac{d\rho_1}{\rho_1^4} \left(\frac{1+\rho_1\Omega^\circ}{2+\rho_1\Omega^\circ}\right)^2 \left( \Re \mathcal U x_{m\mb} + \Re(\mathcal V x_{nm} )\right)
\end{equation}
is determined by $x_{m\mb}$ and $x_{nm}$ in the source free region $r>r_\textrm{max}$ and must be evaluated to determine $f^\circ$.  We find that $I_2$ takes the simple form $I_2 = \frac{(\rho-\rhom)\rhb^2}{\rho^3(\rho+\rhb)} p_5(\rho)$, where $p_5(\rho)$ is an explicitly computed fifth-order polynomial in $\rho$.
We then solve for $f^\circ$ by multiplying the general solution and the particular solution by $1/\rhb^2$ to put them in canonical form and equate coefficients.  The result for $f^\circ$ is
\begin{align}\label{eq:fHeld}
f^\circ = -  \Re \Bigg\{&\frac{1}{2 \rhom^2(1+\rhom\Oh)(2+\rhom\Oh)} \Bigg( 
-2(\Dbar' c^\circ + \Dbar\bar{c}^\circ -2\Nbar a^\circ )  \nls \\
&+\left[ 4\Dbar \Dbar' a^\circ-8 \rho'{}^\circ a^\circ + 4\Nbar b^\circ +8\tabh c^\circ+8\tah \bar{c}^\circ+2\Oh(6\Nbar a^\circ-5\Dbar'c^\circ)\right]\rhom  \nonumber \nls\\
&+\Big[ 2 a^\circ (\bar\Psi^\circ+ \Psih)-8E^\circ_1 +8\Oh\left(\Dbar\Dbar a^\circ+\Nbar b^\circ-2\rho'{}^\circ a^\circ +3\tah\bar{c}^\circ+\tabh c^\circ\right)\  \nonumber \nls\\
&\ \ + 2\Oh{}^2(6\Dbar \bar{c}^\circ-4\Dbar'c^\circ+7\Nbar a^\circ) \Big] \rhom^2 \nonumber \nls\\
&+ \Big[ 2( \Psih +\bar\Psi^\circ) b^\circ + 2 \Dbar \bar{e}^\circ + 2 \Dbar' e^\circ +  4 \tah  \Dbar' b^\circ+ 4 \tabh\Dbar b^\circ +24 \tah{} \tabh a^\circ    \nonumber \nls\\ 
&\ \ +3 \Oh( a^\circ(3\bar \Psi^\circ-\Psih) - 4 \Oh E_1^\circ)   \nonumber \nls\\ 
&\ \ +4\Oh{}^2 ( \Nbar b^\circ + 2 \Dbar \Dbar' a^\circ-2\rho'{}^\circ a^\circ+ 6 \tah \bar{c}^\circ )    + \Oh{}^3(16\Dbar \bar{c}^\circ+9\Nbar a^\circ) \Big]\rhom^3 \nonumber \nls\\
&+ \Big[  8\tah \tabh b^\circ +4 \tabh e^\circ + 4 \tah\bar{e}^\circ 
+  4 \Oh \left( 6 \tah \tabh a^\circ +  (\Psih +  \tabh \Dbar + \tah \Dbar' )b^\circ+\Dbar' e^\circ  \right) \nonumber \nls\\
&\ \ + \Oh{}^2  \left(  (9\Psih-3\bar\Psi^\circ )a^\circ - E^\circ_1 \right) + 4 \Oh{}^3 \left(  \Dbar \Dbar' a^\circ +   2 \tah  \bar{c}^\circ \right)\nonumber \nls \\
& \ \ + 3 \Oh{}^4 ( \Nbar a^\circ   + 2  \Dbar \bar c^\circ ) \Big] \rhom^4 \nonumber \nls\\
 &+ 2 \Oh\left[2 \tabh(e^\circ+\tah b^\circ) + \Oh\left(2 \tah\tabh a^\circ+\Psih b^\circ + \Dbar' e^\circ \right) + \Oh{}^2 \Psih a^\circ  \right]\rhom^5 
 \Bigg)\Bigg\}. \nonumber \nls
\end{align}

In the Schwarzchild limit, where $\Oh=\tah=0$, the coefficients  simplify to
\begin{subequations}\label{eq:coeffs static vals}
\begin{align}
a^\circ &= \half \int^{\rho_\textrm{max}^+}_{\rho_\textrm{min}^-}
d\rho_2  \int_{\rho_\textrm{min}^-}^{\rho_2} 
 d\rho_1  \frac{T_{ll}}{\rho^{4}} -  \half \rho_\textrm{max} \int^{\rho_\textrm{max}^+}_{\rho_\textrm{min}^-} d\rho_1  \frac{T_{ll}}{\rho_1^{4}}, \\
b^\circ &=   \half\int^{\rho_\textrm{max}^+}_{\rho_\textrm{min}^-} d\rho_1  \frac{T_{ll}}{\rho_1^{4}},\\
c^\circ & = -\frac{1}{3}\left(\rhom^2 \Dbar b^\circ +2 \int_{\rho_\textrm{min}^-}^{\rho_\textrm{max}^+}  \frac{ d \rho}{\rho^2}
 (T_{lm} + \mathcal N x_{m\mb})\right),\\
e^\circ  &=   \int_{\rho_\textrm{min}^-}^{\rho_\textrm{max}^+}
\frac{d \rho_2}{\rho_{2}^4} \int_{\rho_\textrm{min}^-}^{\rho_2} \frac{d\rho_1}{\rho_1^2} (T_{lm} + \mathcal N x_{m\mb} ) 
+ \frac{c^\circ}{2\rhom^3} - \frac{\Dbar b^\circ}{6\rhom}, \\
f^\circ &=   \frac{1}{4}\int_{\rho_\textrm{min}^-}^{\rho_\textrm{max}^+}
\frac{d\rho_1}{\rho_1^4} \left(T_{ln}+ \Re( \mathcal U x_{m\mb} )+ \Re(\mathcal V x_{nm} )\right)- \frac{1}{4}(\Psi^\circ+\bar\Psi^\circ) (\rhom b^\circ+a^\circ) \nonumber
 \\ &\quad 
 + \frac{1}{4\rhom^2} \left(\Dbar'c^\circ + \Dbar \bar{c}^\circ - \rhom^3(\Dbar' e^\circ +\Dbar \bar{e}^\circ)\right) \nonumber
 \\ &\quad 
-\frac{1}{2\rhom^2}\left(
(\Nbar+\rhom\Dbar\Dbar'-2\rhom\rho'{}^\circ)a^\circ + \rhom\Nbar b^\circ\right).
\end{align}
\end{subequations}

\vfill
\section*{Glossary}
\label{sec:glossary}

Here we list some commonly used symbols for easier reference. 
Generally, we use the GHP operators $\th$, $\th'$, $\edth$, $\edth'$
for coordinate invariant calculations in Kerr. When making 
expansions or integrations in the NP scalar $\rho$, Held's operators $\tilde{\th}'$, $\tilde{\edth}$, $\tilde{\edth}'$ are more useful. When requiring the 
coordinate form of $\tilde{\edth}$, $\tilde{\edth}'$ acting on 
quantities oscillating as $e^{-i\omega u}$, we use the 
Chandrasekhar operators $\chand{s}{\omega}$, $\chand{s}{\omega}^\dagger$.

\begin{center}
\begin{tabular}{c|c}
Symbol & Defined in\\
\hline\hline\hline\hline
$l^a$, $n^a$, $m^a$ & Eq. \eqref{eq:Kintet up}\\
$\th$, $\th'$, $\edth$, $\edth'$ & Eq. \eqref{eq: GHP ops}\\
$\rho$, $\rho'$, $\tau$, $\tau'$, $\Psi_2$ & Eq. \eqref{eq:GHP coeffs in Kinn}\\
$\tilde{\th}'$, $\tilde{\edth}$, $\tilde{\edth}'$ & Eq. \eqref{eq:Hops}\\
$\tau^\circ$, $\rho^{\circ \prime}$, $\Psi^\circ$ & Eq. \eqref{eq:Held coeffs}\\
$\Omega^\circ$ & Eq. \eqref{eq:OmegaH}\\
$a^\circ$, $b^\circ$ & Eq. \eqref{eq:ab to C1C3}\\
$c^\circ$ & Eq. \eqref{eq:cHeld}\\
$d^\circ_i$ & Eq. \eqref{eq:d from B}\\
$e^\circ$ & Eq. \eqref{eq:eHeld}\\
$f^\circ$ & Eq. \eqref{eq:fHeld}\\
$g^\circ_{\dot{M}}$, $g^\circ_{\dot{a}}$ & Eq. \eqref{def: gdot components}\\
$\chand{s}{\omega}$, $\chand{s}{\omega}^\dagger$ & Eq. \eqref{eq:Chandop}\\
$\mathcal{E}_{ab}$ & Eq. \eqref{eq:E}\\
$\mathcal{O}$ & Eq. \eqref{eq:O}\\
$\mathcal{O}^\dagger$ & Eq. \eqref{eq:Odag}\\
$\mathcal{S}_{ab}$ & Eq. \eqref{eq:S}\\
$\mathcal{S}^\dagger_{ab}$ & Eq. \eqref{eq:Sdagger}\\
$\mathcal{T}_{ab}$ & Eq. \eqref{eq:T}\\
$\mathcal{T}^\dagger_{ab}$ & Eq. \eqref{eq:Tdag}\\
$\swsy{\ess}$ & Eq. \eqref{eq:Ys}\\
$\langle \dots \rangle$ & Eq. \eqref{Averop}\\
$u,\varphi_*$ & Eq. \eqref{eq:EF}\\
$t^a, \varphi^a$ & $\dbd{u},\dbd{\varphi_*}$\\
\end{tabular}
\end{center}

\section*{References}

\bibliography{MyReferences.bib}

\end{document}